\documentclass[PAPER, UKenglish, cernpreprint, PAPER, texmf, orcidlogo]{atlasdoc}
\usepackage{atlaspackage}
\usepackage{atlasbiblatex}

\usepackage[hepparticle=true]{atlasphysics}
\graphicspath{{logos/}{figures/}}

\usepackage{ANA-TOPQ-2023-19-PAPER-defs}

\AtlasTitle{Climbing to the Top of the ATLAS 13 TeV data}

\AtlasAbstract{%
The large amount of data recorded with the ATLAS detector at the Large Hadron Collider, corresponding to 140~fb$^{-1}$ of $pp$ collisions at a centre-of-mass energy of $\sqrt{s}=13$~TeV, has brought our knowledge of the top quark to a higher level.
The measurement of the top--antitop quark pair-production cross-section has reached a precision of 1.8\% and the cross-section was measured differentially up to several TeV in multiple observables including the top-quark transverse momentum and top-quark-pair invariant mass.
Single-top-quark production was studied in all production modes.
Rare production processes where the top quark is associated with a vector boson, and four-top-quark production, have become accessible and cross-section measurements for several of these processes have reached uncertainties of around 10\% or smaller.
Innovative measurements of the top-quark mass and properties have also emerged, including the observation of quantum entanglement in the top-quark sector and tests of lepton-flavour universality using top-quark decays.
Searches for flavour-changing neutral currents in the top-quark sector have been significantly improved, reaching branching-ratio exclusion limits ranging from $10^{-3}$ to $10^{-5}$. Many of these analyses have been used to set limits on Wilson coefficients within the effective field theory framework.
}

\author{The ATLAS Collaboration}

\AtlasRefCode{TOPQ-2023-19}

\PreprintIdNumber{CERN-EP-2024-099}

\AtlasJournal{Physics Report}
\AtlasJournalRef{Phys. Rep. 1116 (2025) 127-183}
\AtlasDOI{10.1016/j.physrep.2024.12.004}

\hypersetup{pdftitle={ATLAS document},pdfauthor={The ATLAS Collaboration}}

\begin{document}

\maketitle

\tableofcontents

%

%

%
\section{Introduction}
\label{sec:intro}

The top quark is the heaviest known elementary particle,
with a mass of about 172.5~\GeV.
Discovered at the Tevatron proton--antiproton collider at Fermilab in 1995~\cite{D0:1995jca,CDF:1995wbb}, it was the subject of a large number of measurements by the ATLAS~\cite{ATLAS:2008xda} and CMS~\cite{CMS:2008xjf} experiments at the Large Hadron Collider (LHC) during both the first data-taking phase called \RunOne, with proton--proton ($pp$) collisions at centre-of-mass energies of $\sqrt{s}=7$ and 8~\TeV, and the subsequent higher-energy phase, referred to as \RunTwo.
During \RunTwo, from 2015 to 2018, the ATLAS detector collected data from $\sqrt{s}=13$~\TeV $pp$ collisions with a total integrated luminosity of 140~fb$^{-1}$.
By climbing this mountain of data, corresponding to about 100~million produced top-quark pairs, the ATLAS Collaboration has had the opportunity to intensify its studies of the top quark,
reaching a better viewpoint for understanding its nature.

\begin{figure}[hbtp]
\centering
\includegraphics[width=0.245\linewidth]{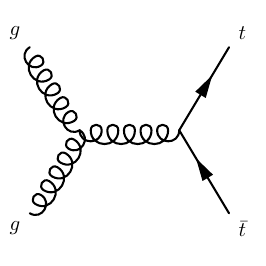}
\includegraphics[width=0.245\linewidth]{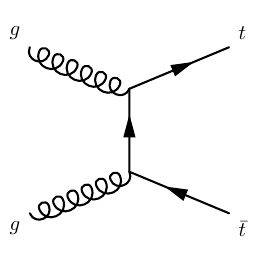}
\includegraphics[width=0.245\linewidth]{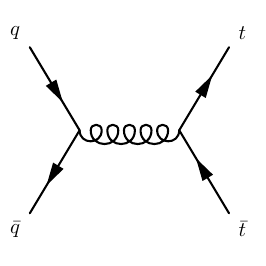}
\caption{Representative Feynman diagrams for \ttbar production at LO in QCD: gluon-initiated fusion in the $s$-channel (left) and $t$-channel (middle), and quark--antiquark annihilation (right).}
\label{fig:ttbar_diagrams}
\end{figure}

\begin{figure}[hbtp]
\centering
\includegraphics[width=0.245\linewidth]{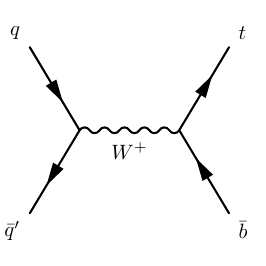}
\includegraphics[width=0.245\linewidth]{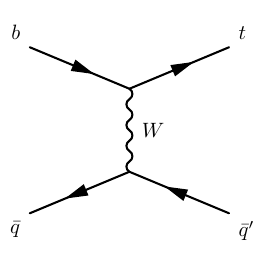}
\includegraphics[width=0.245\linewidth]{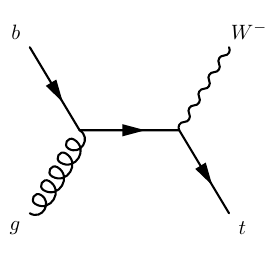}
\includegraphics[width=0.245\linewidth]{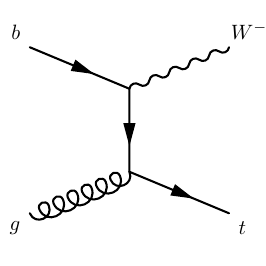}
\caption{Representative Feynman diagrams for single top-quark production at LO: from left to right, $s$-channel, $t$-channel, and the two diagrams for the $tW$ channel.}
\label{fig:singletop_diagrams}
\end{figure}

At hadron colliders, the top quark can be produced either via the strong interaction, as quark--antiquark pairs (\ttbar), or through electroweak (EW) processes, giving rise to single-top-quark events.
Strong \ttbar\ production can be initiated, at leading order (LO) in QCD, by either gluon--gluon fusion processes or quark--antiquark annihilation processes (see Figure~\ref{fig:ttbar_diagrams}), with the former dominating in $pp$ collisions at LHC energies.
Single top quarks are produced mostly through three modes,
which are labelled according to the virtuality of the $W$ boson:
the so-called $t$-channel, $s$-channel and $tW$ production modes (see Figure~\ref{fig:singletop_diagrams}).
At the LHC, single top production is subdominant relative to \ttbar production.

According to the Standard Model (SM), the top quark decays almost exclusively into a $W$ boson and a bottom quark, $b$.
The top-quark lifetime is shorter than the typical hadronisation time scale, so it does not form a bound state with other quarks.
Final-state topologies in top-quark production events are then mainly determined by the $W$-boson decay modes,
with the $b$-quark manifesting itself as a hadronic jet.
Top-quark pair-production events can give rise to three different types of final states.
The fully hadronic final state is characterised by
the production of two $b$-quarks and four light-flavour quarks,
coming from the decays of the top quarks and $W$ bosons respectively,
typically giving rise to six hadronic jets.
The dilepton final state is distinguished by two charged leptons (electrons or muons) and two undetected neutrinos produced in association with a pair of jets originating from the $b$-quarks (referred to as $b$-jets in the following).
Finally, the semileptonic final state, often referred to as single-lepton or lepton-plus-jet, typically has four hadronic jets (two of which are $b$-jets), an electron or a muon, and a neutrino. The $W$-boson decay can also lead to the production of a $\tau$-lepton that is either reconstructed through a hadronic decay or enters one of the two leptonic channels if it decays leptonically.
Analogously, $s$- and $t$-channel single-top-production events can give rise to either fully hadronic or single-lepton final states, with $tW$-production events having dilepton final states as well.

Besides these dominant production mechanisms, $pp$ collisions at the LHC
can give rise to other processes involving the production of top quarks that were mostly not observed at the Tevatron or in \RunOne.
The SM predicts the production of \ttbar\ pairs in association with photons, $W$ or $Z$ bosons, Higgs bosons or even another \ttbar\ pair, while single top production can proceed via $t\gamma$, $tZ$ or $tH$ processes, in addition to the main modes mentioned above.

The interest in studying top-quark physics at the LHC, with the benefits brought by the \RunTwo dataset size and collision energy, is manifold.
The top-quark mass is a particularly important fundamental parameter of the SM, linked to its vacuum stability~\cite{Alekhin:2012py, Espinosa:2015kwx}, and the LHC is the only place to measure it precisely.
Moreover, precise measurements of its couplings, as well as its production and decay properties, are essential in order to fully establish its nature and its role in the SM.
In addition, a number of proposed theories beyond the SM (BSM theories) predict new or modified top-quark production and decay mechanisms, resulting in altered kinematic distributions or even significant enhancements in the rates of very rare processes, such as those mediated by a flavour-changing neutral current (FCNC).
In the SM, FCNC decays such as $t\rightarrow qZ$, $t\rightarrow q\gamma$ or $t\rightarrow qH$ ($q$ being a first- or second-generation up-type quark: $u$ or $c$) are highly suppressed and below the experimental sensitivity.
Direct searches for BSM phenomena are described in Ref.~\cite{ATLAS:2024fdw}.

Experimentally, the identification and study of top-quark events with the ATLAS experiment relies not only on the reconstruction of hadronic jets, and identifying those coming from the fragmentation of $b$-quarks through dedicated $b$-tagging algorithms, but also on the identification of electrons and muons, and the measurement of the missing transverse momentum associated with the presence of undetected neutrinos.
For the all-hadronic final states, which take advantage of the larger hadronic branching fraction of the $W$ boson but are more challenging in terms of background contamination, combinations of multijet and $b$-jet triggers~\cite{TRIG-2016-01, TRIG-2012-01, TRIG-2018-08} are used, while for channels with at least one electron or muon, the online event selection is based on single-lepton triggers~\cite{TRIG-2018-01, TRIG-2018-05}.
High-performance reconstruction and identification algorithms for all these physics objects are essential ingredients for maximising the precision of top-quark measurements.
In \RunTwo, the performance of these algorithms was significantly improved relative to \RunOne, thanks to detector upgrades as well as new identification algorithms and calibration techniques.
In particular, the addition of a new innermost detector layer, the Pixel detector's Insertable B-Layer (IBL)~\cite{ATLAS-TDR-19,PIX-2018-001}, together with the adoption of new machine-learning  (ML) techniques,
allowed the $b$-tagging performance to be dramatically improved
(around 10\% efficiency increase for $b$-jets at the same light-flavour-jet rejection rate)~\cite{FTAG-2019-07,FTAG-2018-01}.
On the other hand, new calibration techniques allowed the systematic uncertainties associated with $c$-jet and light-flavour-jet rejection to be reduced~\cite{FTAG-2019-02, FTAG-2020-08}.
Similarly, improvements in the jet reconstruction algorithms and energy calibration (with about a factor of two reduction in the jet energy scale's uncertainty)~\cite{ATL-PHYS-PUB-2022-021, JETM-2018-05, JETM-2018-02} contributed to the overall gain in measurement precision, beyond that coming from the increase in sample size due to the larger integrated luminosity and production cross-sections.
At the same time, improved trigger algorithms allow the single-lepton transverse momentum thresholds to be kept at reasonable levels (below 27~\GeV) despite the increase in instantaneous luminosity.
Similarly, the introduction of new techniques to mitigate the stronger impact from additional $pp$ collisions in the same or a nearby bunch crossing (\pileup), such as the so-called jet-vertex-tagger (JVT)~\cite{PERF-2014-03}, provided the means to cope with the increased \pileup activity in \RunTwo.
Improved electron and muon identification~\cite{PERF-2017-01, EGAM-2021-01, MUON-2022-01} also provided \pileup mitigation.
Finally, innovative analysis techniques, often relying on modern ML algorithms, as well as refined Monte Carlo (MC) simulation tools were gradually introduced and adopted for the top-quark measurements and searches performed over the past years.

This report reviews a selection of the published ATLAS results using the Run~2 dataset. It is organised as follows.
Section~\ref{sec:general} describes general experimental aspects of the ATLAS \RunTwo top-quark physics analyses, such as typical event and physics object selection criteria, statistical analysis techniques and systematic uncertainties.
The following sections are each devoted to a particular set of measurements.
Section~\ref{sec:ttbar} reports the \ttbar\ cross-section measurements,
while Section~\ref{sec:single_top} describes the single-top-quark measurements.
Section~\ref{sec:associated} describes the measurements of top quarks produced in association
with a boson, as well as four-top-quark production.
Section~\ref{sec:mass} discusses the top-quark mass results and Section~\ref{sec:properties} the
determination of other top-quark properties.
Section~\ref{sec:fcnc} presents the searches for flavour-changing neutral currents in the top-quark sector, and Section~\ref{sec:eft} presents limits on Wilson coefficients within effective field theory.
Section~\ref{sec:conclusion} gives the conclusions of this report.

\section{Event selection, statistical analysis and systematic uncertainties}
\label{sec:general}

\subsection{Data samples}
A set of early \RunTwo top-quark measurements, based on the 2015 $pp$ collision dataset and corresponding to an integrated luminosity
of 3.2~\ifb, were essential for validating the updated detector and software set-ups, as well as the new Monte Carlo simulation settings for the \ttbar\ process and the applicability of the lepton and jet calibrations.
At the completion of the 2016 $pp$ data-taking, the integrated luminosity collected by ATLAS in \RunTwo had increased by a factor of ten to 36~\ifb, allowing an extensive set of new measurements in %
all the top-quark physics sectors.
With the inclusion of the 2017 $pp$ collision dataset, the integrated luminosity reached 80~\ifb.
Finally, with the addition of the data collected in 2018, the full \RunTwo dataset reached an integrated luminosity of 140~\ifb.
The quoted integrated luminosity for measurements released earlier than December 2022~\cite{ATLAS-CONF-2019-021} is 139~\ifb, and was updated to 140~\ifb in accord with the final \RunTwo 13~\TeV $pp$ luminosity measurement~\cite{ATLAS:2022hro}.
This full dataset was used to produce refined results for most of the measurements.

\subsection{Object reconstruction and event selection}
\label{sec:intro_obj}

Events containing top quarks typically produce final states including high-momentum jets, charged leptons and missing transverse momentum.

In \RunTwo, electrons and muons~\cite{ EGAM-2021-01, MUON-2018-03, MUON-2022-01} were typically required to have a transverse momentum (\pt) exceeding a threshold between 25 and 30~\GeV, and to be reconstructed within the geometrical acceptance of the inner detector (i.e.\ with absolute
pseudorapidity\footnote{ATLAS uses a right-handed coordinate system with its origin at the nominal interaction point
(IP) in the centre of the detector and the $z$-axis along the beam line. Observables labelled as transverse are projected onto the $x$--$y$ plane. The
$x$-axis points from the IP to the centre of the LHC ring, and the $y$-axis
points upwards. Cylindrical coordinates $(r, \phi)$ are used in the transverse
plane, $\phi$ being the azimuthal angle around the beam line. The pseudorapidity
is defined in terms of the polar angle $\theta$ as $\eta = -\ln\tan(\theta/2)$, and
the rapidity is defined as $y = (1/2)[(E+p_z)/(E-p_z)]$.
The angular distance $\Delta R$ is defined as
$\Delta R \equiv \sqrt{(\Delta \eta)^2 + (\Delta \phi)^2}$. The transverse momentum is $\pt = p/\cosh(\eta)$.}
$|\eta|<2.5$).
They were required to pass the identification and isolation requirements in Refs.~\cite{EGAM-2021-01,MUON-2018-03}
to improve the rejection of misidentified hadrons faking their signatures and non-prompt leptons from hadron decays or photon conversions, as well as to ensure sufficient precision in the measurement of their energy and momentum.
A multivariate discriminant was developed to further reject non-prompt leptons. This discriminant uses as input the energy
deposits and charged-particle tracks in a cone around the lepton direction, as well as lifetime variables~\cite{ATLAS:2017ztq}.

Hadronic jets were reconstructed either from calorimeter energy-deposit clusters, referred to as topological cell clusters~\cite{PERF-2014-07}, or from combined information from the calorimeters and the inner detector. The latter was assembled by a dedicated particle-flow algorithm~\cite{PERF-2015-09} used for most of the latest measurements.
The jet constituents were then clustered with the anti-$k_t$ algorithm~\cite{Cacciari:2008gp, Cacciari:2011ma}, using a radius parameter $R=0.4$, to form so-called \enquote{small-$R$ jets} (or simply \enquote{jets} in the following).
Kinematic requirements on such small-$R$ jets
include a minimum \pt\ of 20 or 25~\GeV and $|\eta|<4.5$
(although most analyses restrict jets to $|\eta|<2.5$).
A cut was imposed on the JVT output for jets reconstructed within the acceptance of the inner detector and below a \pt\ threshold of 60~\GeV, to reduce the contamination from jets not coming from the hard interaction (\pileup jets).

In addition, analyses targeting particularly high-momentum or \enquote{boosted} top quarks that decay hadronically relied on the reconstruction of larger-radius jets.
Such boosted top quarks are characterised by decay products that are highly collimated and thus difficult to reconstruct as separate small-$R$ jets.
Jets with larger-radius jets, with the $R$ parameter typically set to 1.0, were then built by either directly clustering calorimeter energy deposits or by reclustering of small-$R$ jets~\cite{Nachman:2014kla}.
Such large-$R$ jets were then tagged, relying on jet substructure variables,
and grooming procedures based on trimming or soft-drop were applied~\cite{Krohn:2009th, Dasgupta:2013ihk, Larkoski:2014wba, JETM-2018-06} to mitigate the effect of \pileup and the underlying event.

To identify jets originating from $b$-quarks, dedicated $b$-tagging algorithms were implemented and calibrated.
Several gradually improved implementations were used by ATLAS \RunTwo analyses, all relying on ML techniques based mainly on secondary-vertex reconstruction and charged-track impact parameter measurements~\cite{FTAG-2018-01,FTAG-2019-07}.
For each tagger, a certain number of working points were defined, characterised by $b$-tagging efficiencies between 85\% and 60\% and increasing rejection power against $c$-jets (between 2 and ${\sim}40$) and light-flavour jets (between 40 and ${\sim}1000$).

Missing transverse momentum in an event could indicate the production of neutrinos, which leave no signal in the detector.
It is calculated as the negative vector sum of the \pT\ of the reconstructed and calibrated objects in the event~\cite{ATLAS:2024cmj}. This sum also includes the momenta of the tracks that are matched to the primary vertex but are not associated with any other reconstructed objects.

\subsection{Statistical methods}

Many of the \RunTwo ATLAS measurements used unfolding techniques to correct the detector-level observed distributions in order to obtain results at parton or particle level.
Differential cross-sections were generally measured at particle level in fiducial phase spaces, and often extrapolated to obtain parton-level results in the full phase space as well.
Particle-level objects were defined as stable particles (those with lifetimes longer than 30~ps) produced by the MC generators, and fiducial phase spaces were defined with selection requirements close to those used to select events in the data analysis.
In contrast, parton-level results, relying on observables based on top-quark four-momenta available in MC simulation samples or fixed-order theoretical calculations, were extrapolated from the selected region to the full phase space without any experimental cuts, again by means of MC simulation.
Particle-level results are thus less affected by modelling systematic uncertainties and avoid extrapolations to unmeasured regions of phase space, while parton-level results allow direct comparisons with the most precise predictions.
Besides differential cross-sections, some of the inclusive cross-sections were also measured in fiducial phase spaces, in order to avoid extrapolating the result beyond the event topology and kinematic selection used in the analysis.

In terms of statistical analysis, most of the differential cross-section and top-quark property measurements were based on the well-established regularised unfolding technique known as iterative Bayesian unfolding (IBU)~\cite{DAgostini:1994fjx, DAgostini:2010hil}, while a small number of more recent measurements adopted fully Bayesian unfolding (FBU)~\cite{Choudalakis:2012hz} or binned profile-likelihood-based unfolding (PLU)~\cite{Blobel:2002pu}.
In the last two techniques, systematic uncertainties are encoded in the statistical model as constrained nuisance parameters~\cite{cowan1998statistical}, while in IBU the systematic uncertainties in the unfolded distributions are evaluated by repeating the unfolding procedure for each systematic variation.
Moreover, FBU and PLU facilitate the combination of several signal regions to extract a single differential cross-section, as well as the inclusion of control regions to constrain the main background contributions simultaneously with the unfolding.

For most of the inclusive cross-section measurements, as well as for the extraction of parameters controlling the shape of the signal-process distribution (such as the top-quark mass), maximum-likelihood fits were performed.
Both binned and unbinned likelihood models were used, with an increasing number of measurements being based on binned profile-likelihood fits.
Exclusion limits for BSM processes such as those induced by a FCNC were also based on binned profile likelihoods and computed using the CL$_\textrm{s}$ method~\cite{Read:2002hq} with the asymptotic approximation~\cite{Cowan:2010js}.

\subsection{Systematic uncertainties and Monte Carlo modelling}
\label{sec:systematics}

With the available dataset being larger than in \RunOne, and the consequent reduction of the statistical uncertainty in most of the measurements, systematic uncertainties became more and more important.

In terms of instrumental systematic effects, most of the top-quark measurements and searches are particularly sensitive to uncertainties in jet energy corrections and in the efficiencies of $b$-tagging algorithms.
In \RunTwo, the jet energy was corrected for \pileup effects and further calibrated, based on both MC simulation and data~\cite{JETM-2018-05}.
Uncertainties in the jet energy scale and resolution were then extracted, based on these correction procedures and on considerations of jet flavour, kinematic and generator dependence.
Efficiencies and misidentification rates for the $b$-tagging algorithms were calibrated in data by analysing \ttbar and $Z$+jet events~\cite{PERF-2016-05, FTAG-2020-08, FTAG-2019-02}.
The uncertainties in such measurements were then propagated to each analysis and decomposed into sets of uncorrelated sources of uncertainty, including uncertainties assigned to extrapolations to inaccessible kinematic regimes in the calibrations.
Depending on the specific analysis, other uncertainties may also be relevant; these include uncertainties in the LHC luminosity, beam energy and \pileup conditions, in lepton selection efficiencies and energy--momentum corrections, and in the determination of the missing transverse momentum.

Apart from these instrumental systematic effects, uncertainties in the details of the MC event-generation process, referred to as \enquote{modelling} systematic uncertainties, became increasingly important and were subjected to deeper and deeper studies.
In particular, \RunTwo ATLAS physics measurements adopted a refined approach, learning from the past data results and benefiting from improvements in MC generators~\cite{ATL-PHYS-PUB-2018-009}.

During \RunTwo, the modelling of \ttbar\ and single-top processes relied on MC generators that implement the hard process at next-to-leading order (NLO) in QCD and are interfaced with parton-shower generators to implement perturbative and non-perturbative fragmentation processes, as well as with the underlying-event modelling and hadron decays.
The nominal predictions were initially obtained with \POWHEGBOX[v2]~\cite{Alioli:2010xd} interfaced with \Pythia[6]~\cite{Sjostrand:2006za}, with the \EVTGEN~\cite{Lange:2001uf} package used to better simulate the decay of heavy-flavour hadrons.
After the first set of early \RunTwo results, the \POWPY[6] set-up was replaced by the more recent \POWPY[8]~\cite{Sjostrand:2014zea}, moving from the Perugia 2012 set of tuned parameters (tune)~\cite{Skands:2010ak} to the A14 tune~\cite{ATL-PHYS-PUB-2014-021}, derived by the ATLAS Collaboration from a number of its \RunOne measurements at $\sqrt{s} = 7$~\TeV.
To assess modelling uncertainties in these \enquote{NLO+PS} set-ups,
a number of alternative predictions were considered, obtained either by varying certain internal parameters in the \POWHEG or \Pythia generators or by replacing one of the two with a different generator.

To account for missing higher orders in the hard-process simulation, QCD scale variations were implemented in \POWHEG by scaling the renormalisation and factorisation scales up and down by a factor of two.
Initially, such scale variations were included together with variations of the \Pythia internal parameters %
controlling the amount of initial-state QCD radiation in the parton shower (specifically \enquote{Var3c} in the A14 tune~\cite{ATL-PHYS-PUB-2014-021}, and \hdamp) to constitute a so-called initial-state radiation (ISR) uncertainty source.
This was then refined~\cite{ATL-PHYS-PUB-2018-009} by splitting the ISR uncertainty source into various components in order to deal with the larger dataset and the consequent risk of artificially overconstraining various fits through insufficient flexibility of the systematic uncertainty model.
In addition, uncertainties in the amount of final-state radiation (FSR) were taken into account by varying the effective value of the strong coupling constant, $\alphas^\text{FSR}$, in the parton shower.

A separate uncertainty was then assigned to the choice of model for the parton-shower evolution and hadronisation processes, quantified by comparing samples from the nominal set-up with those generated after replacing \Pythia with the \Herwig parton shower.
These two parton-shower programs implement different radiation-emission-ordering algorithms as well as different hadronisation models.
Similarly to the case of the nominal \POWPY simulation samples, the definition of this systematic variation was gradually refined during \RunTwo, by moving from the \HERWIGpp parton shower to models based on \Herwig[7]~\cite{Bellm:2015jjp}, thereby reducing this uncertainty in most measurements.

Finally, an uncertainty related to the choice of matching scheme between the NLO hard-process generator and the parton shower, referred to as an NLO-matching-scheme systematic uncertainty, was considered.
It was typically evaluated by comparing the nominal-model samples with samples generated using \MGNLO~\cite{Alwall:2014hca} in NLO mode (referred to simply as \AMCatNLO in the following) instead of \POWHEG.\footnote{%
In some particular cases, the \Sherpa~\cite{Sherpa:2019gpd} event generator was used instead.}
Here also, the exact recipe for the systematic variation evolved during \RunTwo.
The comparison between $\AMCatNLO{+}\Pythia[8]$ and a modified \POWPY[8] set-up (suitable for a direct comparison)
was replaced by internal systematic variations within the same generator set-up, \POWPY[8]~\cite{ATL-PHYS-PUB-2023-029}.
In particular, it was broken down into two components:
one obtained by varying the parameter $p_\text{T}^\text{hard}$, regulating how the \Pythia[8] radiation phase space is determined,
and one controlling the shape of the top-quark mass distribution, accessed by comparing the nominal set-up with a set-up where the top-quark decay is handled by \MADSPIN~\cite{Artoisenet:2012st}.

Some measurements took advantage of the available higher-order predictions for top-quark kinematics, employing next-to-next-to-leading-order (NNLO) fixed-order QCD calculations, possibly combined with NLO electroweak corrections~\cite{1705.04105}, to reweight the \ttbar\ MC sample.
This was used either to define an additional uncertainty or to correct the nominal prediction (see Section~\ref{sec:ttbar_differential}).

Uncertainties in parton distribution functions (PDFs) were propagated to the analyses either by considering the envelope of the variations in the \NNPDF[3.0] PDF set~\cite{Ball:2014uwa}, or by implementing the \PDFforLHC \RunTwo procedure~\cite{Butterworth:2015oua},
which aims to capture the differences between individual PDF parameterisations as well as the internal variations of individual PDFs.
Compared to \RunOne, the adoption of these new procedures allowed a factor-of-two reduction of the PDF uncertainty impact in many \RunTwo measurements.

Uncertainties in the modelling of multiple partonic interactions (often referred to as the underlying event) and of colour-reconnection effects in top-quark-pair events were also considered for precision measurements.
These uncertainties were computed from dedicated variations in the \Pythia[8] set of tunable parameters~\cite{ATL-PHYS-PUB-2014-021} and by testing different models for colour reconnection~\cite{ATL-PHYS-PUB-2017-008,Sjostrand:2013cya,Argyropoulos:2014zoa}.

These build a comprehensive set of such systematic variations, including generator comparisons and parameter value changes that evolved over time to more detailed models.

Within the framework of modelling \ttbar\ and single-top processes with NLO+PS generators, an additional uncertainty concerns the procedure adopted to avoid double counting of diagrams between the generated \ttbar\ and $tW$ samples.
These two processes can both lead to $WbWb$ final states and therefore interfere with each other.
In \ttbar production, the two $Wb$ systems are produced on the top-quark mass shell (also called a doubly resonant process) while the $tW$ process is singly resonant.
Two approaches are usually followed to avoid double counting between doubly and singly resonant diagrams~\cite{Frixione:2008yi}:
removing NLO diagrams from the $tW$ amplitude if they overlap with \ttbar\ contributions (DR scheme) or adding subtraction terms built to cancel out doubly resonant contributions in the $tW$ amplitude (DS scheme).
The nominal $tW$ model adopted the DR scheme, while alternative $tW$ samples generated with the DS scheme were used as systematic variations.
As described in Section~\ref{sec:tW}, dedicated measurements were performed by ATLAS to assess the merits of the two alternative models,
with the goal of reducing this significant source of systematic uncertainty.

The modelling of processes other than \ttbar\ and single top production followed these recipes as closely as possible, although different generator set-ups were used for some processes, and not all systematic variations were always technically implementable~\cite{ATLAS:2020esn,ATLAS:2022uiy}.

Further improvements in MC modelling are expected in the future, including
possible refined parton-shower parameter tunings based on ATLAS \ttbar data,
the implementation of generators modelling full \ttbar\ production and decay at NLO (the so-called \enquote{bb4l} set-up~\cite{ATLAS:2021pyq})
and the employment of NNLO matrix-element generators matched to parton showers, as is done in the \textsc{MINNLO}$_{\text{PS}}$ implementation~\cite{Mazzitelli:2021mmm}, already compared with the full \RunTwo ATLAS data in the differential \ttbar cross-section measurement currently being finalised~\cite{ATLAS:2023ubf}.


\section{Top-quark pair production}
\label{sec:ttbar}

%

The unprecedented centre-of-mass energy and large $pp$ collision dataset collected in \RunTwo gave ATLAS the opportunity to exploit the full potential of the LHC as a \enquote{top-quark factory}.
The \ttbar production cross-section in $\sqrt{s} = 13$~\TeV $pp$ collisions is ${\sim}3.5$ times larger than at the highest centre-of-mass energy, 8~\TeV, in \RunOne, and the \RunTwo data sample contains over a hundred~million \ttbar pairs.
With such a large data sample, increasingly precise inclusive and differential \ttbar\ cross-section measurements became possible,
providing valuable input for PDF fits, allowing precision tests of QCD predictions, and probing phase-space regions sensitive to new physics processes.

\subsection{Inclusive top-quark pair cross-section measurements}
\label{sec:ttbar_inclusive}

Increasingly precise predictions for the total \ttbar cross-section have been produced in the past years, reaching NNLO accuracy in the strong coupling constant \alphas~\cite{Czakon:2013goa}, and including EW corrections at NLO~\cite{Bernreuther:2006vg,Kuhn:2006vh,Hollik:2007sw} as well as resummation of next-to-next-to-leading logarithmic (NNLL) soft gluon terms~\cite{Cacciari:2011hy} and even adding soft gluon corrections up to third order, resulting in approximate-N$^3$LO results~\cite{Kidonakis:2023juy}.
At $\sqrt{s}=13$~\TeV, with the top-quark mass value fixed to 172.5~\GeV,
with the strong coupling constant at the $Z$-boson mass $\alphas(M_Z)$ set to 0.118
and using the PDF4LHC21 PDFs~\cite{PDF4LHCWorkingGroup:2022cjn}, the cross-section obtained from the \TOPpp[2.0] program~\cite{Czakon:2011xx}, at NNLO in QCD and including NNLL resummation, is:
\[
834^{~+21}_{\,-30} \text{\,(scale)}^{~+21} _{\,-21} \text{\,(PDF+}\alphas\text{)~pb},
\]
where uncertainties from factorisation and renormalisation scale variations and the impact of uncertainties in the PDFs (calculated using the \PDFforLHC prescription~\cite{Butterworth:2015oua}) and $\alphas$ are quoted separately.
An additional uncertainty of $\pm 23$~pb arises from the uncertainty in the top-quark mass.\footnote{The uncertainty corresponds to a $\mp 1$~\GeV shift in the $m_t$ value set in the computation.}

The most precise measurements of the inclusive \ttbar cross-section were performed by using the cleanest final state, with an opposite-sign electron--muon pair plus at least one $b$-tagged jet.
The measurement using the 2015 dataset~\cite{TOPQ-2015-09} adopted the analysis technique inherited from the corresponding \RunOne measurements~\cite{TOPQ-2013-04}, based on the so-called $b$-tag counting method.
In order to minimise the impact of systematic uncertainties related to hadronic-jet selection (including $b$-tagging), the \ttbar cross-section was extracted simultaneously with an effective value for the $b$-jet selection efficiency by taking as input the numbers of events with exactly one and exactly two $b$-tagged jets.
The cross-section was determined by solving a system of two equations with two unknowns.
The event selection requirements were kept as loose as possible, in order to minimise the impact of many systematic uncertainties.
Finally, by restricting the event topology to opposite-sign $e\mu$ pairs, the contribution from background processes was minimised, with $Z$+jet, $W$+jet, $t$- and $s$-channel single-top, and diboson production contributing very little, and single-top $tW$ events being almost the only ones contributing significantly.
This first measurement, already achieving a precision of 4.4\%, was updated with the larger 36~fb$^{-1}$ dataset~\cite{TOPQ-2018-17}, lowering the uncertainty to 2.4\%, and again with the full \RunTwo dataset~\cite{TOPQ-2018-26}, reaching
an outstanding
precision of 1.8\%.
The last improvement comes almost entirely from the decrease in the luminosity uncertainty~\cite{ATLAS:2022hro}, resulting from improvements in the ATLAS $pp$ luminosity calibration transfer and long-term stability analyses, as well as changes to the van~der~Meer calibration procedures.
The measured values, and the corresponding statistical and systematic uncertainties, are reported in Figure~\ref{fig:ttbarXsec}, together with the other ATLAS \RunTwo inclusive \ttbar cross-section measurements performed in other channels.
All the measurements agree with the theory prediction within the experimental uncertainty.
The statistical uncertainty was already subdominant with the 2015 dataset, despite the small acceptance for the $e\mu$ channel.
Among the systematic uncertainties, the most important contributions were those from uncertainties in the luminosity determination, LHC beam energy, and \ttbar process modelling.
In addition to the reduction in the luminosity uncertainty,
the beam energy uncertainty was also reduced significantly since the first \RunTwo measurement, already reaching a precision of 0.1\% for the measurement based on the 2015+2016 dataset~\cite{Todesco:2017nnk}.
Among the \ttbar process modelling systematic uncertainties, those associated with the choice of parton-shower and hadronisation model constitute some of the most important sources of uncertainty for all the measurements,
even with the most up-to-date tuned \Pythia[8] and \Herwig[7] generators.

\begin{figure}[!tbp]
\centering
\includegraphics[width=0.75\textwidth]{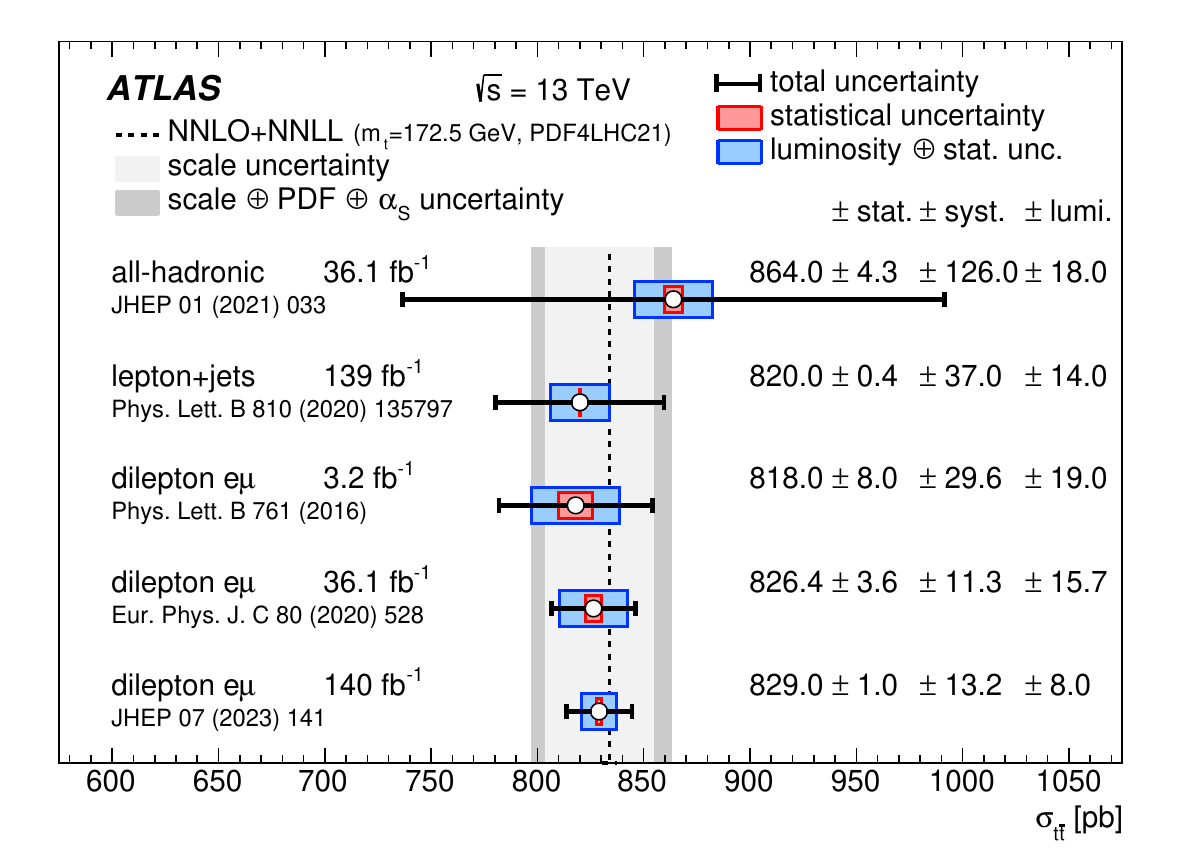}
\caption{Summary of ATLAS inclusive \ttbar cross-section measurements at $\sqrt{s}=13$~\TeV based on \RunTwo data,
compared with the exact NNLO QCD calculation complemented with NNLL resummation (\TOPpp[2.0]),
obtained using the PDF4LHC21 PDF set and
with the QCD renormalisation and factorisation scales set to $\mu_R = \mu_F = m_t = 172.5$~\GeV and the strong coupling constant $\alphas(M_Z)=0.118$.
The theory band represents uncertainties due to renormalisation and factorisation scales, parton distribution functions and the strong coupling constant.
The uncertainties in the experimental measurements are broken down into their statistical and systematic components, quoting the uncertainty related to the integrated luminosity separately.
The measurements and the theory calculation are quoted for $m_t=172.5$~\GeV.
}
\label{fig:ttbarXsec}
\end{figure}

These inclusive \ttbar cross-section measurements were used as one of the inputs for high-precision determinations of ratios of \ttbar and $Z$ production cross-sections at the three centre-of-mass energies where ATLAS measurements are available: $\sqrt{s}=13, 8, 7$~\TeV.
This was already done with the first measurement based on 3.2~fb$^{-1}$ of \RunTwo data~\cite{STDM-2016-02}: single ratios, at a given $\sqrt{s}$ for the two processes and at different $\sqrt{s}$ values for each process, as well as double ratios of the two processes at different $\sqrt{s}$ values, were evaluated and then compared with NNLO calculations using recent PDF sets, demonstrating significant power to constrain both the gluon distribution function for Bjorken-$x$ values near 0.1 and the light-quark sea for $x<0.02$.
The second \ttbar cross-section paper, based on 36~fb$^{-1}$ of \RunTwo data~\cite{TOPQ-2018-17}, included updated computations of these ratios and double ratios of \ttbar and $Z$ cross-sections at different energies.
This inclusive \ttbar cross-section measurement was also used to extract the top-quark mass with an uncertainty of approximately 2~\GeV, as detailed in Section~\ref{sec:mass}.

The inclusive \ttbar cross-section was also measured in other final states, despite not reaching the same precision.
In particular, the total cross-section was measured in the lepton-plus-jet channel~\cite{TOPQ-2020-02}, by a simultaneous profile-likelihood fit of three different binned observables in three different event categories, characterised by different numbers of jets and $b$-tagged jets.
An additional uncertainty, estimated as the difference between the results obtained with the nominal MC generator and a sample reweighted to the NNLO~(QCD) +~NLO~(EW) parton-level prediction~\cite{1705.04105}, was applied (as discussed in Section~\ref{sec:ttbar_differential}).
Moreover, an inclusive \ttbar cross-section was measured in all-hadronic final states~\cite{TOPQ-2018-18}, but with significantly larger uncertainties.
These results are reported and compared with those obtained in the dilepton channel in Figure~\ref{fig:ttbarXsec}.
As can be seen, all the measurements are in good agreement with each other and with the theoretical predictions, with the precision of the experimental determination exceeding that of the QCD NNLO+NNLL computation.

Inclusive \ttbar cross-section measurements were performed by ATLAS with similar techniques at the different LHC $pp$ centre-of-mass energies.
In \RunOne, measurements were performed in both the $e\mu$ dilepton channel~\cite{TOPQ-2013-04} and the lepton-plus-jet channel~\cite{TOPQ-2017-08, TOPQ-2018-40} at $\sqrt{s} = 7$ and 8~\TeV.
The inclusive \ttbar cross-section was also measured by ATLAS at $\sqrt{s} = 5.02$~\TeV, analysing the ${\sim}260$~pb$^{-1}$ of $pp$ data collected during \RunTwo at this reduced centre-of-mass energy in low-\pileup conditions.
This measurement was obtained by combining the lepton-plus-jet and dilepton channels~\cite{TOPQ-2018-40}.
Finally, at the new energy of $\sqrt{s} = 13.6$~\TeV achieved in \RunThr, a measurement was made in the $e\mu$ dilepton channel using the first 29~fb$^{-1}$ of collected data~\cite{TOPQ-2023-21}.
In Figure~\ref{fig:ttbarXsecVsCME}, all these measurements, together with the most precise ones performed at $\sqrt{s} = 13$~\TeV, are shown and compared with QCD NNLO+NNLL predictions as a function of the centre-of-mass energy.
Agreement between the various measurements and the theoretical predictions is remarkable over the whole range of centre-of-mass energies.

\begin{figure}[!tbp]
\centering
\includegraphics[width=0.75\textwidth]{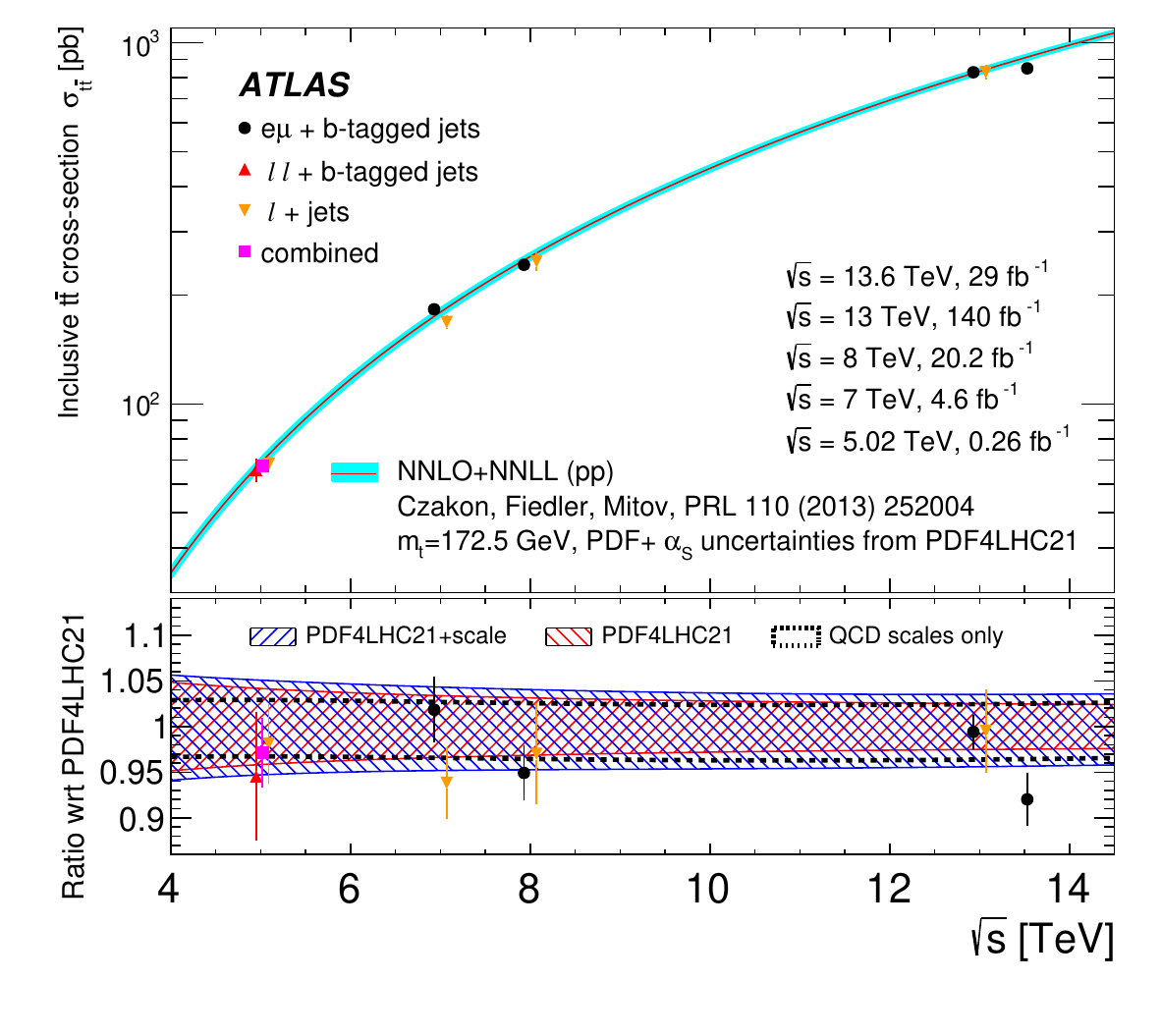}
\caption{
Summary of ATLAS measurements of the top-quark pair production cross-section as a function of the centre-of-mass energy and comparison with the NNLO QCD calculation complemented with NNLL resummation,
as obtained with \TOPpp[2.0] using the PDF4LHC21 PDF set, with the renormalisation and factorisation scales set equal to the top-quark mass and the strong coupling constant set to $\alpha_s(M_Z) = 0.118$.
The theory band represents uncertainties due to renormalisation and factorisation scales, parton distribution functions and the strong coupling constant.
The measurements and the theory calculation are quoted for $m_t = 172.5$~\GeV.
Measurements made at the same centre-of-mass energy are offset slightly for clarity.
The figure was originally published in Ref.~\cite{TOPQ-2023-21}.
}
\label{fig:ttbarXsecVsCME}
\end{figure}

\subsection{Differential top-quark pair cross-section measurements}
\label{sec:ttbar_differential}

Besides the inclusive measurements,
ATLAS used the \RunTwo data to measure a large variety of differential \ttbar cross-sections in different final states and kinematic regimes.
Shortly after the first inclusive cross-section measurement at $\sqrt{s}=13$~\TeV, differential distributions were measured, with the same 3.2~fb$^{-1}$ dataset, in both the dilepton channel~\cite{TOPQ-2015-17, TOPQ-2016-04} and the lepton-plus-jet channel~\cite{TOPQ-2016-01, TOPQ-2017-01}.
With the larger 36~fb$^{-1}$ dataset, a new set of measurements were performed, increasing the number of measured observables and improving the precision in the dilepton~\cite{TOPQ-2018-17} and lepton-plus-jet~\cite{TOPQ-2018-15} topologies, as well as including all-hadronic final states~\cite{TOPQ-2016-09, TOPQ-2018-18}.
Finally, the total \RunTwo dataset was used to refine and update the differential cross-section measurements
in all three channels~\cite{TOPQ-2019-23, TOPQ-2018-11, TOPQ-2018-26}.

Measurements in the different channels can benefit from different opportunities offered by specific topologies.
The \ttbar dilepton channel was used to achieve high-precision measurements as well as to study properties of \ttbar production in an environment characterised by lower hadronic activity than in the other channels.
In Ref.~\cite{TOPQ-2015-17}, quantities sensitive to jet activity in the \ttbar process were measured in events with an $e\mu$ opposite-sign lepton pair and two $b$-tagged jets.
These include the multiplicity of additional jets, their \pt distributions, as well as the so-called \enquote{gap fraction}, i.e.\ the fraction of signal events not containing additional jets in a given rapidity region.
All these quantities are highly sensitive to details of the parton-shower models and it is therefore important to measure them in order to validate the MC generator predictions.
The gap fraction was measured as a function of the \pt threshold for additional jets and for different invariant-mass regions of the $e\mu b\bar{b}$ system.
In Ref.~\cite{TOPQ-2016-04}, using the same sample of \ttbar events, differential cross-sections were measured as a function of the transverse momentum and absolute rapidity of the top quark, and of the transverse momentum, absolute rapidity and invariant mass of the \ttbar system.
These measurements, unfolded to parton level, provide valuable input for testing higher-order QCD predictions.
In order to perform the measurements, the top-quark pair four-momenta were reconstructed using the Neutrino Weighting technique~\cite{D0:1997pjc}, which uses top-quark and $W$-boson mass constraints to infer the two neutrinos' kinematics from only a measurement of the sum of their transverse momenta, given by the event's missing transverse momentum \met.
The most recent measurements in the dilepton $e\mu$ channel~\cite{TOPQ-2018-17, TOPQ-2018-26}, published together with the corresponding inclusive cross-section determinations, followed an alternative and complementary strategy.
No attempt was made to reconstruct the top quarks from their partially invisible decay products, and instead only leptonic kinematic variables were unfolded, at particle level in a fiducial phase space, both individually and as double-differential distributions, resulting in very clean and precise measurements.
Such an approach was pursued previously in \RunOne data~\cite{TOPQ-2015-02}, allowing an indirect determination of the top-quark mass.
The technique used for the inclusive measurement was applied to extract the cross-section in each of the bins.

The lepton-plus-jet and all-hadronic channels offered the advantage of easier reconstruction of the top-quark four-momenta, thanks to the presence of at most one neutrino among the \ttbar decay products.
In addition, they offered the opportunity to probe extreme kinematic regimes by providing a larger number of events and the possibility of adopting \enquote{boosted top-tagging} techniques to reconstruct highly collimated, hadronically decaying top quarks.
In particular, the all-hadronic channel, despite being affected by larger backgrounds, allows full reconstruction of the kinematics of both top quarks in the event without relying on a measurement of the missing transverse momentum.
Moreover, it ensures a higher selection efficiency for high-energy events, where both top quarks have a large boost and their decay products are collimated.\footnote{Dilepton and lepton-plus-jet channel measurements typically need to identify at least one isolated electron or muon,
while leptons in boosted topologies tend to be produced at small angular separations from the $b$-jet coming from the same parent top quark,
resulting in non-isolated charged-lepton signatures, which are hard to identify correctly in hadronic environments.}
Differential cross-sections were measured in the lepton-plus-jet channel in both the \enquote{resolved}~\cite{TOPQ-2016-01, TOPQ-2017-01, TOPQ-2018-15} and \enquote{boosted} topologies~\cite{TOPQ-2016-01, TOPQ-2019-23},
as well as in the all-hadronic channel for resolved~\cite{TOPQ-2018-18} and boosted regimes~\cite{TOPQ-2016-09, TOPQ-2018-11}.
A large number of distributions were unfolded at particle level to fiducial phase spaces, and at parton level to the total phase space.
These distributions included top-quark transverse momentum and rapidity ($y$); \ttbar system invariant mass, \pt and $y$; and other observables related to top-quark kinematics, initial- and final-state radiation emission and the PDFs.
Double-differential distributions were also extracted, in both the resolved and boosted topologies.
Top-quark four-momenta were obtained in boosted topologies by reconstructing large-$R$ jets, with possible identification using top-tagging techniques.
For the resolved topology, top-quark kinematics were retrieved in different ways in the different measurements.
In the lepton-plus-jet channel, the so-called pseudo-top algorithm~\cite{TOPQ-2013-07} was used; it relies on invariant-mass constraints from the known mass of each of the two $W$ bosons and on angular distances between jets and leptons, allowing the hadronically and leptonically decaying top quarks to be reconstructed with very similar efficiency at both detector level and particle level.
Alternatively, measurements performed at parton level used the so-called KLFitter~\cite{Erdmann:2013rxa} package, which provides a constrained fitting algorithm that uses transfer functions to relate the energies of the reconstructed objects to those of the parton-level objects.
In the all-hadronic channel, a $\chi^2$ minimisation, relying on top-quark and $W$-boson mass constraints, was used to reconstruct top quarks in the resolved topology.
In both the lepton-plus-jet and dilepton channels, cross-sections were measured with a precision of about 10\%--20\% for top-quark \pt up to ${\sim}1.5$~\TeV and $m_{\ttbar}$ up to ${\sim}3$~\TeV.

All these measurements were compared with a large set of predictions.
Particle-level results were used to test MC predictions from various generator set-ups, while parton-level ones could also be compared with fixed-order calculations.
The early measurements turned out to be very useful for testing and validating MC generator set-ups for \ttbar production.
In particular, it became possible to rule out certain set-ups because they gave a poor description of the data:
for instance, the prediction from \POWHEGBOX interfaced with \HERWIGpp gave a poor description of most of the inspected distributions, especially the additional-jet multiplicity, and was later replaced with a prediction making use of the more recent \Herwig[7] program.
Measurements based on larger datasets started to systematically reveal limitations in the modelling of certain kinematic distributions by most of the MC generators.
The modelling of the top-quark transverse momentum as well as that of the \ttbar system was found to be particularly poor,
with most of the NLO+PS predictions, including those from the most recent refined MC generator set-ups, overestimating the cross-section at high momenta, especially in boosted topologies (see Figure~\ref{fig:ttbar_diff}~(a)).
Agreement was improved when comparing parton-level measurements with NNLO QCD fixed-order predictions or by reweighting the MC samples to match these predictions for the top-quark kinematics, as can be seen in Figure~\ref{fig:ttbar_diff}~(b).

\begin{figure}[!tbp]
\centering
\subfloat[]{\includegraphics[width=0.49\textwidth]{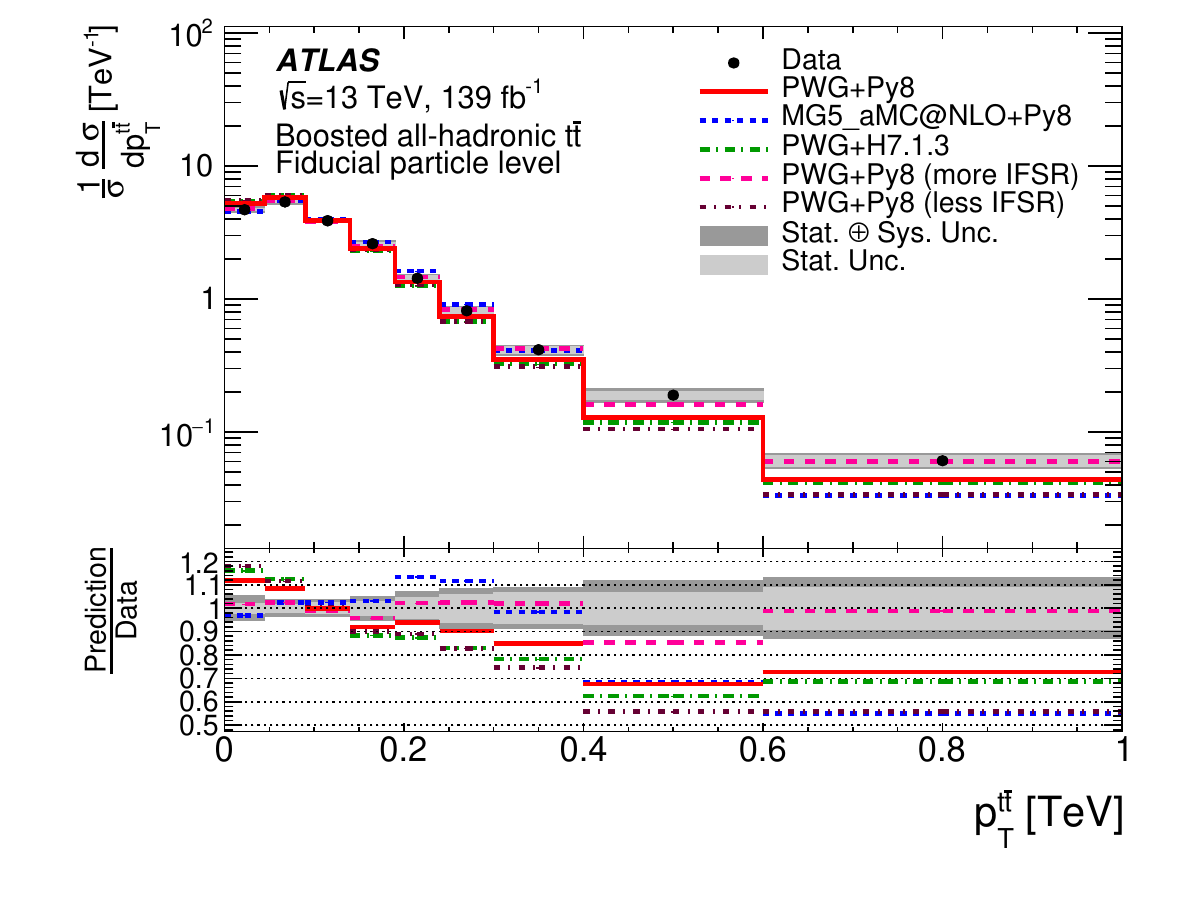}}
\subfloat[]{\includegraphics[width=0.47\textwidth]{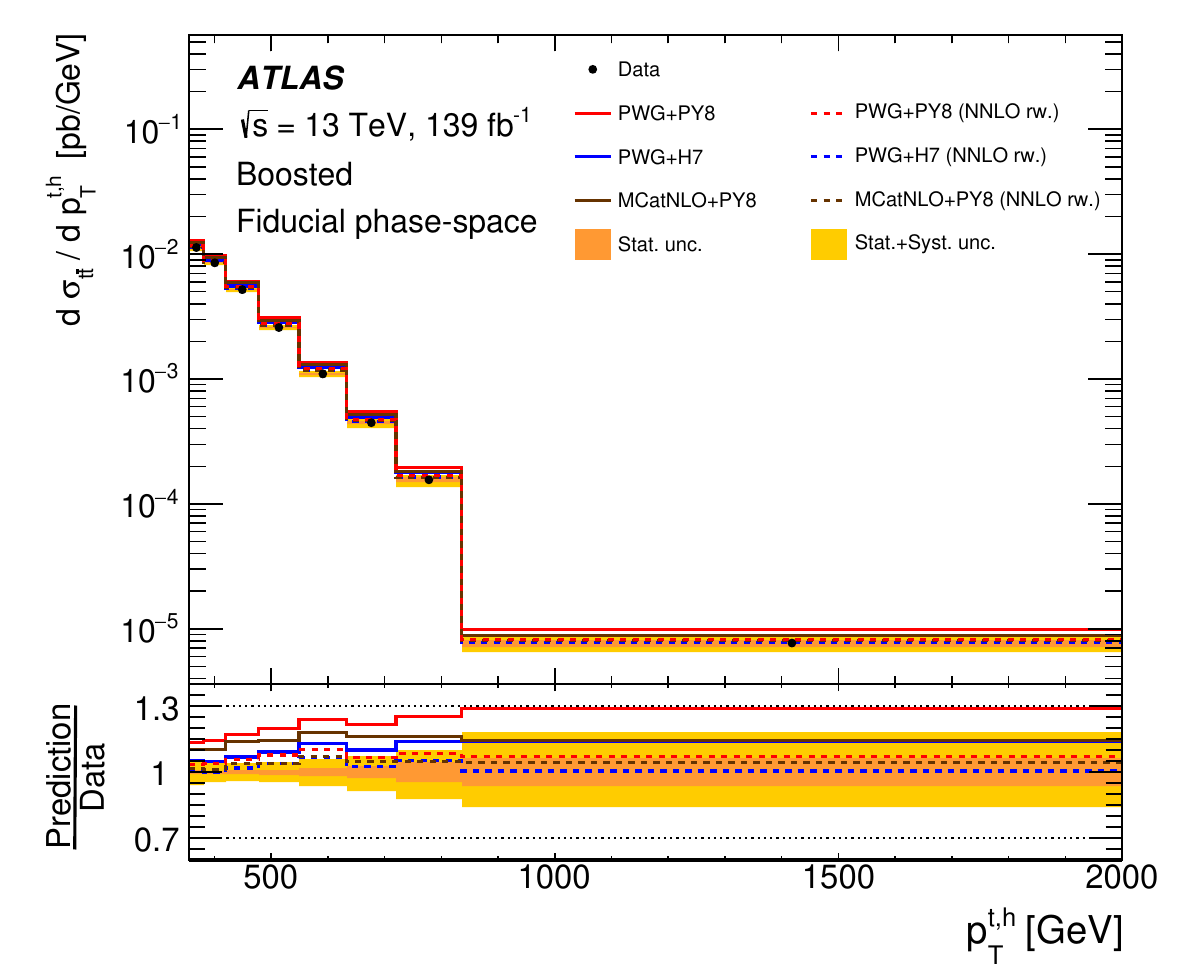}}
\caption{
Two examples of ATLAS differential \ttbar cross-section measurements.
(a) Normalised particle-level \ttbar-system transverse momentum distribution extracted in all-hadronic boosted final states, compared with a number of NLO+PS MC predictions~\cite{TOPQ-2018-11}.
(b) Particle-level transverse momentum distribution of hadronically decaying top quarks  measured in the lepton-plus-jet boosted channel, compared with MC predictions, including predictions where a reweighting of top-quark kinematics is performed to match higher-order fixed-order calculations (see text for more details)~\cite{TOPQ-2019-23}.}
\label{fig:ttbar_diff}
\end{figure}

\subsection{Studies of $b$-jet production in top-quark-pair events}

In addition to inclusive and differential \ttbar cross-section extractions, measurements
of the \ttbar production process with extra $b$-jets were performed with \RunTwo data.
Cross-sections for $\ttbar+\bbbar$ production were measured~\cite{TOPQ-2017-12} with the 36~fb$^{-1}$ dataset.
Inclusive and differential cross-sections were measured in the single-lepton and dilepton channels, specifically for \ttbar events with additional $b$-quark jets defined at particle level in dedicated fiducial phase spaces.
These measurements were compared with predictions from various MC generator set-ups, and validated the background models for measurements of $\ttbar H$ production in the $H\rightarrow \bbbar$ decay channel as well as for new-physics searches in similar final states.
Figure~\ref{fig:tt_bjet}~(a) shows the results of the measurement, in terms of inclusive fiducial cross-sections for the two channels, separately for fiducial phase spaces with ${\geq}3$~$b$-jets or ${\geq}4$~$b$-jets.
These measured inclusive cross-sections are higher than the predictions from the dedicated $\ttbar+b\bbar$ NLO generators matched to parton-shower programs,
both using the so-called five-flavour scheme (with all the quark flavours except the top included in the proton PDFs) and the four-flavour one (with only the four lightest quarks included in the proton PDFs),
but still within the uncertainties.
The differential distribution comparisons do not show significant mismodelling by most of these generators,
beyond the experimental uncertainties of the unfolded results.

\begin{figure}[!tbp]
\centering
\subfloat[]{\includegraphics[width=0.59\linewidth]{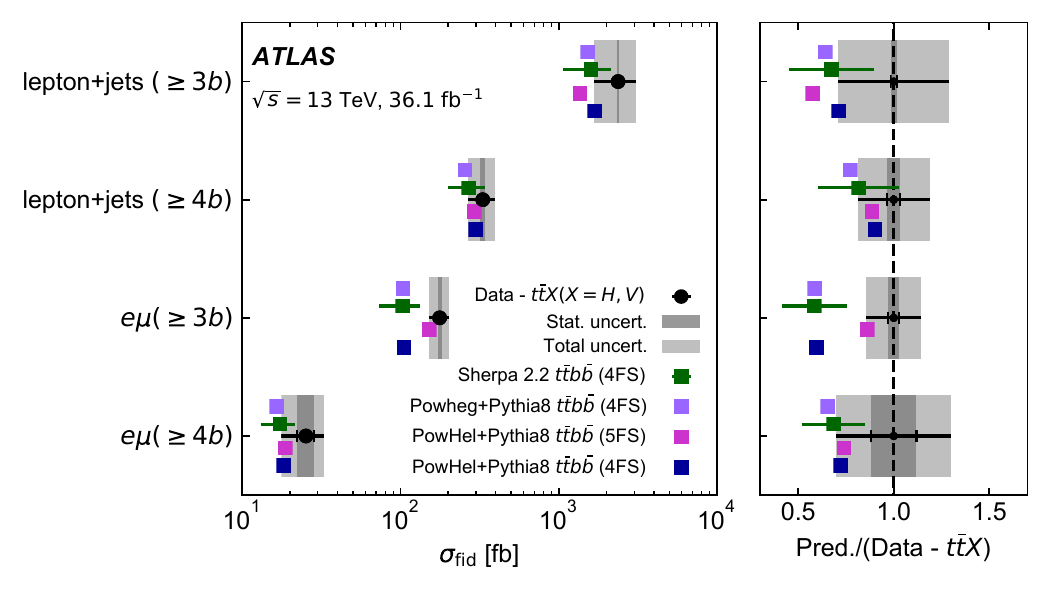}}
\subfloat[]{\includegraphics[width=0.40\linewidth]{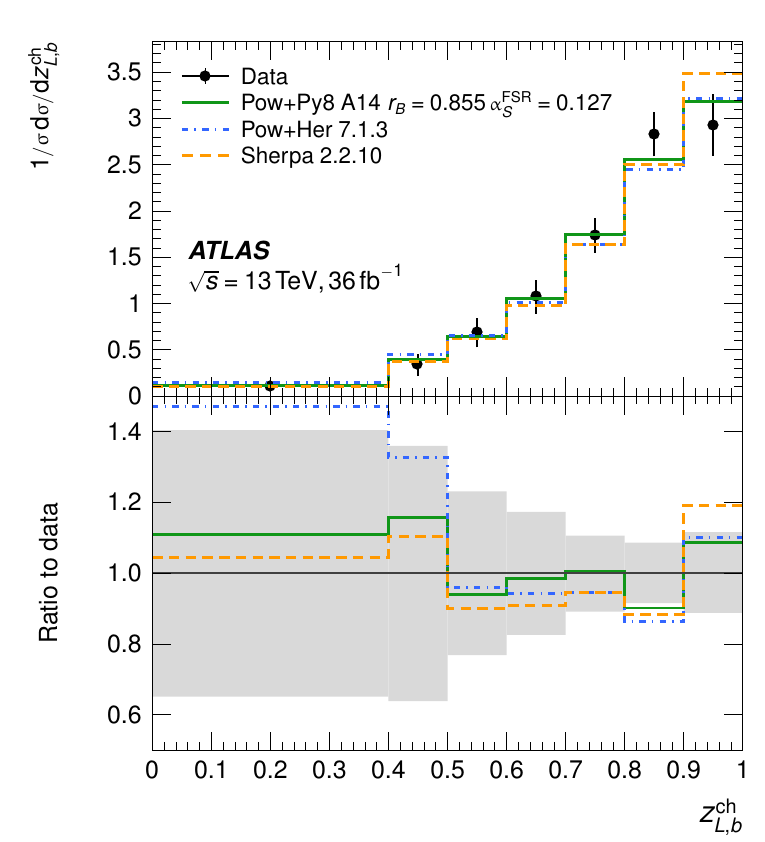}}
\caption{(a) Measured fiducial $\ttbar$\,+\,$b$-jet cross-sections~\cite{TOPQ-2017-12}, with $\ttbar H$ and $\ttbar$-plus-vector-boson contributions subtracted from the data, compared with predictions from different generator set-ups.
(b) Particle-level differential cross-section as a function of $z^{\textrm{ch}}_{\textrm{L},b}$~\cite{TOPQ-2017-19}, compared with predictions from different generator set-ups.
The quantity $z^{\textrm{ch}}_{\textrm{L},b}$ is defined as the fraction of the total charged-particle momentum in a jet carried by the $b$-hadron decay products along the direction of the jet.}
\label{fig:tt_bjet}
\end{figure}

Finally, the same dataset was used to study $b$-jets in \ttbar events~\cite{TOPQ-2017-19}.
A number of observables sensitive to quark fragmentation were measured in jets identified as originating from top-quark decays, thus providing an almost completely pure $b$-quark jet sample.
Top-quark-pair events were selected in the dilepton $e\mu$ channel, and the set of charged-particle tracks associated with jets were separated into those from the primary $pp$ interaction vertex and those from the displaced $b$-hadron decay secondary vertex, in order to construct observables characterising the longitudinal and transverse momentum distributions of $b$-hadrons within $b$-quark jets.
These measurements complement the measurements performed at $e^+e^-$ colliders, in which the $b$-quarks originate from a colour-singlet ($Z/\gamma^*$), allowing the universality of the fragmentation models to be tested.
Figure~\ref{fig:tt_bjet}~(b) shows one of the measured distributions, compared with various MC predictions.
Despite being in overall agreement with the models tuned at $e^+e^-$ colliders, the measured distributions are still affected by large experimental uncertainties or limited by the amount of data in several bins, preventing the current results from constraining the MC models.


%
\section{Single-top-quark production}
\label{sec:single_top}

In the SM, the dominant single-top-quark production process at hadron colliders occurs through the $tWb$ vertex
according to three distinct diagrams at LO in QCD (see Figure~\ref{fig:singletop_diagrams}): the exchange of a virtual $W$ boson in the
$t$-channel, a $W$-boson exchange in the $s$-channel,
and the associated production of a top quark and a $W$ boson (named $tW$).
In $pp$ collisions at LHC energies, the $t$-channel process is dominant.
The cross-section for each of the three single-top-quark production channels is sensitive to the coupling between the $W$ boson and the top quark at the $Wtb$ vertex.
Single top-quark production therefore presents an opportunity to test the structure of this coupling in the SM, as well as to probe classes of new-physics models that can affect the $Wtb$ vertex.
The different single-top production modes are sensitive to different BSM models, so it is important to study them separately~\cite{Tait:2000sh,Cao:2007ea}.

\subsection{Measurements in the $t$-channel}

In the $t$-channel, a light-flavour valence quark from one of the colliding protons interacts with a $b$-quark which can originate from the proton sea (in the five-flavour scheme) or from gluon splitting (in the five-flavour scheme).
After the exchange of a space-like virtual $W$ boson,
the produced top quark or antiquark recoils against a light-flavour quark, referred to as the spectator quark. This quark is preferentially emitted in the forward direction.
Since the density of valence $u$-quarks in the proton is about twice as large as that of valence $d$-quarks,
the single-top-quark production cross-section, $\sigma(tq)$, is about twice as large as the single-top-antiquark production cross-section, $\sigma(\bar{t}q)$.
For $pp$ collisions at $\sqrt{s}=13$~\TeV,  the predicted $t$-channel production cross-sections (for $m_t = 172.5$~\GeV) are
$\sigma(tq) = 134.2 \pm 2.2$~pb and $\sigma(\bar{t}q) = 80.0 \pm 1.6$~pb, computed at NNLO in perturbative QCD with the MCFM program~\cite{Campbell:2020fhf}.
This corresponds to an increase by around a factor of 2.5 compared to the SM $t$-channel cross-sections at $\sqrt{s}=8$~\TeV.
Measurements of $t$-channel production were made first with 3.2~\ifb\ of \RunTwo data~\cite{ATLAS:2016qhd} and later with the full dataset~\cite{ATLAS:2024ojr}.
Separate measurements of top-quark or top-antiquark production were conducted because they provide sensitivity to different PDFs (the $u$-quark and $d$-quark PDFs).
Measurements of their ratio, $R_t = \sigma(tq)/\sigma(\bar{t}q)$, also profit from systematic uncertainties partially cancelling out, allowing even higher sensitivity to the PDFs.
Events were required to have a single isolated electron or muon from the leptonic decay of the top quark or antiquark, and exactly two jets, among which exactly one is $b$-tagged.
The $t$-channel signal was separated from the background by using a neural network (NN).
A binned maximum-likelihood fit to the NN discriminant distribution in the channel with positively or negatively
charged leptons was performed to extract the top-quark or top-antiquark inclusive cross-section, respectively.
The fits yield $\sigma(tq) = 137 \pm 8$~pb,
$\sigma(\bar{t} q) = 84^{\,+6}_{-5}$~pb
and $R_t = 1.636 ^{\,+0.036}_{-0.034}$,
in agreement with SM predictions.

Because of the vector minus axial-vector $(V - A)$ structure of the $Wtb$ vertex,
the single top quarks are highly polarised
along the direction of the momentum of the spectator quark (or opposite to it in the case of
single top antiquark production).
ATLAS used the full \RunTwo dataset and $t$-channel single top production
to probe the polarisation of the top quark and top antiquark~\cite{ATLAS:2022vym}.
The event selection was similar to that used to measure the $t$-channel production cross-section.
Among the two required jets, the non-$b$-tagged jet was assumed to originate from the spectator quark.
The charged lepton from the top-quark decay is the most sensitive probe of the top-quark spin, so
the angular distributions of the charged leptons were used to extract the components of the polarisation vectors.
The polarisation vector was expressed in three orthogonal directions, where the $z'$ direction was chosen to be the momentum direction of the spectator quark~\cite{Aguilar-Saavedra:2014eqa}.
The $W$ boson from the top-quark decay was reconstructed from the lepton kinematics and the reconstructed missing transverse momentum, imposing
a $W$-boson mass constraint. The top-quark candidate was then reconstructed by combining the four-momentum of the $W$ boson with
that of the $b$-tagged jet. Finally, the charged lepton's momentum was boosted into the top-quark rest frame to define
its polar angles in the polarisation coordinate system.
The differential distributions of these polar angles could then be used to extract the three polarisation components $P_{x'}$, $P_{y'}$ and $P_{z'}$ of the top-quark and top-antiquark polarisation vectors,
using a template fit to these distributions in the top-quark rest frame.
The results demonstrate a very high degree of polarisation in $t$-channel production, along the direction of the spectator quark (for top-quark events), or opposite to that direction (for top-antiquark events), in agreement with NNLO QCD predictions.
The polar-angle differential distributions were also unfolded to particle level in a fiducial region.
The normalised unfolded distributions show good agreement with the SM prediction, with a $p$-value close to 1 (see Figure~\ref{fig:singletop}~(a)),
and were used to derive competitive bounds on anomalous $tW$ couplings (see Section~\ref{sec:eft}).

\begin{figure}[!tbp]
\centering
\subfloat[]{\includegraphics[width=0.45\linewidth]{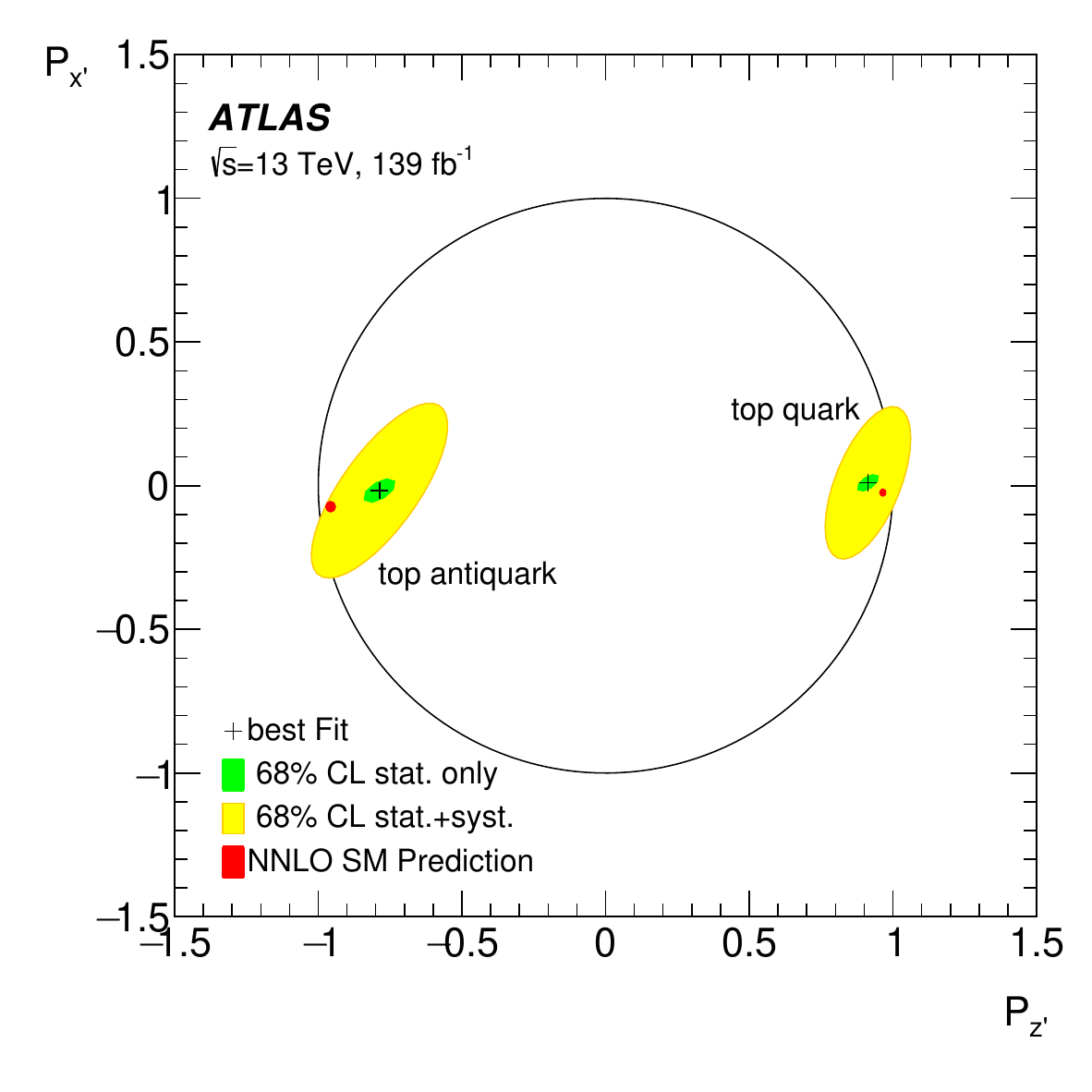}}
\subfloat[]{\includegraphics[width=0.45\linewidth]{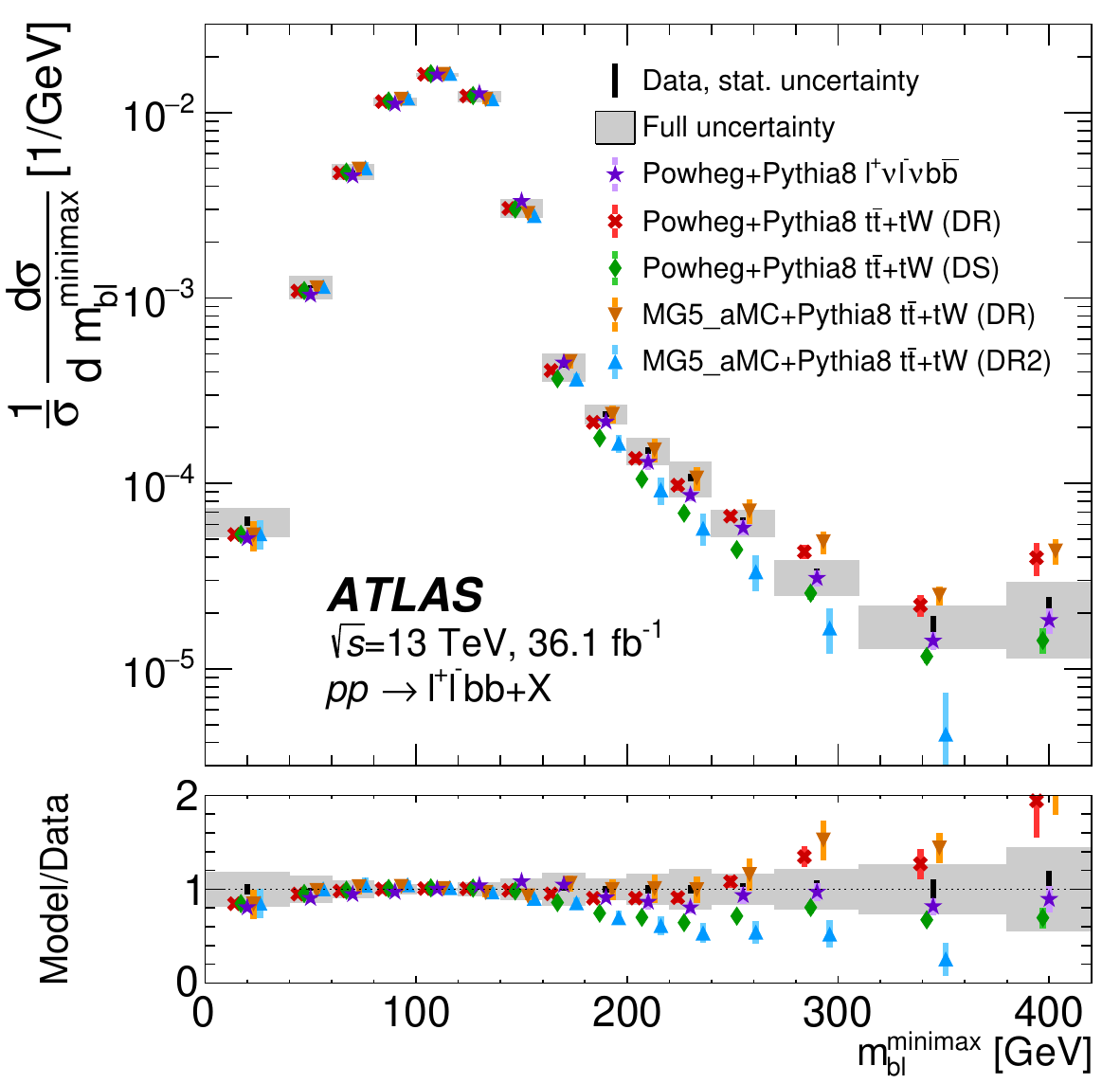}}
\caption{(a) Observed best-fit polarisation measurements with their statistical (green) and statistical+systematic (yellow) uncertainty contours at 68\% CL, plotted on the two-dimensional polarisation parameter space $(P_{z'}, P_{x'})$~\cite{ATLAS:2022vym}.
(b) Unfolded normalized differential $m^{\text{minimax}}_{b \ell}$ cross-section compared with various theoretical models~\cite{ATLAS:2018ivx}.}
\label{fig:singletop}
\end{figure}

\subsection{Measurements in the $tW$ channel}
\label{sec:tW}

At the LHC, the second largest single-top-quark production cross-section %
is the one for single top production in association with a $W$ boson, i.e.\ $tW$, which accounts for approximately 24\% of
the total single-top-quark production rate at $\sqrt{s}=13$~\TeV.
This process was beyond the reach of the Tevatron,
and was first observed in $\sqrt{s}=8$~\TeV data at the LHC~\cite{CMS:2014fut,ATLAS:2015igu}, with cross-section measurements in
good agreement with theoretical predictions.
The expected $\sqrt{s}=13$~\TeV SM $tW$ cross-section at NLO including next-to-next-to-leading logarithms is
$\sigma(tW) = 79.3 ^{~+1.9}_{\,-1.8}~\text{(scale)} \pm 2.2~\text{(PDF+\alphas)}$~pb~\cite{Kidonakis:2021vob}.
ATLAS measured this process using 3.2~\ifb\ of data collected in 2015~\cite{ATLAS:2016ofl}
in the final state with exactly two oppositely charged leptons and at least one $b$-tagged jet,
using the same  boosted decision tree (BDT) technique as in the $\sqrt{s}=8$~\TeV measurement.
Accurately estimating the kinematic distributions of the $tW$ process is difficult since this process is not well-defined
at higher order in QCD because of interference with \ttbar production (see Section~\ref{sec:systematics}).
The difference between results obtained with the DR and DS treatments is then taken as an uncertainty and is sizeable in many analyses.
The $\sqrt{s}=13$~\TeV dataset made it possible to perform measurements that
test this modelling, and ATLAS exploited this in two ways:
measuring the differential $tW$ cross-section distributions~\cite{ATLAS:2017quy} and
using specific variables sensitive to the interference~\cite{ATLAS:2018ivx}.

The differential cross-section was measured~\cite{ATLAS:2017quy} in the dilepton final state,
using 36~\ifb\ of data and requiring the presence of exactly one $b$-tagged jet and no additional jets.
A BDT was constructed to separate the $tW$ signal from the large \ttbar background.
The regions with only untagged jets or more $b$-tagged jets
were used as validation regions.
In this channel, the top quark or the $W$ boson cannot be reconstructed directly because of the undetected neutrinos.
However, some observables are correlated with the kinematic properties of the $tW$ process and its modelling.
The unfolding to particle level was performed within a fiducial phase space and the obtained distributions were normalised.
The largest uncertainties come from the limited size of the data sample as well as \ttbar and $tW$ MC modelling.
In general, most of the MC models show fair agreement with the measured cross-section distributions, although the predicted distributions are softer than the observed ones.
Both the statistical and systematic uncertainties have a significant impact on the result, with total uncertainties ranging from 10\% to 50\% depending on the bins.
The differential $tW$ cross-section measurement is therefore expected to improve significantly as more data is used.

Interference between $tW$ and \ttbar production was specifically probed in the dilepton final state using 36~\ifb\
of data in a phase space with exactly two $b$-tagged jets (and no additional jets passing a
looser $b$-tagging requirement)~\cite{ATLAS:2018ivx}.
The contributions from doubly and singly resonant amplitudes depend on the invariant mass of the $bW$ pairs in the event.
Since the charged-lepton kinematics are correlated with those of the $W$ boson,
the invariant mass of the $b$-jet and the charged lepton is an interesting observable for testing the interference.
As there are some ambiguities in forming this mass, the differential
cross-section was measured as a function of
$m^{\text{minimax}}_{b \ell} = \min \{ \max (m_{b_1 \ell_1}, m_{b_2 \ell_2}), \max (m_{b_1 \ell_2}, m_{b_2 \ell_1}) \}$,
an observable inspired by Refs.~\cite{Lester:1999tx,Barr:2003rg}, with $b_i$ and $\ell_i$ being the two $b$-tagged jets and leptons respectively.
The doubly resonant contribution is suppressed above $\sqrt{m^2_t - m^2_W}$ (where $m_t$ and $m_W$ are the
top-quark and $W$-boson masses), so
the differential cross-section above this kinematic endpoint has more sensitivity to interference effects.
The $m^{\text{minimax}}_{b \ell}$ distribution was unfolded to particle level using an iterative Bayesian unfolding and the normalised unfolded distribution was compared with the predictions (see Figure~\ref{fig:singletop}~(b)).
The modelling systematic uncertainties, which impact the results the most, range from 1\% to 22\% of the unfolded yields, while the statistical uncertainty is as large as 20\%.
The predictions using the DR scheme give a better description of the normalisation of the region $m^{\text{minimax}}_{b \ell} \ge m_t $ but the DS scheme models the $m^{\text{minimax}}_{b \ell}$ shape better in the same region.
In general, the DR and DS predictions bracket the data in the region of large $m^{\text{minimax}}_{b \ell}$, justifying the application of their difference as a systematic uncertainty. Later studies showed that this difference could be narrowed by using dynamic scales.
The full $\ell^+\nu\ell^-\bar{\nu}b\bar{b}$ predictions implemented in \POWHEGBOXRES~\cite{Jezo:2016ujg,Jezo:2015aia}, which includes off-shell top-quark effects at NLO and the interference term, gives the best predictions of $m^{\text{minimax}}_{b \ell}$ over the full distribution.
It is expected to eventually replace the approach where the \ttbar and $tW$ production processes are generated separately.

\subsection{Measurement in the $s$-channel}

Among the three single-top-quark production channels mediated by $tW$ vertices,
the weakest one at the LHC is the $s$-channel, in which a top quark is produced with a bottom antiquark via an $s$-channel $W$-boson exchange.
This process contributed a larger fraction of the single-top events at the Tevatron, where it was observed in proton--antiproton collisions by the CDF and D0 collaborations~\cite{CDF:2014uma}.
At the LHC, ATLAS found evidence for this process in $pp$ collisions at $\sqrt{s}=8$~\TeV with an observed (expected) significance of 3.2 (3.9) standard deviations relative to the background-only hypothesis~\cite{ATLAS:2015jmq}.
Between $\sqrt{s}=8$~\TeV and 13~\TeV, the ratio of the $s$-channel single-top cross-section to
the dominant \ttbar background's cross-section decreases from 2.1\% to 1.2\%, making the analysis more challenging at $\sqrt{s}=13$~\TeV. In the SM, the $s$-channel single-top production cross-section in $pp$ collisions at $\sqrt{s}=13$~\TeV is $\sigma_{s\text{-channel}} = 10.32^{~+0.40}_{\,-0.36}$~pb,
calculated at NLO in QCD with \HATHOR[2.1]~\cite{Kant:2014oha}.

Despite these challenges, ATLAS measured this cross-section at $\sqrt{s}=13$~\TeV in the final state with one lepton and exactly two $b$-tagged jets~\cite{ATLAS:2022wfk}.
To extract the signal from the large background composed of \ttbar,
$W$+jets and $t$-channel single-top events, a discriminant based on the matrix-element method~\cite{Kondo:1988yd,Kondo:1991dw} was used.
This discriminant was built from the likelihood values computed for the hypothesis that a measured event came from a given process.
The likelihood values were computed by integrating the matrix elements for the signal or background processes.
The $s$-channel single-top production cross-section was measured from a binned profile-likelihood fit of this discriminant, which also allowed the normalisation of both the \ttbar and $W$+jets background processes to vary freely.
The analysis yields $\sigma_{s\text{-channel}}= 8.2 \pm 0.6 \text{ (stat.)} ^{~+3.4}_{\,-2.8} \text{ (syst.)}$~pb, in agreement with the SM prediction.
The largest systematic uncertainty comes from the \ttbar normalisation, followed by those from the jet energy scale and signal modelling.
This corresponds to an observed (expected) significance of 3.3 (3.9) standard deviations for $s$-channel production relative to the background-only hypothesis.
The significance is similar to that in the $\sqrt{s}=8$~\TeV analysis, despite the larger data sample, because of the lower signal-to-background ratio at $\sqrt{s}=13$~\TeV.

The $\sqrt{s}=13$~\TeV inclusive single-top cross-section measurements are summarised in Figure~\ref{fig:singletop-summary}.
The $t$-channel cross-section measurement was used to determine that the coupling at the $Wtb$ vertex is $f_\text{LV} \cdot |V_{tb}| = 1.015 \pm 0.031$~\cite{ATLAS:2024ojr}, where $V_{tb}$ is the corresponding element of the Cabibbo--Kobayashi--Maskawa matrix~\cite{Cabibbo:1963yz,Kobayashi:1973fv}, and $f_\text{LV}$ is a possible additional left-handed form factor~\cite{Aguilar-Saavedra:2008nuh} (in the SM, $f_\text{LV}=1$).
Bounds were also placed on possible anomalous couplings within the framework of effective field theory (see Section~\ref{sec:eft}).

\begin{figure}[!tbp]
\centering
\includegraphics[width=0.8\linewidth]{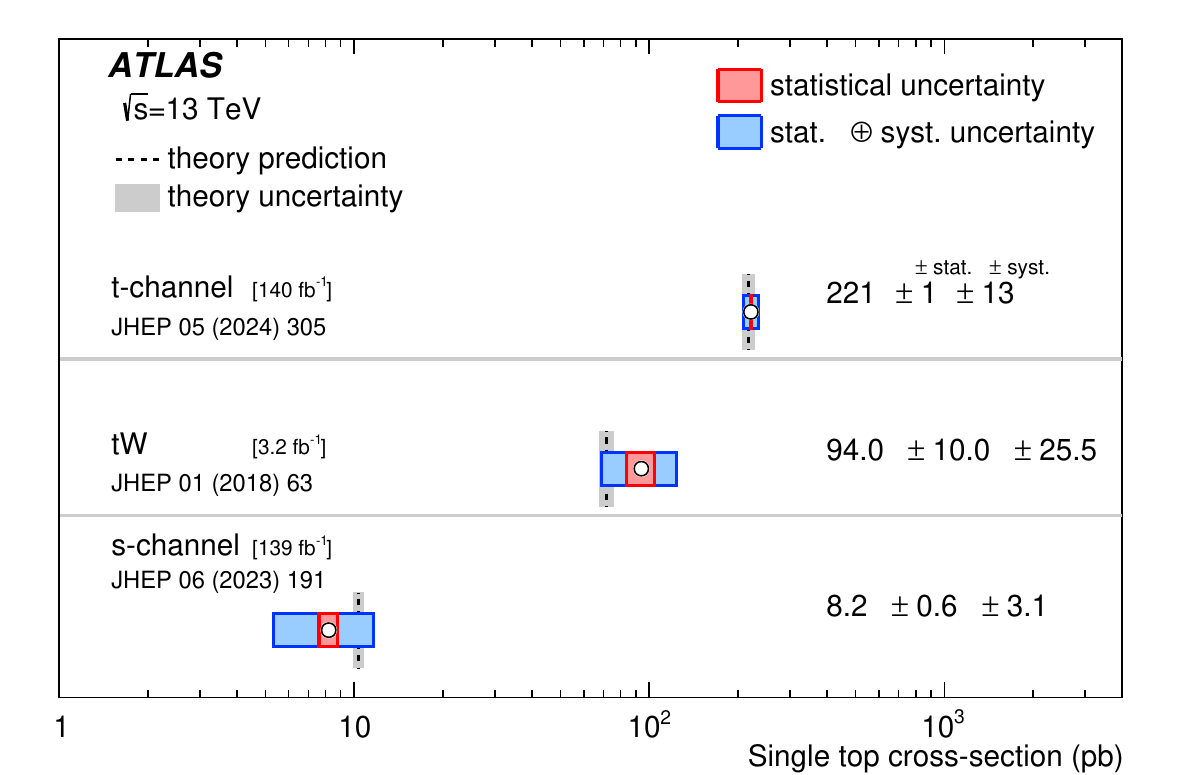}
\caption{\label{fig:singletop-summary} Summary of the $\sqrt{s}=13$~\TeV single-top-quark cross-section measurements, compared with the corresponding theoretical expectations. The values are quoted for $m_t=172.5$~\GeV.
The theory predictions are taken from Refs.~\cite{Campbell:2020fhf} ($t$-channel), \cite{Kidonakis:2015nna} ($tW$) and \cite{Kant:2014oha} ($s$-channel).}
\end{figure}

%
%
%
%
%
%
%
%


%
\section{Associated production of top quarks}
\label{sec:associated}


The high centre-of-mass energy and large \RunTwo data sample have brought the study of rare top-quark production processes to a new level. Such processes include the production of a top quark or \ttbar\ pair in association with a $Z$, $W$ or $\gamma$ (the process where a \ttbar pair is produced with a Higgs boson, \ttH, is described in Ref.~\cite{ATLAS-HIGG-2023-11}).
The cross-section of each of these rare associated-production processes (\enquote{top+$V$} processes, where $V$ is a vector boson) is significantly larger in \RunTwo than at $\sqrt{s}=8$~\TeV, where they were first explored.
With the exception of the $\ttbar+W$ process, these cross-sections
can be related to the coupling of the top quark to the associated boson, allowing the measurement of the top quark's neutral-current coupling or the top quark's Yukawa coupling. The top+$V$ production cross-sections could be altered by physics beyond the SM, such as a vector-like quark~\cite{Aguilar-Saavedra:2009xmz,Aguilar-Saavedra:2013qpa}, a strongly coupled Higgs boson~\cite{Perelstein:2005ka}, or a heavy scalar or pseudoscalar boson~\cite{Dicus:1994bm,Craig:2015jba,Craig:2016ygr}.
Even if such new physics is beyond the energy reach of the LHC for direct observation, it can manifest itself by changing the top+$V$ cross-sections through virtual contributions involving new particles.
Their effect can be parameterised in the context of an effective field theory (EFT) framework
as dimension-six operators that extend the SM Lagrangian~\cite{Degrande:2010kt,Zhang:2017mls} (see Section~\ref{sec:eft}).
At $\sqrt{s}=13$~\TeV, the simultaneous production of four top quarks has a cross-section large enough to make its observation feasible.

As in the case of \ttbar\ production, the experimental challenges differ depending on the decays of the $W$ bosons coming from the top quarks.
The \enquote{golden channel} used to explore many of the top+$V$ processes considers two same-sign (SS) leptons, three leptons or even four leptons. These typically occur when at least one top quark decays leptonically and another lepton comes from the associated boson's decay.
This so-called multilepton channel benefits from low backgrounds from SM processes, which compensates for the small branching fraction.
However, it has rare backgrounds that are challenging to evaluate,
coming from non-prompt leptons produced in hadron decays or jets misidentified as leptons (collectively called \enquote{fake leptons}) and also from prompt electrons that have a misassigned charge (for the channel with SS leptons).
The muon charge-misassignment rate is negligible.

The fake-lepton background in the multilepton channel is usually evaluated using data-driven techniques. A common procedure is to use the \enquote{matrix method}~\cite{EGAM-2019-01}.
It uses two types of events: events with \enquote{loose} selected leptons and events with \enquote{tight} selected leptons, where loose
leptons are defined by loosening or inverting some of the selection criteria so as to increase the fraction of fake leptons.
This method is based on the fact that the numbers of events selected by using either the loose or tight lepton criteria can be expressed as linear combinations of the numbers of events with either prompt or fake leptons, using the fraction of prompt loose leptons meeting the tight criteria (tight lepton efficiency) and the fraction of fake loose leptons also meeting the tight criteria (lepton fake rate).
Knowing this efficiency and fake rate (as well as the numbers of selected loose and tight leptons), the numbers of prompt and fake leptons can be extracted by inverting these relations. The efficiency and fake rate are measured in data using control regions enriched in either prompt or fake leptons. When using this method, the key step is to select control regions that are kinematically representative of the signal region so that the measured
efficiency and fake rate can be applied in the signal region.
Another way to estimate the non-prompt-lepton background is the \enquote{template method}~\cite{EGAM-2018-01}.
This method relies on the simulation to model the kinematic distributions of fake-lepton background processes and on control regions enriched in fake leptons to determine their normalisations.
There are usually several control regions, enriched in fake leptons from different sources (such as electrons from photon conversions or leptons from heavy-flavour hadron decays).
These control regions are then included, together with the signal region, in the fit that extracts the signal, and normalisation factors for the fake-lepton backgrounds are determined simultaneously with the signal strength (defined as the ratio of the measured cross-section to the SM prediction).

The other background usually evaluated from data is the one where the reconstruction assigns the wrong charge sign to an electron in the case where events with two SS leptons are selected. This happens when the electron undergoes a hard bremsstrahlung followed by an asymmetric photon conversion or when the sign of the electron track's curvature is mismeasured.
To suppress this background, another multivariate BDT was used, taking as input the track and energy-cluster properties of the electron candidate~\cite{ATLAS:2017ztq}.
The charge misassignment rate was measured from the fraction of reconstructed $Z \to ee$ data events with a same-charge electron pair. It was parameterised as a function of the electron \pt and $\eta$ and then applied to data events satisfying the signal selection where two leptons with opposite charges are required.

\subsection{Top-quark production in association with a $W$ or $Z$ boson}
Studying the \ttZ process provides a direct probe of the weak couplings of the top quark.
The coupling of the top quark to the $Z$ boson is not yet well constrained,
leaving room for potential new-physics contributions.
Top-quark pair production in association with a $W$ boson
is an irreducible source of SS dilepton pairs. It is charge asymmetric in $pp$ collisions because it is initiated by quark--antiquark initial states, and is unusually complex to predict because of the importance of higher-order QCD and EW corrections.
Although the \ttZ and \ttW processes were not observed individually in \RunOne, the $\ttbar+V$ production process (with $V = Z$ or $W$) was observed at $\sqrt{s}=8$~\TeV~\cite{ATLAS:2015qtq}, reaching a
precision of ${\sim}30$\%.
At $\sqrt{s}=13$~\TeV, the expected SM \ttZ (\ttW) production cross-section increases by a factor of more than 3~(2).
The \RunTwo data sample made precise measurements possible, and also opened the door to differential measurements exploring \ttZ and \ttW kinematic modelling for the first time. ATLAS  performed measurements in the multilepton channel coherently for both the \ttZ and \ttW signals, first using 3.2~\ifb\ of $\sqrt{s}=13$~\TeV data~\cite{ATLAS:2016wgc} and then using 36~\ifb~\cite{ATLAS:2019fwo}. The full \RunTwo dataset was later used to study these processes differentially~\cite{ATLAS:2024moy,2312.04450}.

Using 36~\ifb\ of $\sqrt{s}=13$~\TeV data, the \ttZ and \ttW cross-sections were measured simultaneously~\cite{ATLAS:2019fwo} using SS dilepton events, trilepton events (3L), opposite-sign (OS) dilepton events and four-lepton events (4L).
The OS dilepton region targets \ttZ events where both top quarks decay hadronically and the $Z$ boson decays into a pair of leptons. It suffers from a large background of $Z$+jets and \ttbar events.
BDTs were used to separate signal from background.
The SS dilepton region targets the \ttW process, with one top quark and the $W$ boson decaying leptonically. It was split according to the charge of the selected lepton pairs since \ttW events are preferentially produced with a positively charged $W$ boson. The 3L channel is sensitive to both \ttZ and \ttW events.
The 4L channel targets \ttZ events where both the \ttbar pair and the $Z$ boson decay leptonically.
The \ttZ and \ttW signal strengths were extracted simultaneously using a binned maximum-likelihood fit to all the control and signal regions.
Since they are important backgrounds, the normalisations of the $WZ$, $ZZ$ and $Z$+heavy-flavour-jets processes are determined from data control regions.
The results are $\sigma_{\ttZ} = 950 \pm 80 \text{ (stat.)} \pm 100 \text{ (syst.)} = 950 \pm 130$~fb and $\sigma_{\ttW} = 870 \pm 130 \text{ (stat.)} \pm 140 \text{ (syst.)} = 870 \pm 190$~fb in agreement with SM predictions of $\sigma_{\ttZ} = 863^{~+78}_{\,-89}$~fb from NLO QCD including EW corrections and NNLL resummation~\cite{Kulesza:2018tqz} and $\sigma_{\ttW} = 745.3 \pm 54.9$~fb~\cite{Buonocore:2023ljm} from NNLO QCD with NLO EW corrections.

With the full \RunTwo dataset, differential \ttZ~\cite{ATLAS:2021fzm,2312.04450} and \ttW~\cite{ATLAS:2024moy} measurements were carried out in independent analyses. The latest \ttZ measurement was performed in three final states: the dilepton, 3L and 4L channels.
The $Z$+jets background was estimated using simulation.
However, since the modelling of $Z$+jets with heavy-flavour jets is challenging, the normalisations of the $Z+b$ and $Z+c$ components were obtained in data, simultaneously with
the extraction of the signal strength.
The \ttbar background was estimated using a fully data-driven method, relying on the high
\ttbar purity of a sample requiring one electron and one muon.
A deep neutral network (DNN) was trained in each of the regions to extract the \ttZ signal. Some of the input variables for this DNN were built using the output of the \ttbar-system reconstruction.
In all channels, the background from fake/non-prompt leptons was estimated using the template method.
The inclusive signal strength was extracted by fitting the DNN output distributions simultaneously in all three channels using a profile-likelihood technique.
The cross-section was measured to be:
\[
\sigma_{\ttZ} = 860 \pm 40 \text{ (stat.)} \pm 40 \text{ (syst.)} = 860 \pm 60~\text{fb},
\]
where the largest systematic uncertainty comes from background normalisations.
In addition, normalised and absolute differential distributions sensitive to the \ttZ vertex (and hence interesting for constraining some EFT operators) were extracted in the 3L and 4L channels. The dilepton channel was not used because
of its large background contamination.
To correct for acceptance and detector effects, the differential distributions were unfolded using a profile-likelihood unfolding technique.
The differential observables include the transverse momentum and absolute rapidity of the $Z$ boson (see Figure~\ref{fig:ttZW}~(a)), the transverse momentum or invariant mass of the top or \ttbar system, and the azimuthal angle $\Delta \phi$ or rapidity difference between the $Z$ boson and the leptonically decaying top quark.
Most observables were measured at both parton and particle level.
The compatibility of the unfolded measurements with various predictions was assessed by computing a $\chi^2$ per degree of freedom and its corresponding $p$-value. In all cases, the $p$-values indicate good
agreement between the unfolded data and the predictions.
The particle-level distributions were used to constrain EFT effects in \ttZ production (see Section~\ref{sec:eft}).
Furthermore, detector-level observables sensitive to polarisation and spin correlation of the top quarks were combined to explore spin correlations in \ttZ production.
Simulated templates of \ttZ events with and without spin correlation were used to extract a ratio of the measured spin correlation to the SM prediction in \ttZ events of $f_\text{SM} = 1.20 \pm 0.68$. The total uncertainty is dominated mainly by its statistical component. This result is in agreement with the SM, and represents a $1.8\sigma$ departure from a scenario without top-quark spin correlations.

The full \RunTwo dataset was also used to explore the \ttW process differentially for the first time,
studying the kinematics of \ttW final-state particles and of any associated jets~\cite{ATLAS:2024moy} in the channel with SS dilepton events and the 3L channel.
Beyond the interest in measuring this rare process more precisely, a better understanding of \ttW production is important since indirect measurements in analyses targeting \ttH\ or \tttt\ production have consistently observed larger \ttW yields than the SM predicts in the \ttW phase space with additional jets.
The signal regions were split by lepton charge. An inclusive cross-section was extracted by further splitting them by lepton flavour and both the jet and $b$-tagged-jet multiplicities. Control regions were defined in order to adjust the normalisation of the fake-lepton backgrounds, as well as the diboson and \ttZ backgrounds. These control regions were defined to be orthogonal to the signal regions by applying looser lepton isolation criteria or requiring a different number of jets or $b$-tagged jets. The normalisations obtained from the fit are compatible with unity.
The inclusive cross-section was measured to be:
\[
\sigma_{\ttW} = 880 \pm 50 \text{ (stat.)} \pm 70 \text{ (syst.)} = 880 \pm 80~\text{fb},
\]
with the largest systematic uncertainty coming from the modelling of the \ttW signal. This result is higher than the SM prediction in Ref.~\cite{Buonocore:2023ljm}, but compatible with it at the level of 1.4 standard deviations.
Separate $\ttW^+$ and $\ttW^-$ cross-sections were also measured, together with their ratio and the relative charge asymmetry.
Differential \ttW measurements were performed as a function of observables where discrepancies were observed previously or that are sensitive to NLO corrections. These include the number of jets (see Figure~\ref{fig:ttZW}~(b)), the scalar sum of the transverse momenta of jets, and separately of leptons, and the
azimuthal angle and rapidity difference between the two same-sign leptons.
The corrections to particle level were obtained using a profile-likelihood unfolding.
As expected from the inclusive case, the absolute differential measurements exceed the theoretical predictions. The normalised distributions, however, show rather good $\chi^2$ compatibility with the MC generator predictions.
The total uncertainty, dominated by data statistics, does not currently allow the
modelling performed in the MC simulation to be constrained significantly.

The process with a single $Z$ boson and a top quark in the final state ($tZq$) is another way to probe the coupling of the top quark to the $Z$ boson.
Despite an expected cross-section ten times smaller than that for \ttZ, this process probes two electroweak couplings in a single process: the $t$--$Z$ and $W$--$Z$ couplings.
Evidence for $tZq$ production was seen with the 36~\ifb\ data sample~\cite{ATLAS:2017dsm}, while the full \RunTwo dataset allowed a definitive observation~\cite{ATLAS:2020bhu} as described in the following.
The expected $tZq$ SM cross-section at NLO in QCD (with a dilepton invariant mass $m_{\ell \ell} > 30$~\GeV) is $102^{~+5}_{\,-2}$~fb, computed using MCFM~\cite{Campbell:2013yla}.
The $tZq$ process was searched for in the 3L channel, requiring one $b$-tagged jet and one additional non-$b$-tagged jet, which is expected to be emitted preferentially at high absolute pseudorapidity. A third jet coming from radiation was also allowed. To help separate the signal from the diboson, $Z$+jets and \ttbar backgrounds, both the $Z$-boson and the top-quark invariant masses were reconstructed. Diboson, \ttZ and \ttbar control regions were defined in order to adjust the background normalisations or to help constrain their uncertainties. The contribution from non-prompt-lepton background was estimated by replacing one $b$-tagged jet by a lepton in the \ttbar and $Z$+jets MC event samples. The signal was separated from the background by using a neural network where the most discriminating input variable is the largest invariant mass formed by the $b$-tagged jet and one of the untagged jets. The signal strength was extracted from a maximum-likelihood fit together with the normalisations of the \ttbar and $Z$+jets backgrounds and leads to:
\[
\sigma_{tZq} = 97 \pm 13 \text{ (stat.)} \pm 7 \text{ (syst.)} = 97 \pm 15~\text{fb},
\]
for $m_{\ell \ell}>30$~\GeV, in agreement with the SM prediction.
The statistical significance of the result is well above five standard deviations relative to the background-only hypothesis, which establishes observation of this rare process using $\sqrt{s}=13$~\TeV data.

\begin{figure}[!tbp]
\centering
\subfloat[]{\includegraphics[width=0.45\linewidth]{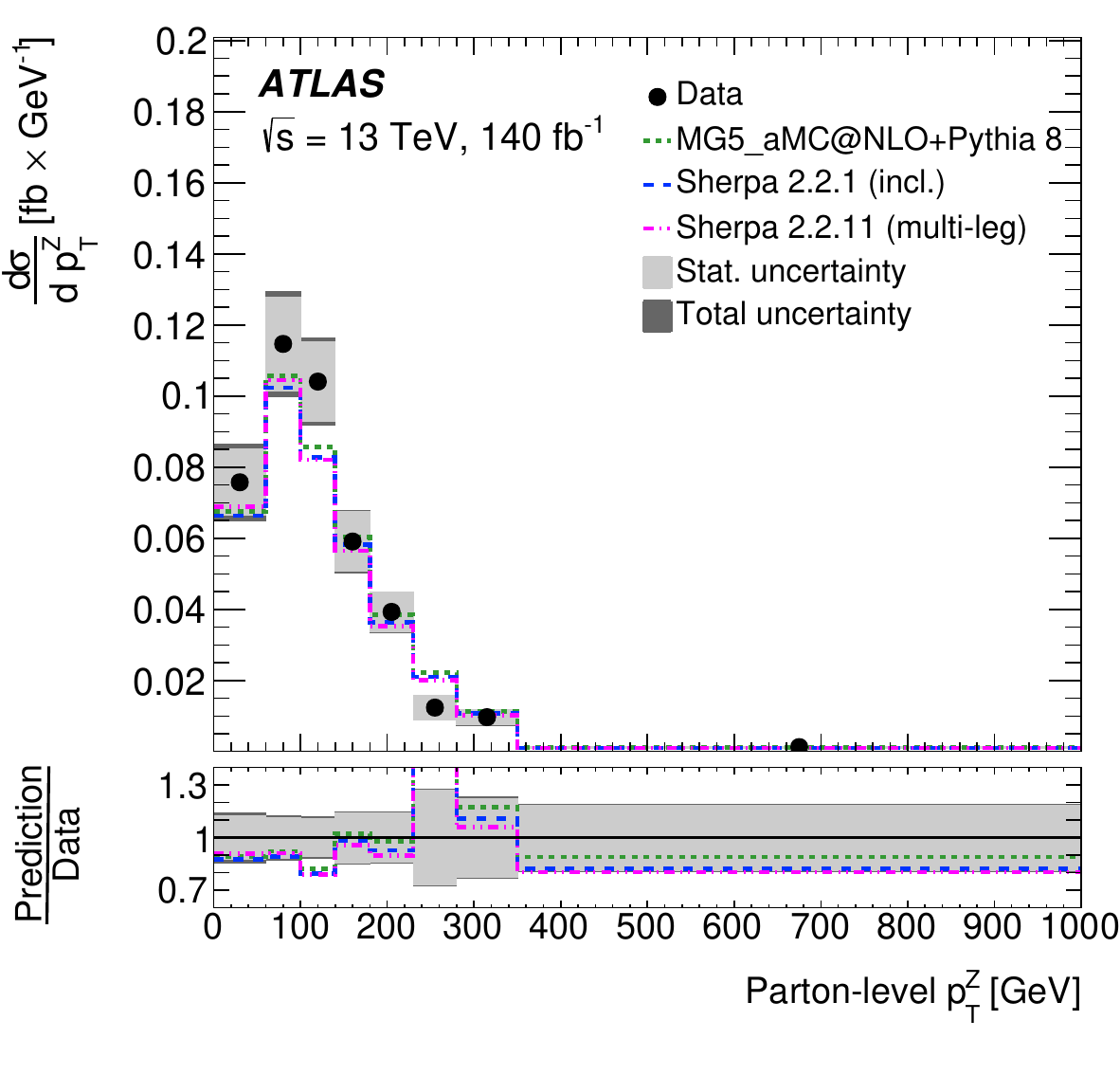}}
\subfloat[]{\includegraphics[width=0.45\linewidth]{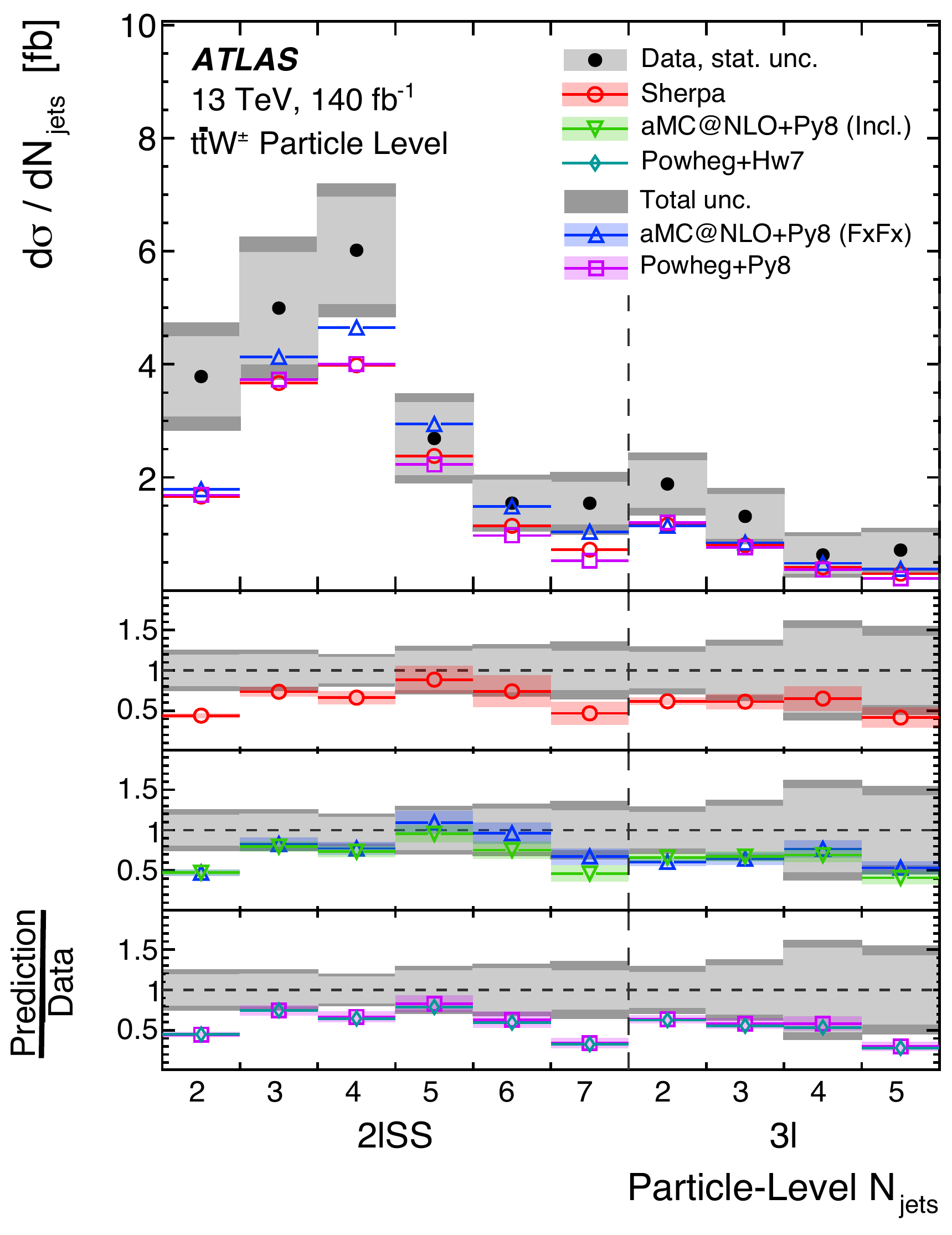}}
\caption{(a) Absolute differential \ttZ cross-section measurement as a function of the transverse momentum of the $Z$ boson~\cite{2312.04450}.
(b) Unfolded distribution of the absolute \ttW cross-section as a function of jet multiplicity~\cite{ATLAS:2024moy}.}
\label{fig:ttZW}
\end{figure}

\subsection{Top-quark production in association with a photon}
The top quark's coupling to a photon is tested
through the measurement of \tty production and its kinematic properties. For instance, the transverse momentum of the photon would be affected by anomalous dipole moments of the top quark~\cite{Baur:2004uw,Bouzas:2012av,Schulze:2016qas} or by EFT operators.
At the LHC, \tty production was already observed in $\sqrt{s}=7$~\TeV $pp$ collisions~\cite{ATLAS:2015jos} and the first differential measurements were performed at $\sqrt{s}=8$~\TeV~\cite{ATLAS:2017yax}.
At $\sqrt{s}=13$~\TeV, benefiting from larger data samples, further differential distributions were explored, first using 36~\ifb\ of data~\cite{ATLAS:2018sos} and then the full \RunTwo dataset~\cite{ATLAS:2020yrp}.
One challenge in the \tty analyses is due to the fact that the photon can originate not only from the top quark
but also from its charged decay products or from initial-state quarks, diluting the information about the $t\gamma$ coupling.
Although separating these sources is difficult, some kinematic variables such as the
angular separation between a lepton and photon in the event
can help. Another challenge in such analyses is the accurate simulation of the signal. Samples could be produced either inclusively without explicitly including a photon in the final state (but where photons are generated through initial- or final-state radiation), or with photons at the matrix-element level, taking the whole decay chain into account at LO. The overlap between these two kinds of samples then needs to be removed. The \tty process also interferes with the singly resonant $tW\gamma$ process, similarly to the interference of the \ttbar and $tW$ processes.
NLO calculations are available in dedicated phase spaces, i.e.\ with specific photon kinematic requirements~\cite{Melnikov:2011ta,Bevilacqua:2018woc,Bevilacqua:2018dny}. Hence, in order to accurately compare the data with theory computations, the analyses need to be performed in the same phase space.
The first \RunTwo analysis, using 36~\ifb, exploited both the semileptonic and dileptonic decays of the \ttbar pair.
The analysis using the full 139~\ifb\ dataset~\cite{ATLAS:2020yrp} focused on the clean $e\mu$ final state so that no multivariate technique had to be used to extract the signal. It also measured the combined resonant \tty and non-resonant $tW\gamma$ production cross-section in order to compare it with the NLO QCD predictions in Refs.~\cite{Bevilacqua:2018woc,Bevilacqua:2018dny}.
Apart from the backgrounds with prompt photons (such as $W\gamma$ and $Z\gamma$), there are two sources of background with misidentified photons: hadron-fakes (photons mimicked by hadronic energy deposits, or non-prompt photons from hadron decays) and electron-fakes (electrons mimicking photon signatures).
In this analysis, they were both estimated using MC samples since the
studies performed with data-driven techniques in the 36~\ifb\ analysis~\cite{ATLAS:2018sos} showed that possible data-driven corrections have a negligible effect on the distribution shapes of relevant observables.
The background from electron-fakes is a minor background contribution in the analysis.
The fiducial inclusive cross-section was extracted using a profile-likelihood fit of the distribution of the scalar sum of all transverse momenta in the event, including leptons, photons, jets and missing transverse momentum. This observable was found to provide good separation between signal and background without too much sensitivity to systematic uncertainties.
The fiducial region's definition required one electron and one muon, as well as two $b$-jets, and one photon with $\ET> 20$~\GeV and $|\eta| < 2.37$ at parton level.
The combined $\tty/tW\gamma$ cross-section was measured to be:
\[
\sigma_{\text{fid}}(\tty \to e\mu b\bar{b} \gamma) = 39.6 \pm 0.8 \text{ (stat.)} ^{~+2.6}_{\,-2.2} \text{ (syst.)}~\text{fb} = 39.6 ^{~+2.7}_{\,-2.3}~\text{fb},
\]
in good agreement with the dedicated theoretical calculation: $38.5^{~+1.2}_{\,-2.5}$~fb~\cite{Bevilacqua:2018woc,Bevilacqua:2018dny}. The systematic uncertainties with the largest impact come from the uncertainties in modelling the signal.
The absolute and normalised differential cross-sections were measured as a function of the kinematics of the photon, the angular separation between the photon and the leptons, and the pseudorapidity difference and azimuthal angle between the two leptons. The last two of these are particularly sensitive to \ttbar spin correlations. The unfolded distributions generally agree well with the predictions, except for the shape of the angular separation between the leptons and the photon or the azimuthal angle between the two leptons, which are not well modelled by the LO MC predictions. The NLO prediction provides a better description of these distributions. The dominant uncertainty in these differential measurements still comes from the limited size of the data sample, at a level slightly below 10\%.

Analogously to the $t$--$Z$ coupling, the coupling between the top quark and a photon can also be studied in the single-top process $tq\gamma$, featuring a forward light-quark jet characteristic of $t$-channel production. Analysing the full $\sqrt{s}=13$~\TeV dataset led to the first observation of this process~\cite{ATLAS:2023qdu}. The $tq\gamma$ cross-section was measured in a fiducial phase space either at parton level, excluding contributions where photons are radiated from the charged decay products, or at particle level, including these contributions. Two signal regions were defined according to the presence or absence of a forward jet. Control regions were included to normalise the large background from \tty production and the $W\gamma$ background. A neural network was trained to separate signal from background by using the reconstructed top-quark mass as the most discriminating input variable. The measured fiducial cross-section (requiring one electron or muon, one $b$-jet, and one photon with \pt greater than 20~\GeV) is:
\[
\sigma_{\text{fid}}(tq\gamma) = 688 \pm 23 \text{ (stat.)} ^{~+75}_{\,-71} \text{ (syst.)}~\text{fb} = 688 ^{~+78}_{\,-75}~\text{fb},
\]
which is 2.1~standard deviations above the SM NLO QCD prediction of $515^{~+36}_{\,-42}$~fb from \MGNLO. The largest systematic uncertainty comes from the modelling of the \tty background. The observed (expected) significance of the $tq\gamma$ signal relative to the background-only hypothesis is 9.3 (6.8) standard deviations, establishing observation of this process.

\subsection{Four-top-quark production}

The \RunTwo data sample also gave access to four-top-quark production, which is one of the rarest and heaviest-final-state processes now accessible at the LHC, with a combined particle rest mass of almost 700~\GeV.
The \RunTwo dataset is expected to contain around 1700 four-top-quark events.
This multiparticle SM process presents a promising avenue to search for signals of new physics beyond the SM.
For example, the \tttt cross-section could be enhanced in top-quark-compositeness models~\cite{Pomarol:2008bh}
or by gluino pair production in supersymmetric theories~\cite{Nilles:1983ge,Farrar:1978xj}.
Within the EFT approach, four-top-quark production is uniquely sensitive to four-top-quark operators.
Because of the existence of electroweak \tttt\ Feynman diagrams where the production of a pair of top quarks is mediated by a Higgs boson, \tttt\ production is also sensitive to the top-quark Yukawa coupling and its $CP$ properties~\cite{Cao:2016wib, Cao:2019ygh}.
Measuring the four-top-quark production cross-section is interesting in its own right since experimental results will challenge the state-of-the-art perturbative QCD calculation techniques.
Within the SM the predicted \tttt\ cross-section
in $pp$ collisions at a centre-of-mass energy of $\sqrt{s}=13$~\TeV is $\sigma_{\tttt}=12.0 \pm 2.4$~fb~\cite{Frederix:2017wme, Bevilacqua:2012em, Jezo:2021smh} at NLO in QCD including NLO electroweak corrections.
This value does not include the effect of threshold resummation at next-to-leading-logarithm
accuracy,
which increases the total production cross-section by approximately 12\% and reduces the scale uncertainty~\cite{vanBeekveld:2022hty}.
This corresponds to about a factor of 10 enhancement relative to the \tttt\ cross-section at $\sqrt{s}=8$~\TeV, demonstrating the new perspective offered by 13~\TeV collisions.

The analysis was first performed in the channel with two SS leptons or at least three leptons.
This channel corresponds to 12\% of the total \tttt\ production cross-section, and offers the best discovery potential.
First, a search for new phenomena targeted vector-like quark and SS top-quark pair production, and placed an upper limit on SM \tttt\ production using 36~\ifb\ of \RunTwo data~\cite{EXOT-2016-16}. Later, dedicated analyses focusing on SM production were developed, finding first evidence for \tttt\ production using the full 139~\ifb\ \RunTwo dataset~\cite{TOPQ-2018-05}.
With the same dataset but with improvements in object reconstruction, calibration and selection criteria, new analysis techniques and a better understanding of major background processes and systematic uncertainties, this process was first observed~\cite{ATLAS:2023ajo} as described in the following.
Backgrounds in the multilepton-channel \tttt\ analysis arise almost entirely from previously described top+$V$ processes (i.e.\ \ttW, \ttZ, and \ttH\ production) when they produce additional jets. These backgrounds with prompt leptons were estimated using MC simulation.
Because the theoretical modelling of the \ttW background at high jet multiplicity suffers from large uncertainties and since, as described above, the measured inclusive \ttW cross-section is higher than the SM expectation, a data-driven estimation of this important background was implemented.
The overall normalisation of the \ttW background and the parameters of the scaling as a function of jet multiplicity were determined from dedicated control regions together with the signal. The fake-lepton backgrounds were evaluated using the template method.
The \tttt signal was separated from the background events using a multivariate discriminant built with a graph neural network (GNN).
The \tttt production cross-section and the normalisation factors for the backgrounds were determined via a binned likelihood fit to the GNN score distribution in the signal region while also using control regions, with systematic uncertainties included as nuisance parameters.
The measured \tttt production cross-section is:
\[
\sigma(t\bar{t}t\bar{t}) = 22.5^{~+4.7}_{\,-4.3} \text{ (stat.)} ^{~+4.6}_{\,-3.4} \text{ (syst.)}~\text{fb} = 22.5 ^{~+6.6}_{\,-5.5}~\text{fb}.
\]
The significance of the observed (expected) signal is found to be 6.1 (4.3) standard deviations relative to the background-only hypothesis, providing the first observation of this process~\cite{ATLAS:2023ajo}.

The measured production cross-section is consistent with the SM prediction to within 1.8 standard deviations. Several limits on four-heavy-flavour-fermion EFT operators were also set using this measurement (see Section~\ref{sec:eft}).
In addition to \tttt\ production being dependent on the top Yukawa coupling,
the \ttH background is a function of the same coupling. The GNN distribution was therefore used to extract limits on the top Yukawa coupling's strength modifier $\kappa_t$, leading to $\kappa_t < 1.8$ (assuming a $CP$-even coupling).
This limit is less stringent than the ones derived from specific Higgs boson studies~\cite{ATLAS:2022tnm,ATLAS:2023cbt} but is less model dependent, without any assumption about the Higgs boson's width.
This \tttt\ analysis also derived 95\% confidence level (CL) intervals for the cross-section of the $t \bar{t} t $ process, which has an experimental signature and kinematic properties very similar to those of \tttt\ events.
This process, composed of two components $t \bar{t} t q$ and $t \bar{t} t W$,
has an expected SM cross-section that is about 10 times smaller than for \tttt, and has not yet been observed. More data and more dedicated analyses will be needed to constrain this very rare process.

Although it is significantly less sensitive, a \tttt\ measurement was also performed in events with a single lepton or two OS leptons, first as a search for heavy new particles~\cite{EXOT-2017-11} using 36~\ifb, and then targeting the SM process with the full $\sqrt{s}=13$~\TeV dataset~\cite{TOPQ-2020-10}. These final states have much higher branching fractions (about 57\% of the \tttt events) but considerable background from top-quark pair production with additional jets.
Since the background in the high jet-multiplicity regions was found to be mismodelled by MC simulation,
a strategy was developed to use data to sequentially reweight the \ttbar MC generation in several observables to obtain a reliable \ttbar{}+jets estimate.
The different \ttbar{}+jets components after reweighting (\ttbar{}+light, \ttbar{}+${\geq}1c$ and \ttbar{}+${\geq}1b$~jets) were further adjusted and constrained in a binned profile-likelihood fit which extracted the signal strength.
In the region most sensitive to \tttt production, BDTs were used to discriminate between signal and background events.
The systematic uncertainties in the \ttbar background prediction were evaluated with special care since these uncertainties have the largest impact on the measurement sensitivity.
The \tttt cross-section was measured to be
$\sigma(t\bar{t}t\bar{t}) = 26 \pm 8 \text{ (stat.)} ^{~+15}_{\,-13}\text{ (syst.)}~\text{fb} = 26^{~+17}_{\,-15}~\text{fb}$,
which corresponds to an observed significance of 1.9 standard deviations relative to the background-only hypothesis (while 1.0 standard deviations was expected). This result is compatible with the result in the multilepton channel.

Now that the \tttt process has been observed, the next step, as already taken for \ttZ and \ttW production, will be to test the SM predictions differentially with \tttt events and to better constrain the small and hard to separate $t \bar{t} t $ process.

The $\sqrt{s}=13$~\TeV inclusive top+$X$ cross-section measurements are summarised in Figure~\ref{fig:topX-summary}.
As described above, a lot of the top+$V$ processes are intertwined, so a strategy to evaluate all of their cross-sections coherently is needed to generically constrain new physics~\cite{ATLAS:2023kfs}.
In order to combine the analyses in the future, this would require the different analyses to harmonise how they define the reconstructed objects used in their selection criteria, harmonise their systematic uncertainties, and harmonise the phase spaces defining their control and signal regions.

\begin{figure}[!tbp]
\centering
\includegraphics[width=0.8\linewidth]{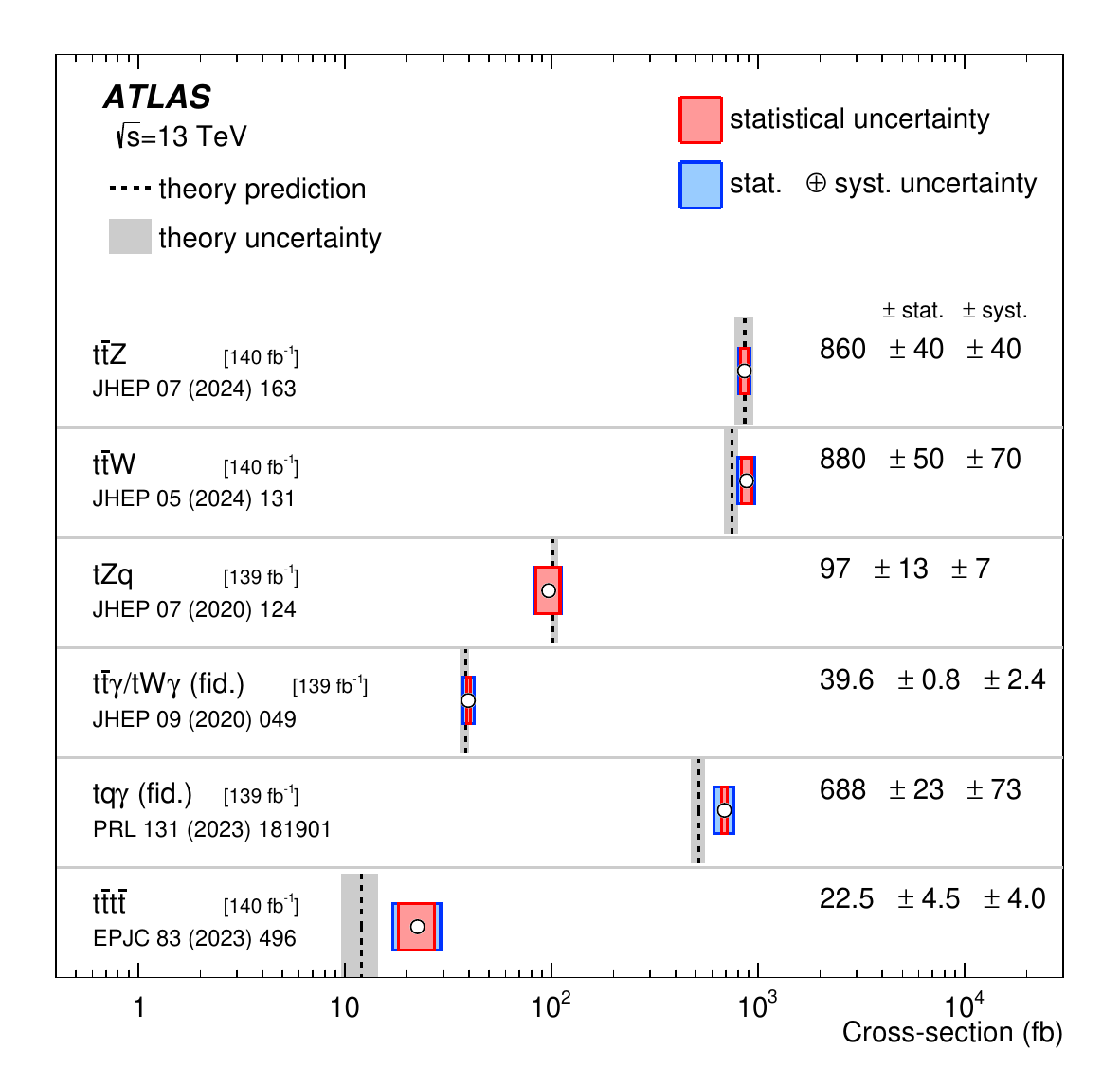}
\caption{\label{fig:topX-summary} Summary of the $\sqrt{s}=13$~\TeV top+$X$ cross-section measurements and comparison with the corresponding theoretical expectations. The fiducial region used to measure the \tty cross-section requires one electron, one muon, two $b$-jets and one photon at parton level. The $tq\gamma$ fiducial region is defined by requiring one electron or muon, at least one photon and at least one $b$-tagged jet at particle level.}
\end{figure}


%
\section{Top-quark mass}
\label{sec:mass}

%

The large value of its mass, $m_t$, is probably the most distinctive property of the top quark.
It is close to the electroweak scale and thus plays an important role in the dynamics of other elementary particles via loop diagrams.
In particular, it significantly affects the radiative corrections to both the Higgs boson's mass and the $W$ boson's mass,  providing a relationship that can be used for precision tests of the consistency of the SM~\cite{PhysRevD.98.030001}.

Measurements of the top-quark mass are typically
categorised into two families, often called \enquote{direct} and \enquote{indirect} measurements.
Measurements relying on the reconstruction of the decay products of the top quark, building partial or total invariant-mass observables, are usually assigned to the first category, and are therefore sometimes referred to as \enquote{measurements from decay}.
In contrast, extractions from total or differential cross-section measurements, relying on the dependence of theoretical predictions on the $m_t$ parameter in the SM Lagrangian, belong to the second category, also referred to as \enquote{measurements from production}.
The direct measurements typically have smaller uncertainties but inevitably rely on predictions from MC generators to relate the considered observable to a top-quark mass value.
Interpreting these measurements in a well-defined renormalisation scheme is subject to additional uncertainties that are challenging to evaluate~\cite{Hoang:2020iah,ATLAS:2021urs}.
In addition, the most precise direct top-quark mass measurements are affected by the relatively large uncertainties in the measurement of hadronic jet energies~\cite{TOPQ-2017-03, CMS-TOP-14-022, 1608.01881}.
Alternatively, indirect measurements can be interpreted in a theoretically cleaner way, but are often affected by even larger uncertainties of both experimental and theoretical nature.

\subsection{Direct top-quark mass measurements}

The most precise experimental $m_t$ determination by the ATLAS Collaboration was obtained through a combination of its \RunOne direct measurements,
with the dominant contributions coming from the 8~\TeV measurements in the dilepton~\cite{TOPQ-2016-03} and lepton+jets~\cite{TOPQ-2017-03} channels.
In order to minimise the impact of the large jet-energy scale and \ttbar modelling systematic uncertainties, the design of the two measurements took orthogonal directions to some extent.
In the dilepton channel, no attempt was made to fully reconstruct the top quark through its decay products, in order to avoid relying on
the determination of the missing transverse momentum for the reconstruction of the two escaping neutrinos in each of the selected events.
The $m_t$ value was then extracted with an unbinned maximum-likelihood fit to the partial top-quark invariant mass formed by a charged lepton and the corresponding $b$-tagged jet, $m_{\ell b}$, after imposing a lower bound on the lepton--$b$-tagged-jet system's transverse momentum, optimised to minimise the uncertainty in the measurement.
On the other hand, the measurement performed in the lepton-plus-jet channel relied on a simultaneous fit of three distributions: the top-quark and $W$-boson mass distributions as reconstructed by a kinematic fit in each event,
and the distribution of the ratio of $b$-jet to light-jet transverse momenta.
Thanks to these three variables, $m_t$ could be extracted at the same time as two overall jet-energy correction factors for $b$-jets and light jets, with a consequent reduction of jet energy scale uncertainties via such in situ constraints.
The combination of the two measurements, which were affected in different ways and in different directions by some of the most relevant sources of systematic uncertainty, could then benefit from this difference in design. The result is:
\[
m_t = 172.69 \pm 0.25 \text{ (stat.)} \pm 0.41 \text{ (syst.)~\GeV},
\]
corresponding to a total uncertainty of 0.48~\GeV.
Combining this with similar measurements by the CMS Collaboration based on the \RunOne data collected at $\sqrt{s} = 7$ and 8~\TeV~\cite{CMS:2023wnd}, yields an even smaller uncertainty:
\[
m_t = 172.52 \pm 0.14 \text{ (stat.)} \pm 0.30 \text{ (syst.)~\GeV}.
\]

With the LHC providing a larger dataset in \RunTwo than in \RunOne, ATLAS had the opportunity to repeat some of the most precise measurements performed in \RunOne and also to investigate new methods for measuring the top-quark mass, using observables less sensitive to hadronic-jet energy determination.
Using 36~fb$^{-1}$, ATLAS measured the top-quark mass using a purely leptonic observable, taking advantage of semileptonic decays of $b$-hadrons in top-quark decays~\cite{TOPQ-2017-17}.
In this analysis the idea is to select \ttbar events where one of the two top quarks decays leptonically ($t \rightarrow Wb \rightarrow \ell \nu b$), and then to require the presence of a relatively soft muon within the hadronic jet formed in the $b$-quark fragmentation process.
In this way, the partial top-quark invariant mass $m_{\ell\mu}$ could be built as the invariant mass of the system composed of the prompt charged lepton $\ell$ from the $W$-boson decay (considering either an electron or muon) and the \enquote{soft} muon from the $b$-jet.
Provided that the prompt lepton and soft muon originate from the same leg of the \ttbar decay process, the $m_{\ell\mu}$ value is strongly correlated with the top-quark mass, while having only a small dependence on the jet reconstruction and energy determination.\footnote{%
A residual jet-energy uncertainty still affects the final result, due to the jet-related selection requirements that need to be applied in the analysis, in terms of event selection and soft-muon identification.%
}
Soft muons were selected by requiring them to have an angular separation of $\Delta R < 0.4$ from a reconstructed jet.
In order to reduce the rate of kaons misidentified as muons, as well as the contribution from $b \rightarrow c \rightarrow \mu$ decay chains (where $b$ and $c$ here represent generic $b$- and $c$-flavoured hadrons), soft muons were required to satisfy tight identification criteria and to have $\pt >10$~\GeV.
The contamination from events where the two leptons come from different legs of the \ttbar decay was mitigated by a cut on the angular distance between the prompt lepton and the soft muon.
Finally, to control the residual $b \rightarrow c \rightarrow \mu$ contribution, due to its lower sensitivity to $m_t$, selected events were separated into two categories, depending on whether the two leptons have equal or opposite electric charge.
The top-quark mass was then obtained from a simultaneous binned profile-likelihood fit of the $m_{\ell\mu}$ distribution in the two event categories, yielding:
\[
m_t = 174.41 \pm 0.39 \textrm{ (stat.)} \pm 0.66\textrm{ (syst.)} \pm 0.25\textrm{ (recoil)~\GeV},
\]
where the statistical uncertainty and the contribution from systematic uncertainties are indicated separately.
This result is complementary to other more traditional direct top-quark mass measurements, as it is largely unaffected by jet energy scale uncertainties (with an impact of $\pm 0.13$~\GeV on $m_t$).
However, uncertainties in both the perturbative and non-perturbative parts of the $b$-fragmentation process (i.e.\ in the so-called parton-shower evolution and fragmentation function) have a larger impact on the final result (${\sim}0.2$~\GeV), despite having been reduced by carefully retuning the parton-shower and hadronisation model in the MC simulation to the most precise $e^+e^-$ data from LEP and SLD~\cite{ALEPH:2001pfo,DELPHI:2011aa,OPAL:2002plk,SLD:1999cuj}.
Moreover, uncertainties in the $b$-hadron decay fractions to different final states make the largest contribution to the total uncertainty in $m_t$ ($\pm 0.4$~\GeV).
In addition, studies have been performed, in contact with the theory community and MC experts, on the impact of the choice of recoil scheme in the simulation of the top-quark decay and successive QCD radiation off the $b$-quark~\cite{Brooks:2019xso}.
The impact of changing the default gluon-recoil scheme, from recoiling against the $b$-quark to recoiling against the $W$ boson or against the top quark, was found to be sizeable.
This affects the modelling of the second and subsequent gluon emissions from quarks produced by coloured-resonance decays and therefore changes both the fraction of jet energy carried by $b$-hadrons and the amount of radiation that fails to be clustered in $b$-jets.
These studies were used to derive a corresponding uncertainty (named the recoil uncertainty), reported as the third contribution to the total uncertainty in the result quoted above.
This uncertainty was not included in \RunOne measurements.

In parallel, ATLAS further developed the technique used for the most precise \RunOne $m_t$ measurement, namely its extraction from the unbinned maximum-likelihood fit in the dilepton channel~\cite{ATLAS:2022jpn}.
By taking advantage of the full 140~fb$^{-1}$ \RunTwo dataset, the analysis could be refined by optimising both the final-state reconstruction and the event selection in order to maximise the resolution of the defined observable and to minimise the impact of jet energy scale and \ttbar modelling systematic uncertainties.
In particular, a deep neural network (DNN) was trained to identify the best $b$-jet candidate to be assigned to each of the two charged leptons for the partial top-quark reconstruction.
After this DNN-based lepton--$b$-jet ($\ell b$) pairing selects the two pairs, only the one with higher \pt is used to build the final observable, following the same rationale as was behind the optimised event selection in the \RunOne analysis.
Similarly to \RunOne, top-quark-mass templates were built from simulated \ttbar events for the $m_{\ell b}$ observable.
An unbinned maximum-likelihood fit to the observed data events was used to extract the central value of $m_t$, with systematic uncertainties being estimated by repeating the fit on varied pseudo-data samples.
The resulting measurement is:
\[
m_t = 172.21 \pm 0.20 \text{ (stat.)} \pm 0.67 \text{ (syst.)} \pm 0.39 \text{ (recoil)~\GeV},
\]
with the last component of the uncertainty representing the impact of the choice of recoil scheme in the top-quark decay, as in the previously discussed analysis.
The other systematic uncertainties are dominated by jet energy scale and resolution uncertainties (with an impact of ${\sim}0.41$~\GeV on $m_t$) and by uncertainties in the matching scheme used between the NLO hard-scattering and parton-shower MC generators ($\pm 0.4$~\GeV).
Other important uncertainties are those from the initial- and final-state QCD radiation ($\pm 0.17$~\GeV) and colour-reconnection modelling ($\pm 0.27$~\GeV).
The impact of the choice of recoil modelling is large and similar to the case of the previously described analysis, as it directly affects the fraction of energy carried by the undetected neutrino in the leptonic top-quark decay.

Figure~\ref{fig:mass} shows a comparison of the ATLAS \RunTwo $m_t$ measurements with the results of the combination of the ATLAS direct $m_t$ measurements in \RunOne and of the combination of the ATLAS and CMS measurements.
The measurements are affected by mostly uncorrelated sources of uncertainty, which is reflected in the large spread.
A possible gain can be foreseen from repeating the measurements on larger datasets or in different final states, as well as from performing an updated combination, including \RunOne measurements.

\begin{figure}[!tbp]
\centering
\includegraphics[width=0.75\linewidth]{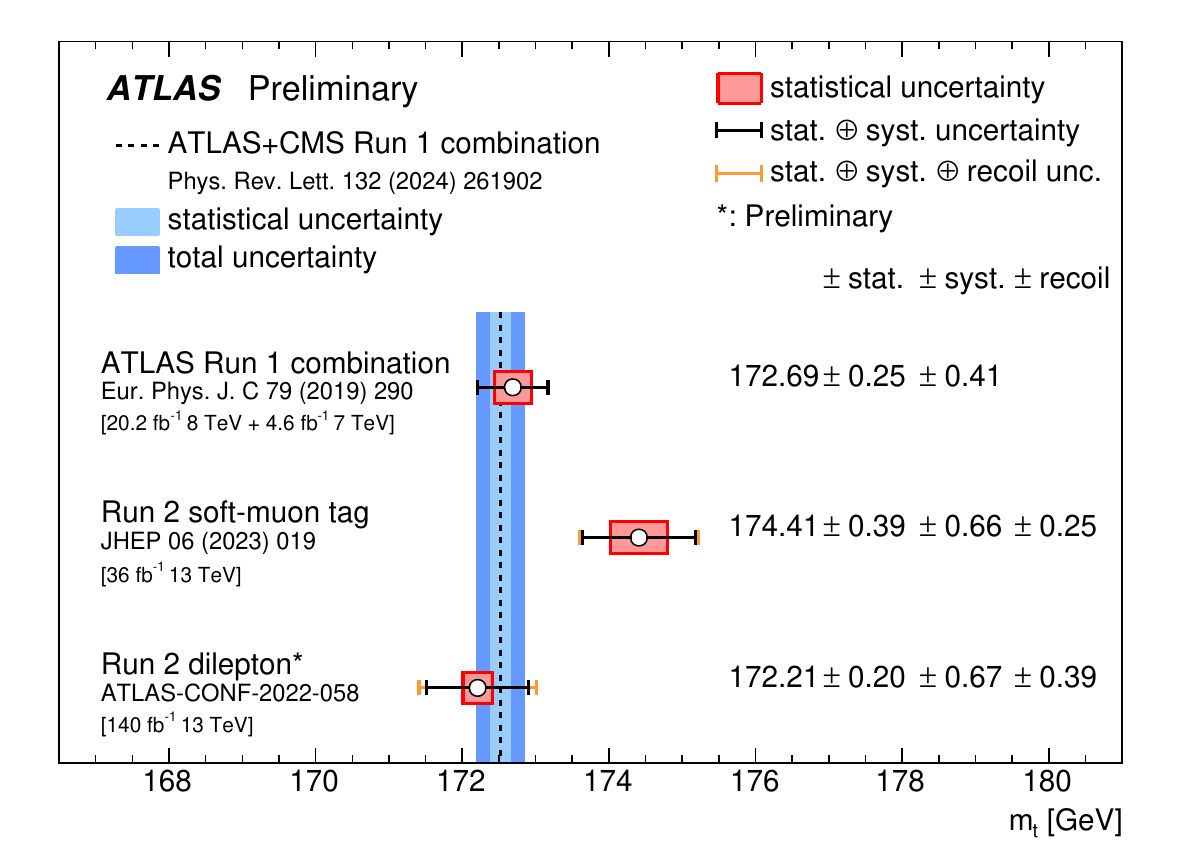}
\caption{Summary of the most precise top-quark mass measurements performed by ATLAS in \RunOne and \RunTwo.
The reference value shown by the vertical dashed line, with the blue bands indicating the statistical and total uncertainties, comes from the combination of the ATLAS and CMS \RunOne top-quark mass measurements.
For each of the measurements, the uncertainty is broken down into a statistical uncertainty and a systematic uncertainty component, with the effect of the recoil modelling indicated separately for the \RunTwo measurements (see text for more details).}
\label{fig:mass}
\end{figure}

\subsection{Indirect top-quark mass measurements}

For the indirect measurements, ATLAS took advantage of the improved precision in the determination of the inclusive \ttbar cross-section (see Section~\ref{sec:ttbar}) relative to \RunOne.
As reported in Ref.~\cite{TOPQ-2018-17}, the inclusive \ttbar cross-section measurement in the dilepton $e\mu$ channel based on 36~\ifb\ of \RunTwo data was used to extract the top-quark pole mass via its effect on the predicted cross-section, yielding $m_t^{\text{pole}}=173.1^{\,+2.0}_{-2.1}$~\GeV.
The result, obtained with CT14 as the reference PDF set, is dominated by uncertainties in the theoretical cross-section evaluated through PDF+\alphas and QCD-scale variations.
Therefore, improved experimental measurement precision (such as that achieved in the updated cross-section measurement based on the full \RunTwo dataset) would not have significant effects on this determination.
Instead, extractions from differential cross-section measurements, especially when using the same technique as in the \RunOne top-quark mass measurement from \ttbar{}+1-jet production~\cite{TOPQ-2017-09} which obtained a precision of ${\sim}1$~\GeV, have the potential to improve the precision of indirect determinations.


%
\section{Top-quark properties}
\label{sec:properties}

%
In addition to its very large mass, the top quark has other unique properties.
Its decay width ($\Gamma_t \approx 1.4$~\GeV~\cite{ParticleDataGroup:2022pth}) is larger than the QCD hadronisation scale $\Lambda \approx 250$~\MeV, so the top-quark decay time is shorter than the typical hadronisation time, unlike other quarks.
This unique feature allows its properties to be studied through its decay products.

Precisely predicted properties of top-quark decays are measured as a stringent test of the SM and as a probe of contributions from BSM physics.
These include the polarisation of the $W$ bosons produced in top-quark decays, the spin correlations and entanglement of top-quark pairs, and their forward--backward asymmetry.
Additionally, top-quark decays are used both to test the modelling of QCD effects such as colour connection and as a source of $W$ bosons to test the universality of their couplings to different lepton families.

These properties were studied extensively in \RunTwo data, benefiting from a data sample much larger than the one from \RunOne.
As it is cleaner than the other channels, the dilepton channel was often used for these studies
despite the more complex reconstruction of top-quark kinematics in this channel.

\subsection{Top-quark decay angular properties}

Since the top quark decays almost exclusively into a $W$ boson and a $b$-quark, its decay products are naturally useful for studying the $Wtb$ interaction vertex. This vertex and the masses of the interacting particles define the fractions of $W$ bosons with longitudinal ($f_0$), left-handed ($f_\text{L}$), and right-handed ($f_\text{R}$) polarisation from the top-quark decays. Because of the vector minus axial-vector $(V-A)$ nature of the $Wtb$ interaction, $f_\text{R}$ is expected to be very small, making it particularly sensitive to signs of new physics.
The helicity fractions are predicted at NNLO in QCD to be
$f_0 = 0.687 \pm 0.005$, $f_\text{L} = 0.311 \pm 0.005$ and $f_\text{R} = 0.0017 \pm 0.0001$~\cite{Czarnecki:2010gb}, with $f_0+f_\text{L}+f_\text{R}=1$.
These fractions could be altered by new physics processes, and
deviations from the SM expectation can be parameterised in terms of
dimension-six EFT operators affecting the $Wtb$ vertex.
The full \RunTwo data sample was used to measure these fractions~\cite{ATLAS:2022rms} in the dilepton \ttbar final state.
They are accessible through the normalised differential distribution of the cosine of the polar angle $\theta^*$, defined as the angle between the momentum of the charged lepton from the $W$-boson decay and the reversed momentum of the $b$-quark from the top-quark decay, both calculated in the $W$ boson's rest frame.
Measuring $\cos \theta^*$ requires the reconstruction of the \ttbar kinematics,
which is achieved by using the Neutrino Weighting method.
Using top-quark and $W$-boson mass constraints, this method scans over the neutrino pseudorapidities to find two possible kinematic solutions that are the most compatible with
the measured \met of the event.
Unlike the $\sqrt{s}=8$~\TeV results, which were obtained using a template method applied to detector-level distributions, the $W$ helicity fractions in the $\sqrt{s}=13$~\TeV analysis were extracted by unfolding the normalised differential $\cos \theta^*$ distribution to parton level and
fitting the unfolded cross-section distribution, minimising the $\chi^2$ value. In the fit, the $f_0$ parameter was set to $f_0 = 1 - f_\text{L} - f_\text{R}$. The results are
$f_0 = 0.684 \pm 0.005 \text{(stat.)} \pm 0.014 \text{(syst.)}$,
$f_\text{L} = 0.318 \pm 0.003 \text{(stat.)} \pm 0.008 \text{(syst.)}$, and
$f_\text{R} = - 0.002 \pm 0.002 \text{(stat.)} \pm 0.014 \text{(syst.)}$
in agreement with the SM predictions to within one standard deviation.
The systematic uncertainty dominates the total uncertainty for all three helicity fractions. The largest systematic uncertainty arises from the uncertainty in \ttbar modelling, so better understanding of the \ttbar MC simulation would be needed to improve this measurement.

Top-quark pair production at the LHC is mostly mediated by the parity-invariant strong interaction, so the top quarks and antiquarks are predicted to be produced unpolarised in the SM, while the spins of the top quark and top antiquark are expected to be correlated.
Since the spin correlation is transferred to their decay products, and almost maximally to the two leptons~\cite{Brandenburg:2002xr}, the study can be performed with the leptons from the top-quark and top-antiquark decays.
The dilepton channel is particularly relevant for measuring spin correlations because,
in addition to their sensitivity to spin correlations, charged leptons are easy to identify in hadron collisions.
This correlation has been observed experimentally at the LHC using both $\sqrt{s}=7$~\TeV~\cite{ATLAS:2012ao,ATLAS:2014aus} and 8~\TeV~\cite{ATLAS:2014abv,ATLAS:2016bac} collisions, showing slightly stronger correlation
than expected, although with rather large experimental uncertainties.
Studies of spin correlation at $\sqrt{s}=13$~\TeV with large datasets are therefore particularly relevant.
Using 36~\ifb\ of $\sqrt{s}=13$~\TeV data, ATLAS performed a measurement in the channel with one electron and one muon~\cite{ATLAS:2019zrq}.
The full spin information of the top quarks is encoded in the spin density matrix~\cite{Bernreuther:2015yna}.
The simplest observable sensitive to spin correlation is the azimuthal opening angle $\Delta \phi$ between the electron and the muon in the transverse plane, measured in the laboratory frame.
The spin correlation measurement can be performed inclusively, but also in different \ttbar invariant-mass bins since the degree of correlation is expected
to vary with the \ttbar invariant mass. The $\sqrt{s}=13$~\TeV analysis also utilised the difference between the pseudorapidities of the two charged leptons
as an additional observable, $\Delta \eta$. This observable is less sensitive to spin correlation than $\Delta \phi$ but, in addition to $\Delta \phi$,
is useful when searching for the presence of supersymmetric top squarks
with a mass close to the top-quark mass.
The \ttbar\ invariant mass was reconstructed using the Neutrino Weighting method.
The $\Delta \phi$ and $\Delta \eta$ distributions were corrected to parton level and particle level using an iterative Bayesian unfolding, and
the resulting absolute and normalised cross-sections were compared with predictions from NLO MC generators.
The dominant systematic uncertainty in this measurement comes from the modelling of initial- and final-state radiation.
The comparison revealed several shape effects, with data tending to be higher than the expectation at low $\Delta \phi$ or high $\Delta \eta$ (see Figure~\ref{fig:spin-entanglement}~(a)).
The compatibility with the SM prediction was assessed using a template fit to the normalised parton-level cross-sections,
with one template from dilepton \ttbar events with SM spin correlation and one where spin correlation had been removed.
Using the inclusive $\Delta \phi$ distribution, the extracted spin-correlation fraction is
$f_\text{SM} = 1.249 \pm 0.024 \text{ (stat.)} \pm 0.061 \text{ (syst.)} ^{~+0.067}_{\,-0.090} \text{ (theo.)}$.
When including the template's theoretical uncertainties, the measurement is 2.2~standard deviations higher than the SM expectation of $f_\text{SM}=1$.
The value of $f_\text{SM}$ is observed to increase slightly as a function of the \ttbar\ invariant mass, \mttbar, but no bin shows a
significant deviation from the prediction, due to the still large statistical and systematic uncertainties, and the relatively poor \mttbar\ resolution.
Several cross-checks were performed to understand the sensitivity of the result to the limitations of the \ttbar\ modelling, such as
the impact of the narrow-width approximation, the impact of NNLO corrections, or the use of expanded NLO predictions.
None of these alternative predictions agree completely with the measurements, even though including higher-order effects brings the predictions closer to the data.
Studies with the full \RunTwo dataset or during the future LHC runs, as well as improvements in the predictions, should shed further light on this difference.
The double-differential distributions of $\Delta \phi$ in bins of $\Delta \eta$ were used to search for the
pair production of supersymmetric top squarks.
In the absence of a signal, top squarks with a mass between 170~\GeV and 230~\GeV were excluded for most of the allowed neutralino mass range.

\subsection{Quantum entanglement}
Precise measurements of the \ttbar spin density matrix in a very restricted phase space were recently proposed as a new laboratory to study
quantum information properties, especially entanglement~\cite{Afik:2020onf,Fabbrichesi:2021npl,Severi:2021cnj,Aguilar-Saavedra:2022uye,Ashby-Pickering:2022umy}.
Entanglement is a feature of quantum mechanics, where two particles cannot be described independently of each other. This has been observed in many systems but only at lower energy scales.
The LHC provides the opportunity to study this effect at high energies and in systems composed of other elementary particles.
The spins of the top and anti-top quarks form a two-qubit system.
Entanglement in top-quark pairs can be studied via the spin correlation between the produced top quark and top antiquark, using leptons from their decays as spin-analysing particles.
When produced close to their production threshold (i.e.\ $\mttbar \approx 2 m_t$), the \ttbar pairs produced through gluon--gluon fusion are in a spin-singlet state.
In this case, maximum entanglement among the spins of the top quark and top antiquark is expected.
When the \ttbar system has a larger mass, the entanglement is reduced
(but increases again at very high top-quark \pt above around 500~\GeV).
A simple observable can signal the presence of entanglement close to the production threshold~\cite{Afik:2020onf}: $D = - 3 \langle \cos \varphi \rangle$, where $\varphi$ is the angle between the charged-lepton directions in the respective parent top-quark and top-antiquark rest frames. The existence of an entangled state is then demonstrated if $D < -1/3$.
ATLAS performed the first study of quantum entanglement at high energy using \ttbar events in the channel with one electron and one muon using the full \RunTwo dataset at $\sqrt{s}=13$~\TeV~\cite{ATLAS:2023fsd}.
The measurement of $D$ requires the reconstruction of the top-quark and top-antiquark kinematics. The main method used in this analysis was the Ellipse method~\cite{Betchart:2013nba}, which analytically calculates the unmeasured neutrino momenta through a geometrical approach. If the Ellipse method failed, the Neutrino Weighting method was used instead. The optimal window for the signal region was determined to be $340 < \mttbar < 380$~\GeV at particle level. Two validation regions where entanglement is expected to be small were also defined to validate the method used for the measurement: one close to the threshold region ($380 < \mttbar < 500$~\GeV) and one at higher \mttbar\ ($\mttbar > 500$~\GeV). The observed $\cos \varphi$ distribution at reconstruction level was corrected, after background subtraction, for distortions from the detector response and event selection by using a simple calibration curve that relates the reconstructed values to the corresponding particle-level values.
The resulting value of $D$ at particle level in a fiducial phase space is $D = -0.547 \pm 0.002 \text{ (stat.)} \pm 0.021 \text{ (syst.)}$. This value is compared with the entanglement limit at particle level, which was obtained by converting the parton-level bound of $D = -1/3$ to the particle-level equivalent taking into account parton-shower effects from the \POWPY or \POWHER generators (see Figure~\ref{fig:spin-entanglement}~(b)).
Despite the large discrepancy between the results obtained using the two generators, the observed $D$ value is well below the entanglement limit, beyond five standard deviations, establishing the discovery of an entangled \ttbar state.
It is important to note that the measured value for $D$ is also significantly smaller than the predictions from the \POWPY and \POWHER generators.
This could be at least partially explained by non-relativistic QCD processes, such as Coulomb bound state effects~\cite{Kiyo:2008bv}, that are known to affect the \ttbar production close to the production threshold and are not accounted for in the MC generators.
Nevertheless, this measurement constitutes the first observation of entanglement of spin in a quark--antiquark pair, paving the way for further studies of fundamental quantum mechanics at the LHC, such as measurements of quantum discord or the steering ellipsoid~\cite{Afik:2022dgh} or testing the Bell inequality in \ttbar events~\cite{Fabbrichesi:2021npl, Afik:2022kwm}.

\begin{figure}[!tbp]
\centering
\subfloat[]{\includegraphics[width=0.45\linewidth]{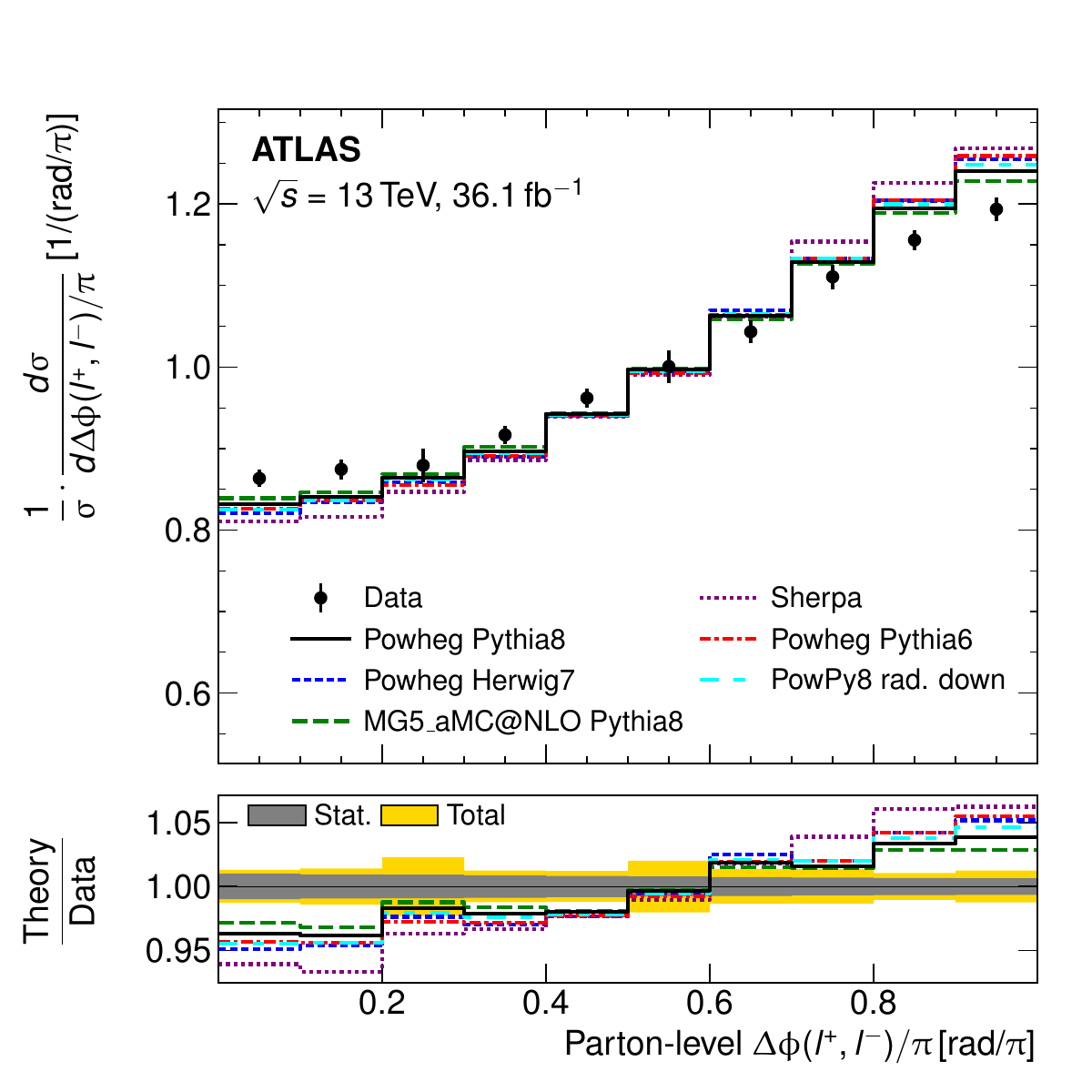}}
\subfloat[]{\includegraphics[width=0.45\linewidth]{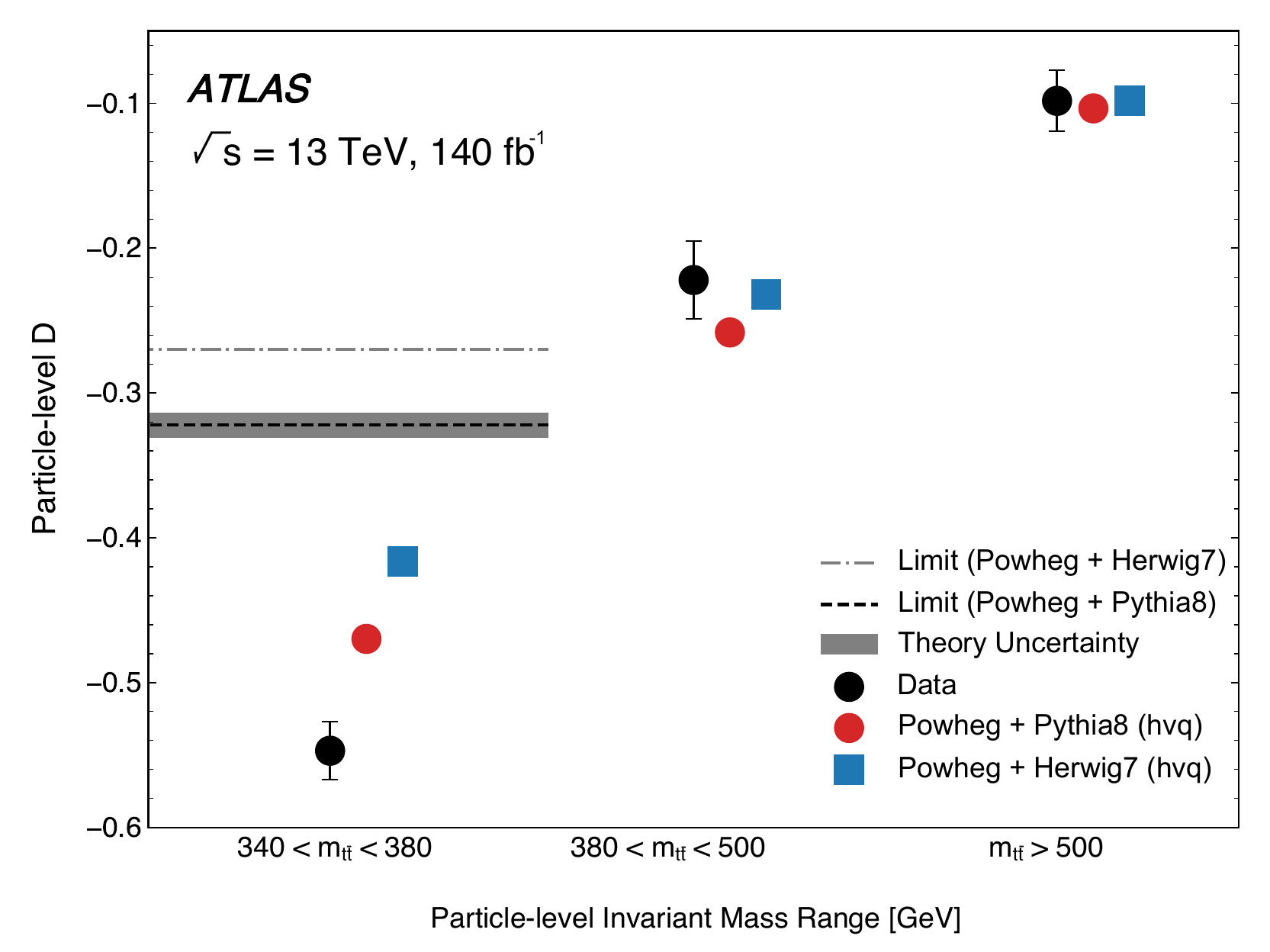}}
\caption{(a) The parton-level normalised $\Delta \phi$ differential cross-section compared with various predictions~\cite{ATLAS:2019zrq}.
(b) Particle-level results for the measurement of the quantum entanglement observable $D$ in the signal and validation regions, compared with various MC models~\cite{ATLAS:2023fsd}.}
\label{fig:spin-entanglement}
\end{figure}

\subsection{Asymmetry measurements}

At leading order in QCD, \ttbar production is symmetric under charge conjugation.
However, at NLO, an asymmetry occurs between the top quark and the top antiquark.
This asymmetry comes from the
interference of initial-state and final-state radiation (ISR and FSR) diagrams and from the interference of the Born and box diagrams for quark--antiquark initial states: $q \bar{q} \to \ttbar $.
As a result, the top quark (top antiquark) is preferentially emitted in the direction of the incoming quark (antiquark).
Production via gluon--gluon fusion, $gg \to \ttbar$, which is dominant at the LHC, is charge symmetric at all orders.
The asymmetry in $q\bar{q}$ manifests itself by the top antiquark being produced more centrally while the top quark is produced at more forward rapidities.

The charge asymmetry at the LHC can then be defined as:
\begin{equation*}
A^{\ttbar}_\text{C} = \frac{N(\Delta|y| > 0) - N(\Delta|y| < 0)}{N(\Delta|y| > 0) + N(\Delta|y| < 0)},
\end{equation*}
with $\Delta|y|= |y_{t}| - |y_{\bar{t}}|$.
Since \ttbar production by gluon--gluon fusion, which is symmetric, dominates at the LHC, the inclusive asymmetry is expected to be small, of the order of 1\%.
For events where the top quark and antiquark decay leptonically, the leptons from the top quark or antiquark inherit the directions of their parent quarks, so a similar asymmetry can be defined using the pseudorapidities of the leptons via the $\Delta|\eta|= |\eta_{\ell^+}| - |\eta_{\ell^-}|$ observable.
This leptonic charge asymmetry \All\ is slightly diluted compared to $A^{\ttbar}_\text{C}$ but has the advantage of not requiring the kinematic reconstruction of the \ttbar\ pair.
It is also interesting to measure the asymmetries differentially as a function of kinematic variables of the \ttbar system, such as the transverse momentum \ptttbar, the invariant mass \mttbar, and the boost of the \ttbar system in the $z$ direction, \betattbar.
In the SM, the charge asymmetries are expected to be enhanced for high values of some of these variables.
Several BSM models predict a modification of the asymmetry, for instance by anomalous vector or axial-vector couplings (e.g.\ axigluons or a heavy $Z'$ boson~\cite{Antunano:2007da,Frampton:2009rk,Rosner:1996eb,Aguilar-Saavedra:2011bsa}).
These modifications could also be studied with EFT~\cite{Rosello:2015sck}.
In particular, BSM effects are expected to be enhanced in specific kinematic regions, such as in the phase space with large
\betattbar\ or large \mttbar~\cite{Aguilar-Saavedra:2011dxk}.

Asymmetry measurements in \ttbar\ events were first performed at $\sqrt{s}=8$~\TeV~\cite{ATLAS:2017gkv}, although with rather large uncertainties, around 60\% of the predicted asymmetry.
The fraction of $q\bar{q}$-initiated top-quark pair production decreases with increasing centre-of-mass energy, and hence the \ttbar charge asymmetry decreases as well.
Despite this disadvantage, ATLAS measured $A^{\ttbar}_\text{C}$ in both the single-lepton and dilepton channels, and \All\ in the dilepton channel, at $\sqrt{s}=13$~\TeV~\cite{ATLAS:2022waa}.
The lepton+jets channel was further split into resolved and boosted topologies.
In the resolved case, the assignment of the jets to the corresponding partons from the decaying top quarks was assessed with a BDT that aims to discriminate between signal and the combinatorial background, separately for events with one or two $b$-tagged jets.
In the boosted topology, the selected large-radius jet was taken to be the hadronically decaying top quark. In both cases, the semileptonically decaying top quark's four-vector was
reconstructed from the lepton and a small-radius jet, calculating the neutrino four-vector from the missing transverse momentum and the $W$-boson mass constraint.
In the dilepton channel, the \ttbar system was reconstructed using the Neutrino Weighting method.
The differential $\Delta |y|$ distributions were corrected for acceptance and detector effects using the FBU method, where systematic uncertainties that affect the measurements are treated as nuisance parameters.
The combined inclusive $A^{\ttbar}_\text{C}$ asymmetry from single-lepton and dilepton events was measured to be $0.0068 \pm 0.0010 \mathrm{(stat.)} \pm 0.0010 (\mathrm{syst.})$, in agreement with the SM calculation of $0.0064^{\,+0.0005}_{-0.0006}$ at NNLO accuracy in the strong coupling with NLO electroweak corrections~\cite{Czakon:2017lgo}. The SM computation was performed by expanding the numerator and denominator to a given order in perturbation theory.
The measurement differs from zero by 4.7~standard deviations, providing strong evidence for \ttbar charge asymmetry at the LHC. The precision of the result is dominated by the lepton+jets channel because of its smaller statistical uncertainty.
The $A_\text{C}^{\ell \bar{\ell}}$ asymmetry, measured in the dilepton channel only, is $0.0054 \pm 0.0012 \mathrm{(stat.)} \pm 0.0023 (\mathrm{syst.})$, while the SM calculation at NLO in QCD, including NLO EW corrections, predicts $0.0040^{\,+0.0002}_{-0.0001}$~\cite{Bernreuther:2012sx}.
The combined $A^{\ttbar}_\text{C}$ results were interpreted in terms of EFT using new operators for four-quark interactions with different coupling chiralities (see Section~\ref{sec:eft}).
Differential $A^{\ttbar}_\text{C}$ measurements were performed as a function of \mttbar, \ptttbar\ and \betattbar, with the binning at larger values being finer than was possible for $\sqrt{s}=8$~\TeV data (see Figure~\ref{fig:asymmetry-substructure}~(a)).
Differential measurements of \All\ were presented as a function of the invariant mass, transverse momentum and longitudinal boost of the dilepton pairs.
The results were found to be compatible with the SM predictions.

Another way to study the \ttbar charge asymmetry is to use an observable linked to the energy difference between the top quarks and antiquarks, $\DeltaE = E_t - E_{\bar{t}}$.
The energy asymmetry~\cite{Berge:2013xsa} occurs mainly through the $q g \to \ttbar q$ process, which is a more abundant source of events than the $q\bar{q} \to \ttbar$ process at the LHC. It is then expected to be larger than asymmetries based on rapidity.
The presence of an additional jet allows QCD effects at leading order to be investigated,
while the asymmetry in $q\bar{q} \to \ttbar$ only appears at NLO.
In the $pp \to t\bar{t}j$ process, the energy asymmetry can be defined as a function of the jet angle \thetaj:
\begin{equation*}
A_E(\thetaj) = \frac{ \sigma_{\ttj}(\thetaj | \DeltaE > 0) - \sigma_{\ttj}(\thetaj | \DeltaE < 0)}{ \sigma_{\ttj}(\thetaj | \DeltaE > 0) + \sigma_{\ttj}(\thetaj | \DeltaE < 0) },
\label{eq:Ae-def}
\end{equation*}
where $\sigma_{\ttj}(\theta_j)$ is the differential $\ttj$ cross-section as a function of $\theta_j$.
Both $\DeltaE$ and $\thetaj$ are defined in the \ttj rest frame, which corresponds to the partonic centre-of-mass frame in tree-level processes.
ATLAS measured this energy asymmetry differentially using 139~\ifb\ of $\sqrt{s}=13$~\TeV data~\cite{ATLAS:2021dqb}.
The analysis was performed in the semileptonic \ttbar decay channel.
The number of events observed at detector level was corrected for detector effects to particle level using the FBU method.
The uncertainty in the measurement is dominated by the statistical component.
The measured differential distribution
was found to be in good agreement with the SM expectation, with a $p$-value of 0.80.
In the most sensitive bin, $\pi/4 \le \thetaj \le 3\pi/5$, the asymmetry is measured to be: $-0.043 \pm 0.02$ for a SM expectation of: $-0.037 \pm 0.02$.
In the first bin, $0 \le \thetaj \le \pi/4$, the measured asymmetry differs from zero by 2.1~standard deviations.
The sensitivity of this energy asymmetry measurement to new physics was investigated in the context of EFT (see Section~\ref{sec:eft}).
The energy asymmetry is particularly sensitive to the chirality and colour charges of the involved operators. It complements the constraints from asymmetries built using rapidities.

The large \RunTwo data sample allows asymmetries to be measured in rarer processes where \ttbar\ is produced with an associated gauge boson.
Some of these processes are predicted to have a larger asymmetry than in \ttbar production.
For instance, the \ttW process is initiated at LO by a $q \bar{q}'$ initial state. These \ttW events can then serve as an interesting tool for measuring the \ttbar\ charge asymmetry since it is expected to be larger than in \ttbar\ production~\cite{Maltoni:2014zpa,Bevilacqua:2020srb}.
In addition, the $W$ boson in this process can be radiated from the
$q \bar{q}'$ initial state, thereby serving as a way
to polarise the $q \bar{q}'$ pair and thus also the \ttbar pair.
This polarisation further enhances the asymmetry between the decay products of the top quarks and top antiquarks, leading to an enhanced leptonic asymmetry.
The drawback of using the \ttW process is, however, its much smaller cross-section in comparison with \ttbar production.
Despite the challenges, ATLAS probed the asymmetry in \ttW events using the full \RunTwo dataset~\cite{ATLAS:2023xay}.
The measurement was performed in the 3L channel at detector level and also at particle level after unfolding.
In order to compute \All,
the two opposite-sign leptons from the top-quark decays need to be separated from the one from the $W$ decay.
This was addressed by using a BDT.
The second lepton needed to compute \All\ was taken to be the lepton with charge opposite to that of the lepton selected by the BDT.
A profile-likelihood fit was used to extract the signal, together with the normalisation of each of the most relevant background processes,
i.e.\ \ttZ, non-prompt electrons and muons, and electrons from $\gamma$-conversions.
The leptonic charge asymmetry in \ttW events was measured to be
$\All (t\bar{t}W) = -0.12 \pm 0.14 \mathrm{(stat.)} \pm 0.05 \mathrm{(syst.)}$.
The SM prediction from the \Sherpa simulation~\cite{Sherpa:2019gpd} in this phase space is $-0.084^{\,+0.005}_{-0.003} \mathrm{(scale)} \pm 0.006 \mathrm{(MC~stat.)}$.
With the current precision,
the result is compatible with zero and not yet sensitive to the SM asymmetry.
The asymmetry was also unfolded to particle level in a fiducial phase space.
The results are limited by statistical uncertainty, so they are expected to improve in the years to come.

The \tty final state is also relevant for asymmetry measurements~\cite{Aguilar-Saavedra:2014vta}.
When the photon is emitted from the initial state, the process benefits from an enhanced fraction of quark--antiquark-initiated \ttbar production relative to symmetric production via gluon--gluon fusion.
In this process, the asymmetry arises through the interference of QED initial-state radiation and final-state radiation, which yields a negative asymmetry.
The overall asymmetry in \tty events at $\sqrt{s}=13$~\TeV is predicted in the SM to be between $-1\%$ and $-2\%$ depending on the phase space~\cite{Pagani:2021iwa,Bergner:2018lgm}.
The main challenge of studying the charge asymmetry in \tty events comes from the fact that the asymmetry is only present for events where the photon is radiated from the initial-state parton or from the final-state top quark or antiquark. It is diluted by \tty events where the photon is emitted from any of the charged decay products of the \ttbar final-state system.
ATLAS measured this charge asymmetry in \tty production for the first time using the full 139~\ifb\ dataset recorded at $\sqrt{s}=13$~\TeV~\cite{ATLAS:2022wec} and the semileptonic \ttbar decay channel.
The top-quark and top-antiquark kinematic properties, in particular their rapidities, were reconstructed using the KLFitter package.
The estimation of background events with prompt or misidentified photons followed the methods developed for the measurement of the \tty cross-section described in Section~\ref{sec:associated}.
A neural network was used to separate the \tty signal from the background processes.
The asymmetry value was extracted from the $\Delta |y_t|= |y_t|-|y_{\bar{t}}|$ distribution in a fiducial region at particle level after performing
a maximum-likelihood unfolding.
After the fit, the asymmetry was found to be
$A^{\tty}_\text{C} = -0.003 \pm 0.024 \text{(stat.)} \pm 0.017 \text{(syst.)}$ (assuming a SM charge asymmetry in \ttbar events of $A^{\ttbar}_\text{C} = 0.0064$). The measured value is compatible with the result from NLO simulation in the same phase space: $A^{\tty}_\text{C} = - 0.014 \pm 0.001$.
The dependence of the measured $A^{\tty}_\text{C}$ on the asymmetry in \ttbar events was also studied and found to be linear.
Here also the result is still limited by the statistical uncertainty and so is expected to be improved in the future.

The different asymmetry measurements are summarised in Figure~\ref{fig:asymmetry-summary}.
They are in agreement with the SM predictions.
The uncertainties in the SM predictions are much smaller than the current experimental uncertainties,
which are dominated by statistical uncertainties, as systematic effects largely cancel out.
Therefore, improvements in the experimental set-up, as well as the analysis of the larger datasets available after the completion of LHC \RunThr and at the HL-LHC, will lead to more valuable comparisons.

\begin{figure}[!tbp]
\centering
\includegraphics[width=0.8\linewidth]{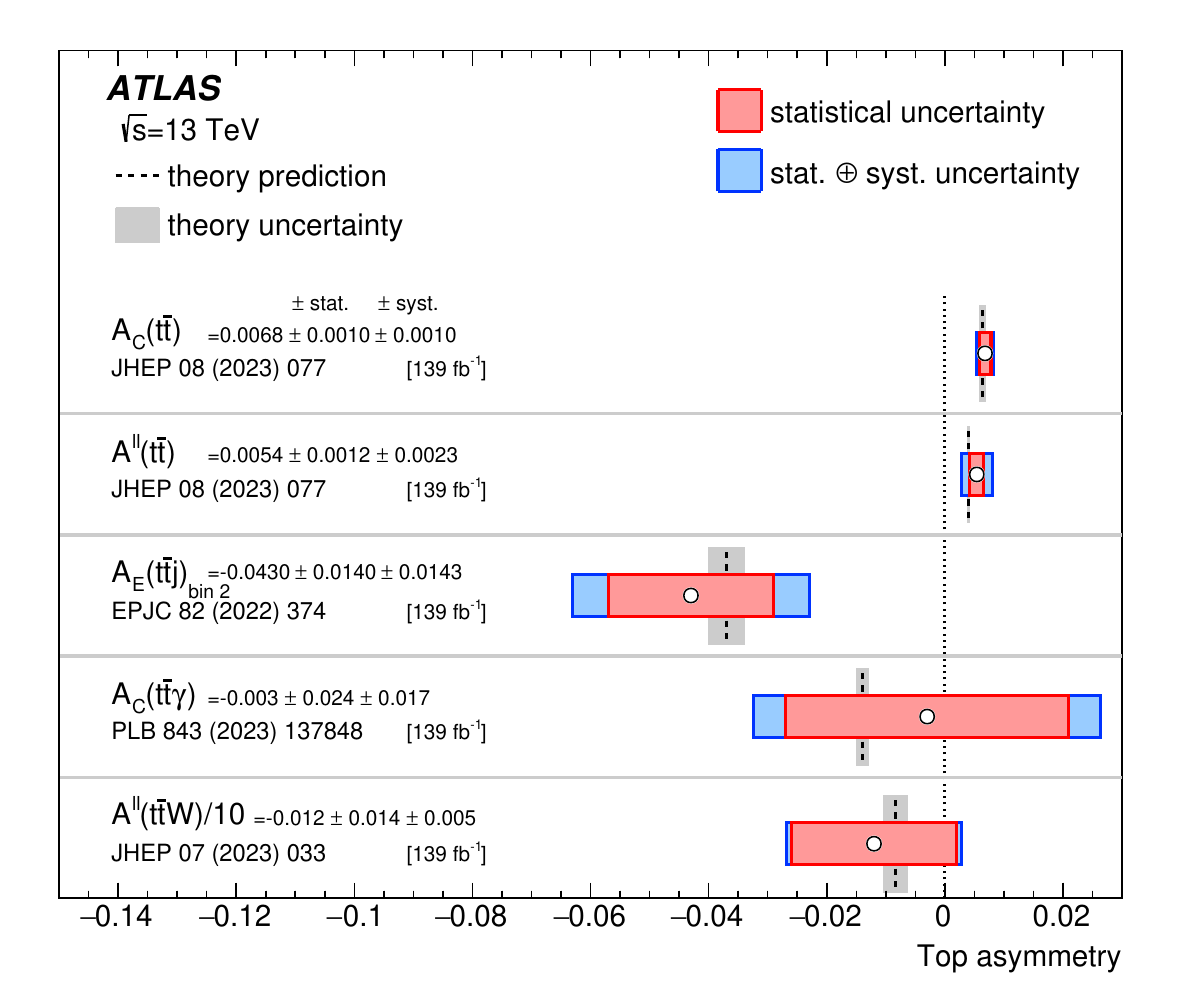}
\caption{\label{fig:asymmetry-summary} Summary of the $\sqrt{s}=13$~\TeV top asymmetry measurements using the full \RunTwo dataset: \ttbar\ asymmetry, lepton-based asymmetry, energy asymmetry in the bin with the largest expected asymmetry, and asymmetry in \tty and \ttW events. Each of the results is compared with their corresponding theoretical expectation. The value of the \ttbar asymmetry in \ttW events is shown divided by 10 for display purposes.}
\end{figure}

\subsection{Tests of QCD}

Top pair production also serves as a unique environment to test QCD and tune the modelling of QCD effects in MC generators.
One of the aspects of QCD that can be tested with \ttbar\ events is the modelling of radiation through colour connection and colour propagation, i.e.\ colour flow, which is modelled by colour strings between quarks and gluons.
Experimentally, quarks and gluons appear as jets.
The radiation emission pattern between particles, governing the structure of the jets, is affected by their colour connections.
In particular, QCD predicts that there is more radiation between particles that are colour-connected.
Experimental measurements are needed to validate the phenomenological description of such predictions.
In \ttbar\ production where only one of the top quarks decays leptonically, there are four quarks as decay products.
The two quarks from the decay of the $W$ boson, which is a colour-neutral object, are expected to be colour-connected.
The two $b$-quarks from the top-quark decays carry the colour of their parent top quarks but are not expected to share any colour connection.
These features can be tested experimentally.
Building on the $\sqrt{s}=8$~\TeV analysis~\cite{ATLAS:2015ytt}, ATLAS performed a measurement of colour flow using 36~\ifb\ of $\sqrt{s}=13$~\TeV data~\cite{ATLAS:2018olo}.
The observable used to encode the colour information was the so-called jet-pull vector $\vec{\cal{P}}$~\cite{Gallicchio:2010sw}.
It is defined as the sum of the transverse momenta of the jet constituents $i$ weighted by their distance from the jet axis, $\Delta \vec{r}_i = (\Delta y_i, \Delta \phi_i)$, in the
$y{-}\phi$ space and normalised by the jet \pt: $\vec{\cal{P}} = (1/\pt)\sum_i |\Delta \vec{r}_i | \pt^i\;\!\Delta \vec{r}_i$.
With two jets $j_1$ and $j_2$ in the $y{-}\phi$ space, one can construct the jet-pull angle $\theta_{\cal{P}}$ between the jet-pull vector $\vec{\cal{P}}(j_1)$
and the vector connecting $j_1$ and $j_2$: $(y_{j_2}- y_{j_1}, \phi_{j_2} - \phi_{j_1})$.
If the two jets $j_1$ and $j_2$ are colour-connected,
$\vec{\cal{P}}(j_1)$ and $\vec{\cal{P}}(j_2)$ are aligned with the connecting vector and so
a bias toward $\theta_{\cal{P}} \sim 0$ is expected, whereas if they are not connected,
the distribution of the jet-pull angle is expected to be uniform.
Applied to \ttbar events, three of the four observables measured are the jet-pull angles, $\theta_{\cal{P}}(j^W_1, j^W_2)$ and $\theta_{\cal{P}}(j^W_2, j^W_1)$, of the jets from the $W$-boson decays, $j^W_1$ and $j^W_2$, taken
to be the two leading non-$b$-tagged jets, and the magnitude of the pull-vector from $j^W_1$.
These are expected to show colour connections.
The jet-pull angle of the two leading $b$-tagged jets was also measured but is not expected to show colour connection.
In this analysis, the jet-pull vectors were calculated only from tracks in the inner detector.
All four reconstructed observables were unfolded to particle level using the IBU method and normalised, so as to study
only their shape and mitigate the impact of systematic uncertainties.
The results were compared with several generator predictions that differ in their hadronisation modelling (the string model for \Pythia and the cluster model for \Herwig).
Agreement with the \Herwig modelling was found to be better.
The unfolded distributions are also used to test the prediction of an exotic model
that implements flipped colour flow.
The data disfavours this model, with $p$-values of at most 0.002 for all the tested variables.
These unfolded data can be used to tune MC generators to better model the effects of colour connections between partons.

The top-quark sector is also useful for tuning colour models that can not be derived from QCD first principles.
In MC generators, the colour information is traced using the leading-colour approximation~\cite{Nagy:2008ns,Buckley:2011ms}, where each quark is connected to only one other parton, while
each gluon, which carries a colour and an anticolour, is connected to two other partons.
The multiple parton interactions overlaid on the hard-scattering process add additional colour lines, which could potentially lead to phase-space overlaps.
The colour-reconnection models~\cite{Sjostrand:2013cya} reassign colour connections between partons in order to resolve these overlapping colour lines.
There are several mechanisms that could be introduced and that should be tuned to data.
The exact involvement of the top quark and its decay products in these processes is not known. The current models involved only
the top quark itself in the colour-reconnection mechanisms, and not its decay products, so it was important to test these models on data.
With 139~\ifb\ of $\sqrt{s}=13$~\TeV data, ATLAS used \ttbar\ events in the channel with one electron and one muon, and two or three jets (including exactly two $b$-tagged jets) to measure
distributions sensitive to colour reconnection~\cite{ATLAS:2022ctr}.
The chosen observables were the charged-particle multiplicity outside the jets (excluding leptons from the top-quark decays), the scalar sum of the transverse momenta
of these charged particles, and the double-differential cross-section in these two quantities.
These observables need to be corrected for %
tracks from \pileup and from
secondary vertices, and for tracking inefficiencies.
The corrected observables were unfolded to particle level using the IBU method to obtain normalised differential cross-sections.
The measurements have a precision of 2\%--10\% in the central bins and up to 50\% in the outer bins.
Agreement between the measured differential cross-sections and various models implemented in MC generators was assessed by means of a $\chi^2$ test.
The \Herwig generator describes the data well for the observable built with the \pt scalar sum, while the prediction from \Pythia is better for the charge-particle multiplicity.
The results were also compared with predictions from different models of multiple parton interactions, since the chosen observables are also sensitive to these.
The results could be used for future tuning of the MC parameters for both colour reconnection and multiple parton interactions, which should be performed simultaneously.

As discussed in Sections~\ref{sec:intro_obj} and~\ref{sec:ttbar_differential}, hadronically decaying top quarks with sufficiently high transverse momentum can be reconstructed as single large-radius jets.
These boosted top-quark jets are characterised by their distinctive substructure, which is often used to separate them from energetic jets arising from lighter quarks or gluons~\cite{JETM-2018-03,CMS-JME-18-002,Kasieczka:2019dbj}.
It is therefore important to test the modelling of these features in MC simulation, through a comparison with observed data.
Moreover, deviations of top-quark-jet substructure from the SM predictions could also serve as tests of BSM effects that may not be detectable with inclusive cross-section measurements.
ATLAS measured substructure properties of jets emerging from hadronically decaying boosted top quarks, using the full \RunTwo dataset, in \ttbar events in both the semileptonic and all-hadronic channels~\cite{ATLAS:2023jdw}.
Top-quark jets were reconstructed either in the all-hadronic channel as large-$R$ jets, using calorimeter energy deposits as input, or in the semileptonic channel as large-$R$ reclustered jets, using small-$R$ particle-flow jets as input.
In both cases, the anti-$k_t$ clustering algorithm with $R=1$ was used, and top-quark candidate jets were required to have $\pt>350$~\GeV.
In the all-hadronic channel, in order to discriminate between \ttbar events and QCD multijet background events, two large-$R$ jets were selected, with one required to be tagged by a top-quark-tagging algorithm~\cite{JETM-2018-03} using several substructure variables as input, and a second one without this requirement, to avoid biasing the measured observables.
Differential cross-sections were measured as a function of eight substructure variables, defined using only the charged components of the jets (see Figure~\ref{fig:asymmetry-substructure}~(b)).
Double-differential distributions were also measured, with two of these substructure variables (namely $\tau_{32}$ and $D_2$) measured in bins of top-quark-jet \pt and mass.
Results unfolded to particle level were compared with a number of MC predictions.
The nominal prediction from \POWPY[8] was found to properly model measurements of the broadness and the two-body structure, but to poorly model variables sensitive to the three-body structure of the top-quark jets, with the predicted jet substructure being more three-body-like than observed.
On the other hand, alternative predictions, such as those from $\AMCatNLO{+}\Pythia[8]$ and \POWHER[7], as well as those from \POWHER[8] with less FSR, were found to better model some of these distributions.
Overall, the measurement indicates the need to improve the models used to predict the substructure of boosted top-quark jets.

\begin{figure}[!tbp]
\centering
\subfloat[]{\includegraphics[width=0.45\linewidth]{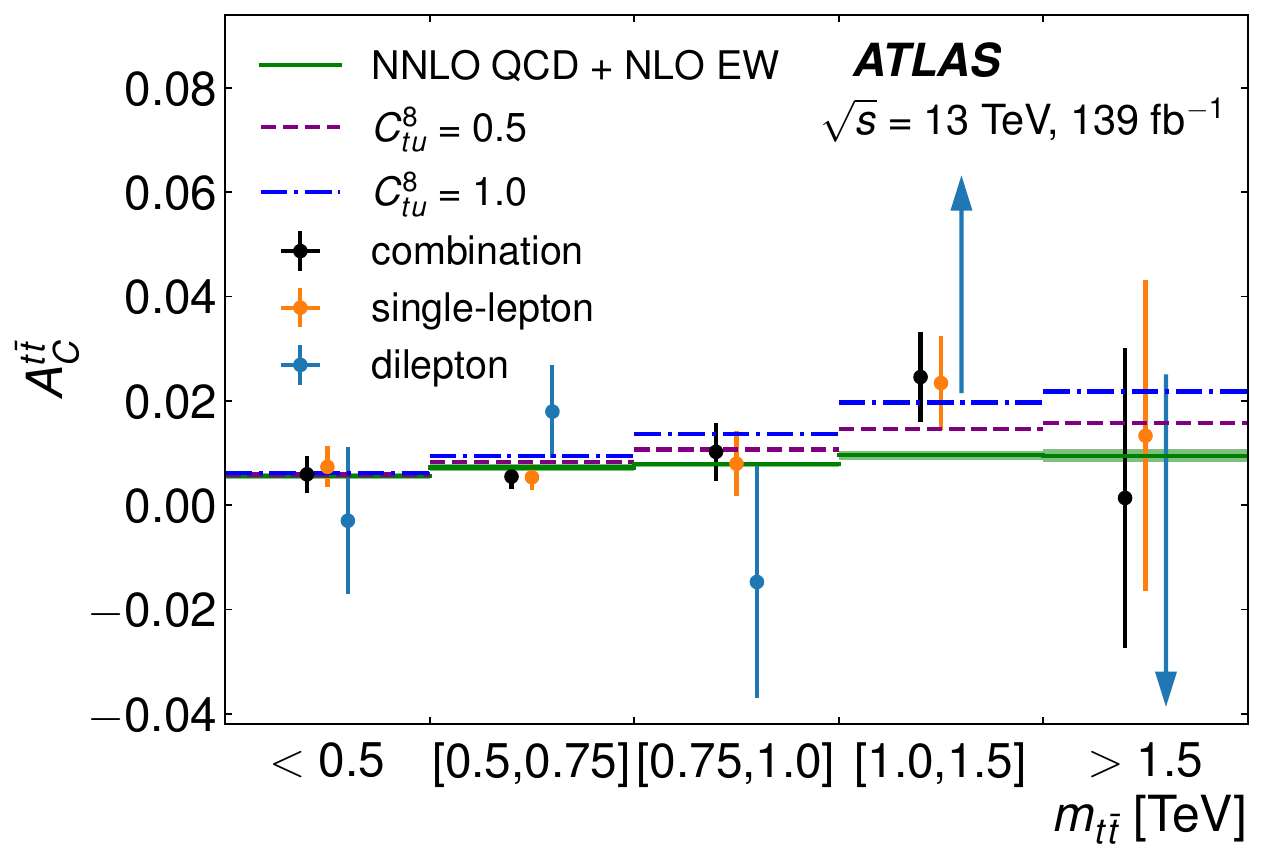}}
\subfloat[]{\includegraphics[width=0.45\linewidth]{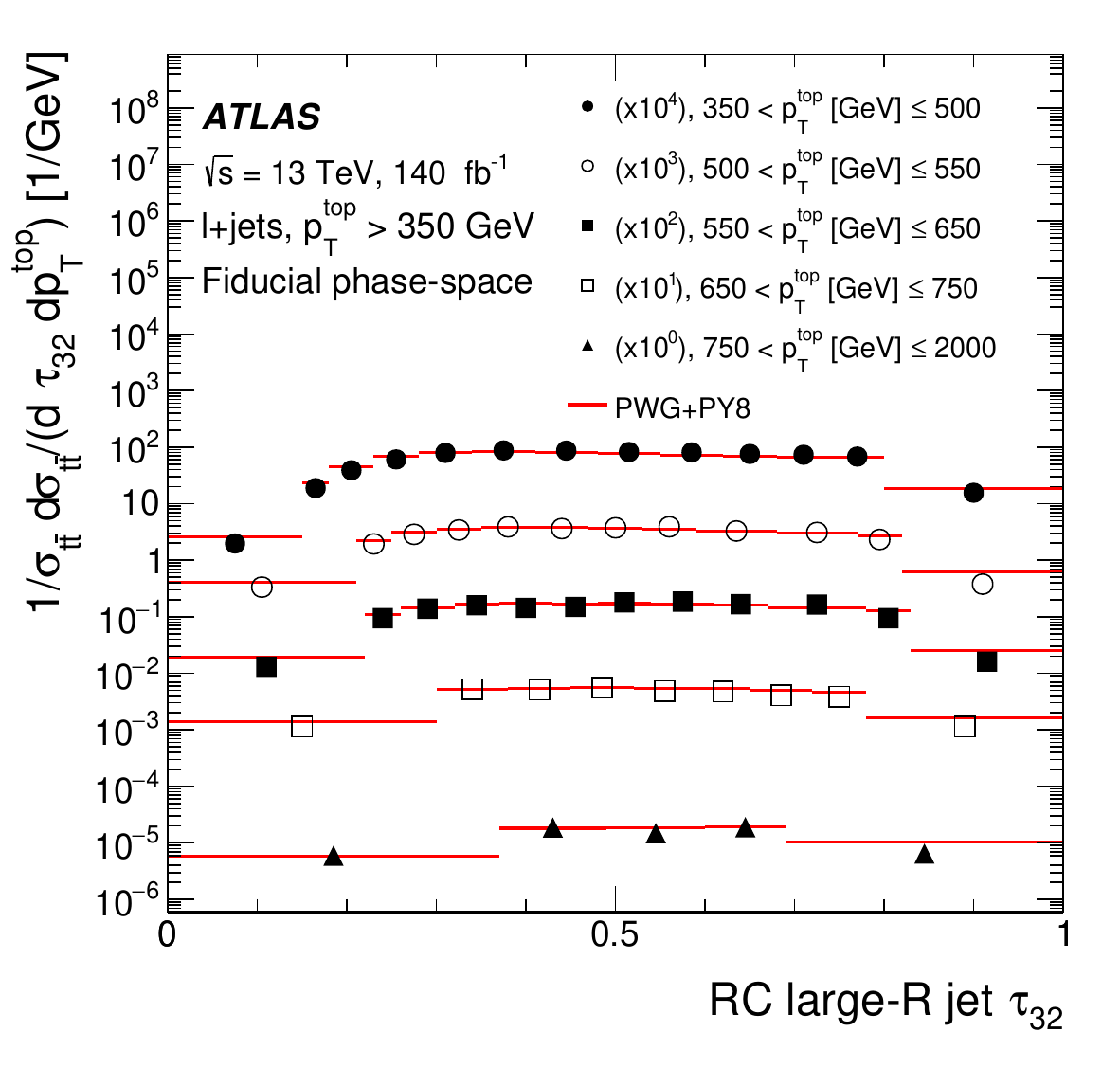}}
\caption{(a) Unfolded differential charge asymmetry $A^{\ttbar}_\text{C}$ as a function of the invariant mass of the reconstructed top-quark pair~\cite{ATLAS:2022waa}.
(b) Particle-level differential \ttbar production cross-section as a function of $\tau_{32}$ for several intervals of top-quark transverse momentum, comparing data measurements with predictions~\cite{ATLAS:2023jdw}.}
\label{fig:asymmetry-substructure}
\end{figure}

\subsection{Test of lepton-flavour universality}

Top-quark production and decay can also be used as a tool to study generic properties of the SM, taking advantage of the large \ttbar sample available at $\sqrt{s}=13$~\TeV to perform precise measurements. One interesting example is a test of lepton-flavour universality in top-quark decays. Lepton-flavour universality is the SM axiom stating that the couplings of the electroweak gauge bosons to the three charged leptons are equal.
A large and unbiased sample of $W$ bosons can be selected in \ttbar events and used to test this axiom by measuring the ratio of $W$-boson decay rates into different charged leptons, $\ell, \ell' = e, \mu, \tau$:
$R_W(\ell/\ell') = B(W \to \ell \nu_\ell) / B(W \to \ell' \nu_{\ell'})$.

The ratio $R_W(\mu/\tau)$  is particularly relevant because a measurement in $e^+ e^- \to W^+ W^-$  events at LEP found a deviation from unity:
$R_W(\tau/\mu) = 1.070 \pm 0.026$~\cite{ALEPH:2013dgf}.
ATLAS measured this ratio by analysing the full \RunTwo dataset, identifying the $\tau$-leptons through their decay into muons~\cite{ATLAS:2020xea}.
Hence the key step in the analysis is to distinguish these muons from those coming directly from $W$-boson decay (prompt muons).
This was achieved by using the distinct features of the $\tau$-lepton:
its significant lifetime, giving rise to a track with a large impact parameter, and its multibody decay into a muon and two neutrinos, leading to an average muon transverse momentum that is lower than in direct $W \to \mu \nu$ decays.
The measurement was performed in both the $e\mu$ channel and the $\mu\mu$ channel,
with one lepton used to probe the $W$-boson branching ratio and the other lepton to trigger the event and ensure a high-purity selection of top-pair events.
The fraction of muons that are prompt ($W \to \mu \nu_\mu$) and the fraction that come from non-prompt $\tau$-lepton decay ($W \to \tau \nu_\tau \to \mu \nu_\mu \nu_\tau$) were determined from a fit using templates that exploit shape differences between those fractions' distributions of both the muon's transverse impact parameter $|d_0|$ (i.e.\ the distance of closest approach of the muon track to the beam line in the $x$--$y$ plane) and the muon's transverse momentum \pt. The $|d_0|$ shape for the prompt muons was calibrated using $Z \to \mu \mu$ events. The normalisations of the main backgrounds, i.e.\ prompt muons coming from $Z \to \mu \mu$ decays and non-prompt muons coming from $b$- or $c$-hadron decays, were extracted using control regions. The ratio $R_W(\tau/\mu)$ was measured from a two-dimensional profile-likelihood fit of the muon $|d_0|$ and \pt distributions, where the overall \ttbar normalisation was allowed to vary freely, leading to
$R_W(\tau/\mu) = 0.992 \pm 0.007 \text{(stat.)} \pm 0.011 \text{(syst.)}$.
The result agrees with the SM expectation of lepton-flavour universality
and constitutes the most precise determination of this ratio. %
It demonstrates the ability of the LHC experiments to perform high-precision measurements.

Moreover, the large \ttbar event sample collected during \RunTwo was used to precisely measure the ratio of $W$-boson decay rates into electrons and muons~\cite{ATLAS:2024tlf}.
The analysis selected \ttbar events in the $ee$, $\mu\mu$ and $e\mu$ dilepton final states, relating the relative difference between channels to the ratio $R_W(\mu/e)$, and extracting the ratio by means of a maximum-likelihood fit.
The measurements in the three channels were obtained with a technique similar to that used for the inclusive \ttbar cross-section measurement in the $e\mu$ channel (see Section~\ref{sec:ttbar_inclusive}),
with the dilepton invariant mass $m_{\ell\ell}$ being exploited in the $ee$ and $\mu\mu$ channels to separate signal events from the dominant $Z$+jets background.
In order to reduce the sensitivity to uncertainties in the electron and muon identification efficiencies, the result was normalised to the square root of the ratio of $Z$-boson decay rates to electron and muon pairs, $R_Z(\mu\mu/ee)$, which was measured simultaneously in data.
The precise value of $R_Z(\mu\mu/ee)$ from the LEP and SLD experiments (with a ${\sim}0.3\%$ uncertainty~\cite{ParticleDataGroup:2022pth, ALEPH:2005ab}) was then used to extract the final result:
$R_W(\mu/e) = 0.9995 \pm 0.0022 \text{ (stat.)} \pm 0.0036 \text{ (syst.)} \pm 0.0014 \text{ (ext.)}$,
where the last uncertainty refers to the external measurement of $R_Z(\mu\mu/ee)$.
The SM assumption of lepton-flavour universality in $W$-boson decays into electron--neutrino and muon--neutrino pairs was thus confirmed at the 0.5\% level.


%
\section{Search for flavour-changing neutral currents in the top-quark sector}
\label{sec:fcnc}

%

In the SM, flavour-changing neutral-current (FCNC) processes are forbidden at tree level and are strongly suppressed in loops by the GIM mechanism~\cite{Glashow:1970gm}.
In the top-quark sector, the rare decay channels in which the top quark decays into a neutral boson (a Higgs or $Z$ boson, a photon or a gluon) and a $c$- or $u$-quark belong to this family and are predicted to have branching ratios (BRs) ranging from $10^{-12}$ to $10^{-17}$~\cite{Aguilar-Saavedra:2004mfd}.
In addition, the same interaction vertices can give rise to single-top FCNC production processes, where a top quark (or antiquark) is produced in association with a neutral boson. These processes are predicted to have negligible cross-sections in the SM, but
a number of BSM scenarios predict enhancements, increasing the FCNC top-quark decay BRs to $10^{-4}$ in some cases.
These SM extensions include the quark-singlet model~\cite{Aguilar-Saavedra:2002phh}, the two-Higgs-doublet model~\cite{Atwood:1996vj}, the Minimal Supersymmetric Standard Model (MSSM)~\cite{Cao:2007dk}, the MSSM with R-parity violation~\cite{Yang:1997dk}, models with warped extra dimensions~\cite{Agashe:2006wa}, and extended mirror-fermion models~\cite{Hung:2017tts}.

During LHC \RunTwo, ATLAS investigated essentially all the accessible channels for such FCNC searches.
Signal processes with a Higgs boson in the final state, i.e.\ involving the $tHq$ vertex with $q = c$ or $u$, were the most intensely studied.
This kind of process gives rise to a number of possible final-state topologies, depending on the Higgs boson's decay mode, each requiring a dedicated analysis strategy and optimisation.
On the other hand, these analyses could benefit from the experience gained through the studies of the Higgs boson production and decay modes, particularly in the context of $\ttbar H$ production.
In addition, searches for processes involving neutral gauge bosons, i.e.\ with a $tgq$, $t\gamma q$ or $tZq$ vertex, were performed.
Both FCNC top-quark decays and single-top FCNC production processes were investigated, by eventually adopting a global approach as suggested in Ref.~\cite{Durieux:2014xla}, relating FCNC decay branching ratios to FCNC production cross-sections via EFT operators.

\subsection{Searches for top-quark FCNC processes involving a Higgs boson}

The first published \RunTwo searches for FCNC top-quark decays into a $c$- or $u$-quark and a Higgs boson, separately focusing on the $H \rightarrow \gamma\gamma$~\cite{HIGG-2016-26}, $H \rightarrow W^+W^-$~\cite{TOPQ-2017-15} and $H \rightarrow b\bar{b}/\tau^+\tau^-$~\cite{TOPQ-2017-07} decay channels, were all based on the dataset collected in 2015 and 2016, corresponding to an integrated luminosity of 36~fb$^{-1}$.
All three analyses looked for $\ttbar$ events where one of the two top quarks decays into $bW$ and the other via a FCNC into $cH$ or $uH$.
With the larger dataset available at the end of \RunTwo, updated results were released for most of these main Higgs boson decay channels.
These new analyses considered FCNC processes not only in the top quark's decay but also in its production, by searching for events with
a single top quark plus a Higgs boson $(tH)$ that were initiated by a $u$- or $c$-quark.
The results were reported in dedicated papers for the $H\rightarrow\gamma\gamma$~\cite{TOPQ-2019-04}, $H \rightarrow W^+W^-$~\cite{ATLAS:2024mih} and $H \rightarrow \tau^+\tau^-$~\cite{TOPQ-2019-17} channels, while for Higgs boson decay into a pair of bottom quarks, ATLAS took advantage of the full \RunTwo dataset to perform a more general search for FCNC top-quark decays into a generic scalar boson and an up-type quark~\cite{HDBS-2020-16}, in a new-scalar-boson mass range from 20 to 160~\GeV, thus including the SM Higgs boson.

Searches using the Higgs boson's diphoton decay channel~\cite{HIGG-2016-26,TOPQ-2019-04} looked for final states with a pair of energetic isolated photons with an invariant mass peaking at the Higgs boson mass, a top quark (decaying either leptonically or hadronically), and possibly a hadronic jet from a light quark or $c$-quark.
In the partial-dataset analysis, events were separated into categories depending on the presence of a charged lepton and on the compatibility of the reconstructed final states with $\ttbar$ decays.
With the full dataset, the analysis was further improved by using a dedicated charm-tagging algorithm to split the event categories more finely, and using a BDT-based selection for their definition.
The result was then extracted by fitting the diphoton invariant-mass spectra with a resonant signal function centred around the Higgs boson mass and a background function, with constraints mainly from $m_{\gamma\gamma}$ bands on either side of each signal region.

The analyses reported in Ref.~\cite{TOPQ-2017-15} and Ref.~\cite{ATLAS:2024mih} focused on multilepton final states, targeting Higgs boson decays giving rise to at least one electron or muon, such as $H\rightarrow WW$ with at least one of the two $W$ bosons decaying leptonically, but also $H\rightarrow ZZ$ or $H\rightarrow \tau\tau$.\footnote{The analysis vetoes events with at least one reconstructed hadronic $\tau$-lepton decay, so as to avoid statistical overlap with the dedicated search described before. Therefore, the analysis is sensitive to $H\rightarrow \tau\tau$ only when both $\tau$-leptons decay to an electron or muon.}
When the SM leg of the top-quark or top-antiquark decay gives rise to a charged lepton as well, these processes can produce events with two same-sign electrons or muons, or even three-lepton events, for which the background from SM processes is significantly smaller.
This analysis is characterised by final states similar to those for the $\ttbar H$ measurement in the multilepton channel~\cite{HIGG-2017-02} and therefore shares with it a good fraction of the event selection, event categorisation and background estimation.
However, dedicated multivariate discriminants were employed in the signal regions to separate the SM processes from the FCNC signal.

The $H\rightarrow \tau^+\tau^-$ analyses~\cite{TOPQ-2017-07,TOPQ-2019-17} categorised events according to the numbers of reconstructed hadronic $\tau$-lepton decays ($\tau_{\text{had}}$) and light charged leptons from leptonic $\tau$-lepton decays ($\tau_{\text{lep}}$), to separately target $\tau_{\text{had}}\tau_{\text{had}}$ and $\tau_{\text{lep}}\tau_{\text{had}}$ events.\footnote{Higgs boson decays with two $\tau_{\text{lep}}$ were targeted by the multilepton search.}
While the partial-dataset analysis focused on hadronic decay in the SM leg of the \ttbar process, the full-dataset analysis also included single-top topologies and dedicated event categories targeting \ttbar events with leptonic top-quark decay.
Events in the various categories were processed through dedicated kinematic reconstruction algorithms, aiming to deduce the four-momenta of the invisible decay products for each $\tau$-lepton decay.
Finally, dedicated BDT discriminants were built in each category, separately for the $tcH$ and $tuH$ couplings, using as input a number of topological and kinematic final-state variables, including ones with values coming from the kinematic reconstruction.

The $H\rightarrow b\bar{b}$ analyses~\cite{TOPQ-2017-07,HDBS-2020-16} were also heavily based on studies and analysis techniques developed in the context of the searches for the $\ttbar H$ process, in this case in the $b\bar{b}$ decay channel during \RunOne~\cite{HIGG-2013-27} and \RunTwo~\cite{HIGG-2017-03}.
They relied on the presence of a high-momentum electron or muon in the SM leg of the \ttbar decay and categorised selected events according to jet and $b$-tagged-jet multiplicities (4--6 jets, and 2--4 $b$-tagged jets), with the goal of retaining events with $b$-quark jets that fail the $b$-tagging requirements, $c$-quark jets that are $b$-tagged, jets that are outside the acceptance or produced from hard QCD radiation, as well as including background-enriched data samples, useful for placing in situ constraints on some of the associated systematic uncertainties.
In the first analysis, a dedicated likelihood discriminant was constructed in each of the categories to assess the compatibility of the observed final states with signal processes by considering invariant masses of jet and lepton combinations.
The second analysis implemented a discriminating variable based on a NN in each signal region for each scalar mass hypothesis.

None of these searches for FCNC top-quark decays observed any significant excess over the SM expectations, with the largest deviation from the background-only hypothesis found in the case of the full-dataset analysis in the $\tau\tau$ channel, yielding a $2.3\sigma$ significance.
Exclusion limits were then set on the branching ratios for the $t\rightarrow Hc$ and $t\rightarrow Hu$ processes, from each of the individual channels as well as from their combinations, and reported in Ref.~\cite{TOPQ-2017-07} and Ref.~\cite{ATLAS:2024mih} for the searches based on 36~fb$^{-1}$ and 140~fb$^{-1}$ respectively.
For the partial-dataset analyses, the exclusion upper limits at 95\% CL ranged from $1.5 \times 10^{-3}$ to $5 \times 10^{-3}$ depending on the channel, with combined limits of around $1 \times 10^{-3}$ for both up-type-quark flavours.
The corresponding 95\% CL exclusion limits obtained by combining the full-dataset analyses instead yielded:
\[
B(t\rightarrow uH) < 2.6 \times 10^{-4} \text{ and } B(t\rightarrow cH) < 3.4 \times 10^{-4}.
\]
These results are shown in Figure~\ref{fig:fcnc_higgs_140fb}, together with those of the individual input analyses.

\begin{figure}[!tbp]
\centering
\subfloat[]{\includegraphics[width=0.49\textwidth]{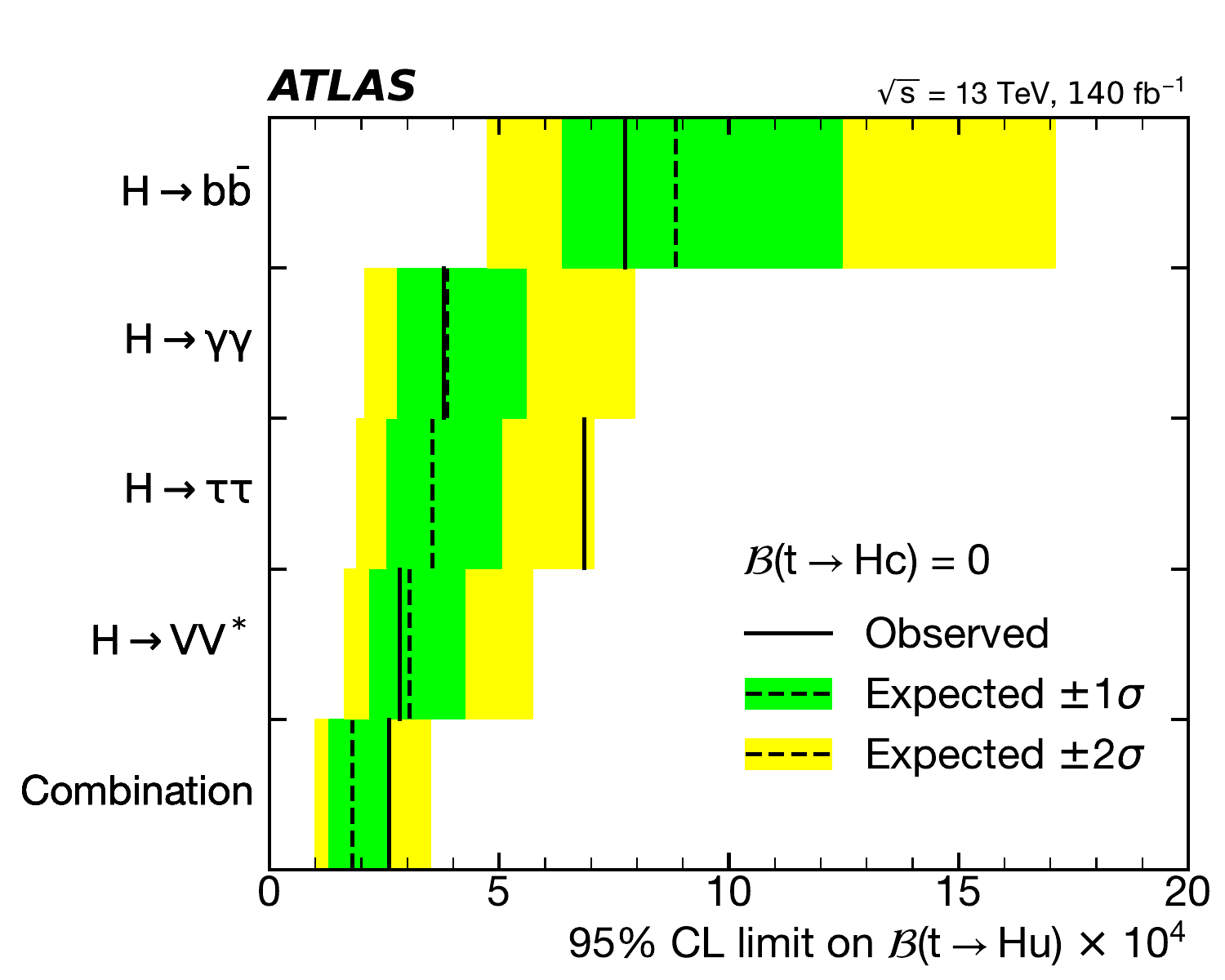}}
\subfloat[]{\includegraphics[width=0.49\textwidth]{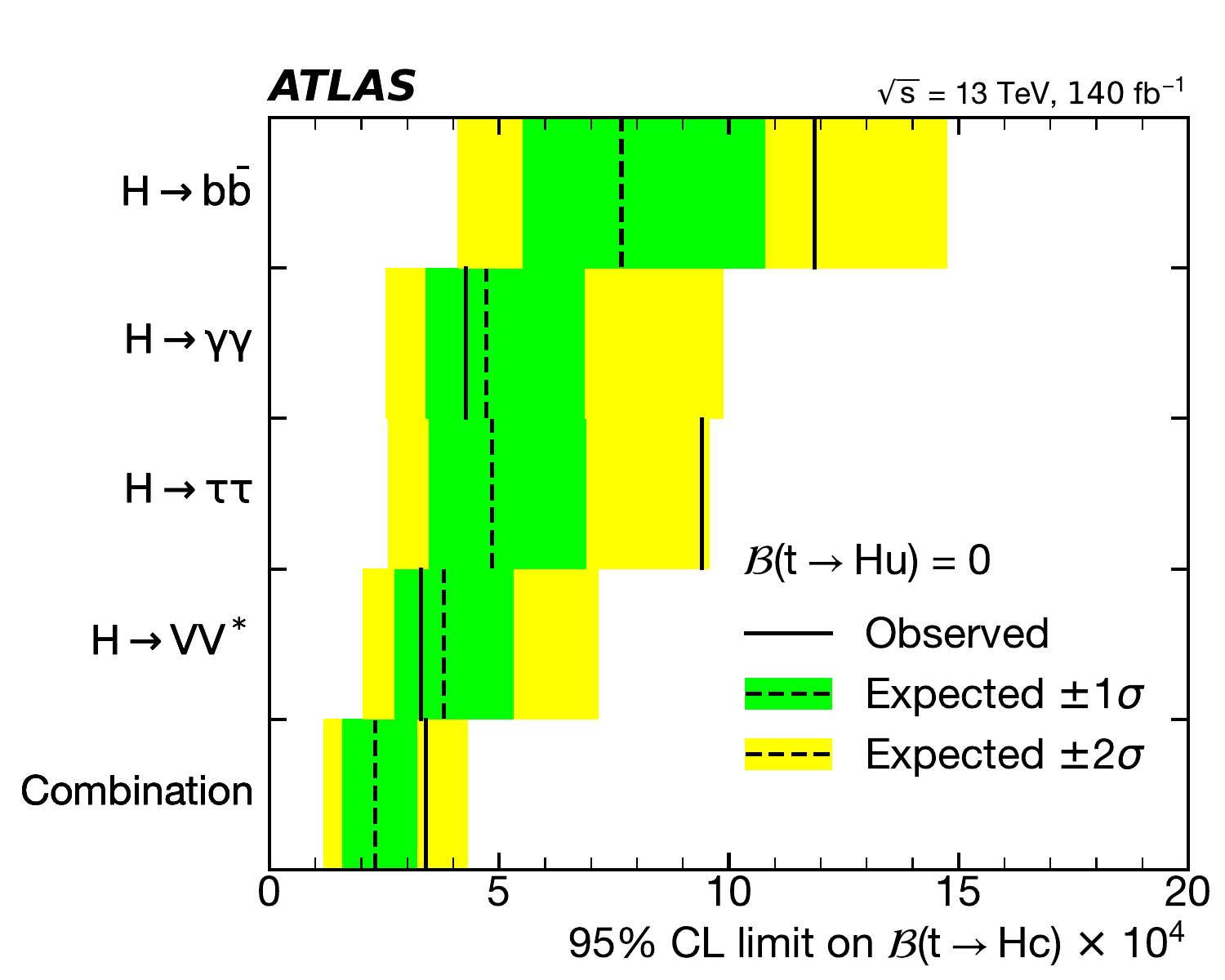}}
\caption{95\% CL upper limits on (a) $B(t\rightarrow Hu)$ and (b) $B(t\rightarrow Hc)$ for the individual ATLAS searches based on the full \RunTwo dataset, as well as their combination (assuming (a) $B(t\rightarrow Hc)=0$ and (b) $B(t\rightarrow Hu)=0$.
The observed limits (solid lines) are compared with the expected (median) limits under the background-only hypothesis (dotted lines).
The surrounding shaded bands correspond to the 68\% and 95\% CL intervals around the expected limits, denoted by $\pm 1 \sigma$ and $\pm 2 \sigma$, respectively.
Figures taken from~Ref.~\cite{ATLAS:2024mih}.}
\label{fig:fcnc_higgs_140fb}
\end{figure}

\subsection{Searches for top-quark FCNC processes involving neutral gauge bosons}

With the full \RunTwo dataset, ATLAS also performed searches for top-quark FCNC couplings to the gluon, the photon and the $Z$ boson.
In particular, limits on the top-quark FCNC interactions involving a gluon were set by searching for the single-top-quark production processes $ug\rightarrow t$ and $cg\rightarrow t$~\cite{TOPQ-2018-06}.
This analysis relied on the selection of a sample of events compatible with the production of a single top quark, without the accompanying jet characterising the SM single-top-quark production channels (the $t$-, $s$- and $tW$-channels, see Section~\ref{sec:single_top}).
The event selection was defined by requiring the presence of an electron or muon, large missing transverse momentum, and exactly one central jet passing a particularly tight $b$-tagging requirement corresponding to a nominal 30\% efficiency for $b$-quark-initiated jets and a rejection rate of around 900 (30\,000) for $c$-quark-initiated (light-quark-initiated) jets.
Two NN discriminants were built to discriminate between signal events and background events, including the irreducible background from $t$-channel single-top-quark production, based on a number of kinematic properties of final-state leptons, jets and the missing transverse momentum.
One of these discriminants was trained specifically to isolate FCNC top-quark-production events initiated by a valence quark and a gluon, namely $ug \rightarrow t$, characterised by kinematic properties different from those of events from sea-quark-initiated processes such as $\bar{u}g \rightarrow \bar{t}$, $cg \rightarrow t$ and $\bar{c}g \rightarrow \bar{t}$.
No significant excess of events compatible with FCNC-production kinematics was observed, and exclusion limits were set on the production cross-sections of the two signal processes.
These were then turned into 95\% CL limits on top-quark FCNC-decay branching ratios:
\[
B (t \rightarrow ug) < 0.61 \times 10^{-4} \text{  and  } B (t \rightarrow cg) < 3.7 \times 10^{-4}.
\]
Limits on the $c$-quark-initiated process are significantly weaker than those on the $u$-quark-initiated process.
This is mainly due to the predicted cross-section being lower for the $cg \rightarrow t$ process because it can only be initiated by sea quarks, which typically have a lower momentum fraction than the valance quarks contributing to the $ug \rightarrow t$ process.
However, $c$-quark-initiated processes are still phenomenologically relevant, especially for two-Higgs-doublet models, which predict stronger FCNC couplings between top and charm quarks.

FCNC coupling between a top quark and a photon was also investigated deeply using \RunTwo data.
A search for FCNCs in the production of a top quark with a photon, $gu\rightarrow t\gamma$ or $gc\rightarrow t\gamma$, was performed using 80~fb$^{-1}$ of data~\cite{TOPQ-2018-22}.
Events with an energetic isolated photon and the typical final-state signature of a leptonically decaying top quark (an isolated electron or muon, a $b$-tagged jet and missing transverse momentum), and possibly additional jets, were selected, and a NN based on kinematic variables was used to discriminate between signal and SM background processes.
The analysis was updated to use the full \RunTwo dataset~\cite{TOPQ-2019-19}, and extended to cover both the production and decay processes, $gq\rightarrow t\gamma$ and $t\rightarrow \gamma q$ (in \ttbar events) with $q$ being an $u$- or $c$-quark, as target signal.
As done in previous analyses, the search for a FCNC in top-quark decay was limited to \ttbar-production events, so the two signal processes were characterized by similar event topologies.
A multi-class DNN discriminant was built to classify events as coming from one of the two signal processes (FCNC in production or FCNC in decay) or as background events.
Dedicated control samples were selected in order to control the normalisation of the main background processes, $\ttbar \gamma$ and $W\gamma$, while data-driven scale factors were obtained to correct the simulation of the background component induced by the misidentification of a hadron or an electron as a photon.
The DNN was trained separately for $t\gamma u$ and $t\gamma c$ couplings,
because of the different kinematics for the $gu\rightarrow t\gamma$ and $gc\rightarrow t\gamma$ processes, induced by the differences between the up- and charm-quark PDFs,
and the different $b$-tagging probabilities for \ttbar events with a $t\rightarrow \gamma u$ decay or $t\rightarrow \gamma c$ decay.
The final results of the two searches, extracted with a binned profile-likelihood fit to the DNN discriminant in the signal region and to the photon \pt distributions in the background control regions, didn't show any significant excess over the SM expectations, allowing exclusion limits to be set on the signal production cross-sections and decay branching ratios, separately for right-handed (RH) and left-handed (LH) $t\gamma q$ couplings, as well as on the relative effective coupling constants~\cite{Aguilar-Saavedra:2008nuh, Durieux:2014xla, Grzadkowski:2010es, Aguilar-Saavedra:2018ksv}.
The exclusion limits on the branching ratios range from $2.8 \times 10^{-5}$ to $22 \times 10^{-5}$ for the former analysis, and
from $0.85 \times 10^{-5}$ to $4.5 \times 10^{-5}$ for the latter analysis, depending on the light quark's flavour and on the coupling chirality, with the full-\RunTwo results improving on the previous limits by factors of 3.3 to 5.4.
For a LH $t\gamma q$ coupling, the 95\% CL exclusion limits obtained are:
\[
B (t \rightarrow u \gamma) < 0.85 \times 10^{-5} \text{  and  } B (t \rightarrow c \gamma) < 4.2 \times 10^{-5},
\]
while for a RH coupling, they are:
\[
B (t \rightarrow u \gamma) < 1.2 \times 10^{-5} \text{  and  } B (t \rightarrow c \gamma) < 4.5 \times 10^{-5}.
\]

Finally, FCNC coupling between a top quark and a $Z$ boson was studied in Ref.~\cite{TOPQ-2017-06} and Ref.~\cite{TOPQ-2019-06}.
The former analysis, based on 36~fb$^{-1}$, relied only on the top-quark-pair events, with one top quark decaying through the $t \rightarrow Zq$ channel, while the latter analysis, based on the full \RunTwo dataset, included single-top-quark production via $gq \rightarrow tZ$ (with $q=u,c$) as a signal process.
Since the final-state topologies are similar to those investigated in SM $\ttbar Z$ and $tZ$ associated-production processes (see Section~\ref{sec:associated}), the analysis naturally relied on the selection of events with three charged leptons, two of them coming from the $Z$-boson decay and the other from the top-quark decay, plus a $b$-tagged jet and possibly additional jets.
In the first analysis, the top quarks were kinematically reconstructed from the final-state leptons and jets, using $\chi^2$ minimisation across all the possible jet and lepton permutations in each event.
The result of the $\chi^2$ minimisation was used as a discriminating variable in the three-lepton signal region, and a number of control regions were included in the analysis to specifically target the $\ttbar Z$, $WZ$, $ZZ$ and non-prompt-lepton backgrounds.
For the full-\RunTwo search, multivariate discriminants were introduced to improve the signal sensitivity beyond the gain from the larger analysed dataset.
In particular, after separating selected events into two orthogonal signal regions, individually optimised for single-top-quark and $\ttbar$ production signal processes, gradient boosted decision trees (GBDTs) were trained to discriminate between FCNC signals and SM backgrounds, using the outputs of the kinematic reconstruction as input variables.
In the signal region dedicated to single-top-quark production, two separate GBDTs were built to separately target $u$- and $c$-quark-initiated processes.
Finally, the presence of signal events from RH and LH FCNC couplings was tested separately.
The full-\RunTwo analysis improved the sensitivity to the branching ratios $B(t\rightarrow Zq)$ by a factor of two, and set the following 95\% CL exclusion limits for the LH coupling scenario:
\[
B (t \rightarrow u Z) < 6.2 \times 10^{-5} \text{  and  } B (t \rightarrow c Z) < 1.3 \times 10^{-4}.
\]
Similar limits were set for the RH coupling hypothesis ($6.6 \times 10^{-5}$ and $1.2 \times 10^{-4}$ for the $tZu$ and $tZc$ couplings, respectively).

\subsection{Summary of FCNC process constraints}
A summary of the exclusion limits set by the ATLAS \RunTwo searches for FCNCs in the top-quark sector is shown in Figure~\ref{fig:fcnc}.
As can be seen, exclusion limits on photon-mediated FCNC processes are the strongest, followed by those on transitions mediated by $Z$ bosons and gluons, while weaker limits are set on processes involving Higgs bosons.
In all cases, the sensitivity is stronger for $t\rightarrow u$ than for $t\rightarrow c$ transitions.
All of these stringent limits on top-quark FCNC interactions were also interpreted as constraints on Wilson coefficients in an EFT framework,
as detailed in Section~\ref{sec:eft}.

\begin{figure}[!tbp]
\centering
\includegraphics[width=0.75\linewidth]{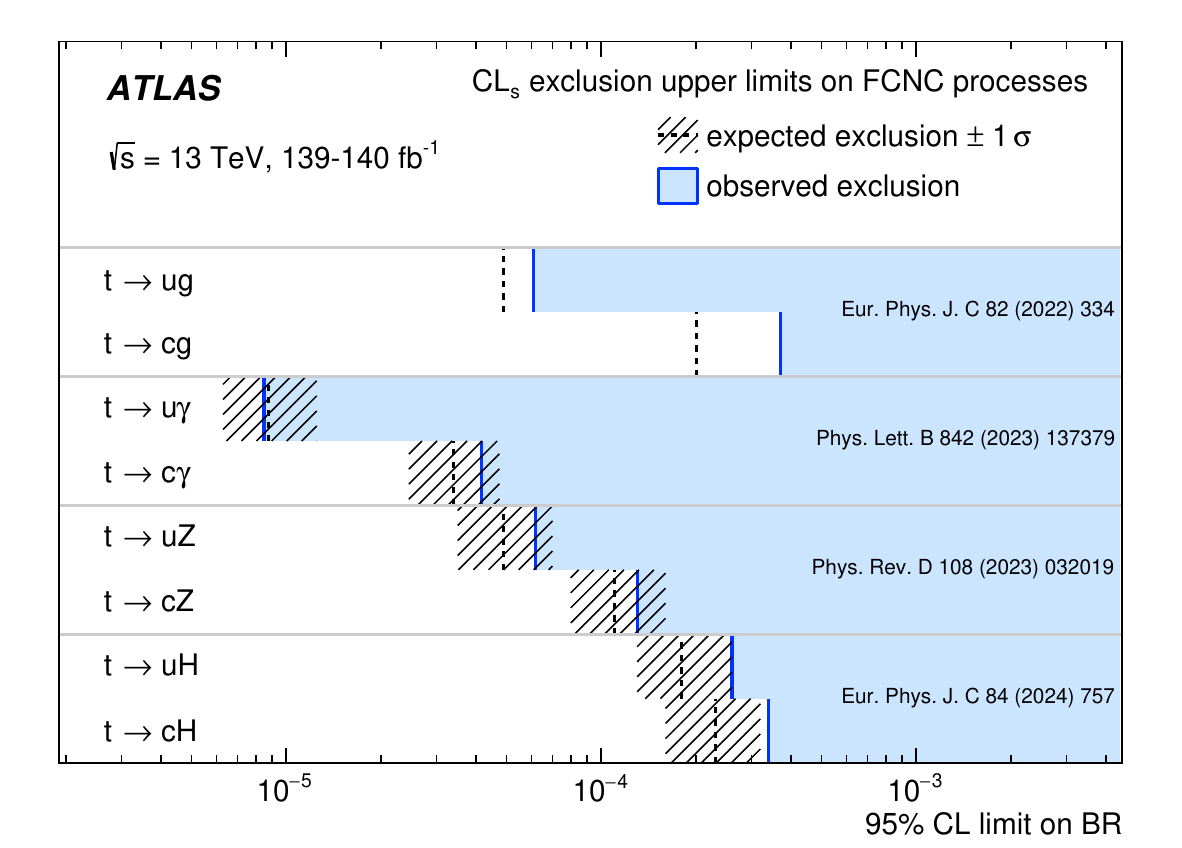}
\caption{Summary of the 95\% CL observed and expected limits on the branching ratios of the top-quark decays via flavour-changing neutral currents (FCNC) to a quark and a neutral boson $t \rightarrow qX$ ($X = g, Z, \gamma$ or $H$; $q = u$ or $c$) set by the ATLAS Collaboration using its full \RunTwo dataset.
The hatched area represents the $\pm 1 \sigma$ band around the expected exclusion limit.
It is not shown in the case of the $tgq$ coupling, as the numbers are not quoted in the publication.
The quoted upper limits refer to the left-handed chirality hypothesis, and expected limits are computed assuming zero Standard Model contribution to the FCNC processes.}
\label{fig:fcnc}
\end{figure}


\section{Limits on Wilson coefficients within effective field theory}
\label{sec:eft}


New physics effects from BSM theories characterised by a mass scale higher than the energy accessible at the LHC can be investigated in the framework of the Standard Model effective field theory (SMEFT), without specifying a particular BSM model.
The SMEFT extension of the SM Lagrangian is:
\begin{equation*}
\mathcal{L}_{\mathrm{SMEFT}} = \mathcal{L}_{\mathrm{SM}} + \sum_i \frac{C_i}{\Lambda^2} O_i + ... ,
\end{equation*}
where $\mathcal{L}_{\mathrm{SM}}$ represents the SM Lagrangian, $O_i$ are effective dimension-six operators and $C_i$ the corresponding Wilson coefficients.
Here, $\Lambda$ corresponds to the cut-off scale of the effective theory and is usually set to $\Lambda=1$~\TeV.
Dimension-five operators induce baryon- and lepton-number-violating terms and are therefore usually ignored.
Higher-dimension operators are $\Lambda$-suppressed and usually also ignored.
The operators can be expressed in different bases.
The Warsaw basis~\cite{Grzadkowski:2010es} is chosen for the results presented here.

Top-quark pair production is sensitive to the operator $O_{tG}$, which modifies the top--gluon vertex and primarily changes the overall production rate, and to the 14 operators characterising four-quark interactions: $O_{tu}^{(1)}$, $O_{tu}^{(8)}$, $O_{td}^{(1)}$, $O_{td}^{(8)}$, $O_{tq}^{(1)}$, $O_{tq}^{(8)}$, $O_{Qu}^{(1)}$, $O_{Qu}^{(8)}$, $O_{Qd}^{(1)}$, $O_{Qd}^{(8)}$, $O_{Qq}^{(1,1)}$, $O_{Qq}^{(3,1)}$, $O_{Qq}^{(1,8)}$, $O_{Qq}^{(3,8)}$.
In particular, $O_{tq}^{(8)}$ can significantly alter differential \ttbar cross-sections in the high-energy tails.
The $t$-channel single-top-quark cross-section is particularly sensitive to the operator $O^{3,1}_{Qq}$.
The $Wtb$ vertex could also be altered by the operators $O^3_{\phi Q}$ and $O_{tW}$.
The \ttZ cross-section is sensitive to the 14 four-quark operators as well as to the operators modifying the top--boson vertices ($O_{tW}$, $O_{tB}$, $O_{tG}$, $O_{HQ}^{(1)}$, $O_{HQ}^{(3)}$, $O_{Ht}$).
The \tttt cross-section is specifically sensitive to the four heavy-flavour fermion operators ($O^1_{QQ}$, $O^1_{Qt}$, $O^1_{tt}$, $O^8_{Qt}$).
The \ttbar charge asymmetry $A^{\ttbar}_\text{C}$ is also sensitive to the 14 four-quark operators but to different linear combinations of these, compared to \ttbar production, which helps to disentangle the contributions when global fits are performed.
It is also sensitive to the top--gluon operator $O_{tG}$.
The differential $A^{\ttbar}_\text{C}$ measurements lead to tighter limits than obtained from the inclusive $A^{\ttbar}_\text{C}$ measurement, so the differential constraints are reported here.
The energy asymmetry $A_E(\thetaj)$ is particularly sensitive to the chirality and the colour charges of the involved quark fields, and therefore to six specific four-quark operators ($O_{tu}^{(1)}$, $O_{tu}^{(8)}$, $O_{tq}^{(1)}$, $O_{tq}^{(8)}$, $O_{Qq}^{(1,1)}$, $O_{Qq}^{(1,8)}$).

Finally, limits on top-quark FCNC couplings can be interpreted as constraints on the Wilson coefficients of the EFT operators inducing tree-level FCNC transitions~\cite{Durieux:2014xla}.
Exclusion limits on top--Higgs FCNC couplings can be translated into constraints on the coefficients $C^{13}_{u\phi}$ and $C^{23}_{u\phi}$, those on the top--gluon coupling into constraints on $C^{13}_{uG}$ and $C^{23}_{uG}$,
while limits on both photon- and $Z$-boson-induced FCNC interactions translate into constraints on linear combinations of the coefficients $C^{i3}_{uB}$ and $C^{i3}_{uW}$, with $i=1,2$.

The 95\% CL limits on Wilson coefficients extracted for $\Lambda=1$~\TeV from the measurements and searches described in this paper, assuming one coefficient at a time to be non-zero, are summarised in Tables~\ref{tab:eft-boson}, \ref{tab:eft-fourq} and~\ref{tab:eft-fcnc}. These bounds are obtained using a parameterisation that includes the dimension-six-squared terms.

\begin{table}[!tbp]
\caption{Summary of the 95\% CL intervals for top--boson Wilson coefficients within SMEFT set by the top-quark \RunTwo measurements and searches. These limits are set for $\Lambda=1$~\TeV and when one coefficient at a time is assumed to be non-zero. $\Re{C_{tW}}$ and $\Im{C_{tW}}$ represent the real and imaginary parts of the complex $C_{tW}$ coefficient.}
\begin{center}
\begin{tabular}{cccc}
\toprule
Top--boson coefficient & Limits & Input measurement & Reference \\
\midrule
$C^3_{\phi Q}$ & [--0.87, 1.42] & $t$-channel cross-section & \cite{ATLAS:2024ojr} \\[4pt]

$\Re{C_{tW}}$  & [--0.9, 1.4]   & $t$-channel polarisation  & \cite{ATLAS:2022vym} \\
& [--0.84, 1.0]~~  & $\ttbar Z$ cross-section  & \cite{2312.04450} \\[4pt]

$\Im{C_{tW}}$  & [--0.8, 0.2]   & $t$-channel polarisation  & \cite{ATLAS:2022vym} \\
& [--1.0, 1.0]   & $\ttbar Z$ cross-section  & \cite{2312.04450} \\[4pt]

$C_{HQ}^{(1)}$ & ~~[--1.4, 0.84]  & $\ttbar Z$ cross-section  & \cite{2312.04450} \\[4pt]
$C_{HQ}^{(3)}$ & [--0.95, 2.0]~~  & $\ttbar Z$ cross-section  & \cite{2312.04450} \\[4pt]
$C_{Ht}$       & [--2.2, 1.6]   & $\ttbar Z$ cross-section  & \cite{2312.04450} \\[4pt]
$\Re{C_{tB}}$  & [--1.7, 1.6]   & $\ttbar Z$ cross-section  & \cite{2312.04450} \\[4pt]
$\Im{C_{tB}}$  & [--1.9, 1.9]   & $\ttbar Z$ cross-section  & \cite{2312.04450} \\[4pt]
$\Re{C_{tG}}$  & [--0.23, 0.34] & $\ttbar Z$ cross-section  & \cite{2312.04450} \\
& [--0.75, 0.14] & \ttbar charge asymmetry   & \cite{ATLAS:2022waa} \\
& [--0.52, 0.15] & \ttbar cross-section      & \cite{TOPQ-2019-23} \\[4pt]
$\Im{C_{tG}}$  & [--0.32, 0.33] & $\ttbar Z$ cross-section  & \cite{2312.04450} \\
\bottomrule
\end{tabular}
\label{tab:eft-boson}
\end{center}
\end{table}

\begin{table}[!tbp]
\caption{Summary of the 95\% CL intervals for four-quark Wilson coefficients within SMEFT set by the top-quark \RunTwo measurements and searches. These limits are set for $\Lambda=1$~\TeV and when one coefficient at a time is assumed to be non-zero.}
\begin{center}
\begin{tabular}{cccc}
\toprule
Four-quark coefficient & Limits & Input measurement & Reference \\
\midrule
$C^{3,1}_{Qq}$      & [--0.37, 0.06] & $t$-channel cross-section & \cite{ATLAS:2024ojr} \\
& [--0.34, 0.23] & $\ttbar Z$ cross-section  & \cite{2312.04450} \\
& [--0.34, 0.43] & \ttbar charge asymmetry   & \cite{ATLAS:2022waa} \\[3pt]
$C_{Qd}^{(1)}$      & [--0.47, 0.46] & $\ttbar Z$ cross-section  & \cite{2312.04450} \\
& [--0.39, 0.42] & \ttbar charge asymmetry   & \cite{ATLAS:2022waa} \\[3pt]
$C_{Qd}^{(8)}$      & ~~[--1.8, 0.62]  & $\ttbar Z$ cross-section  & \cite{2312.04450} \\
& [--0.96, 1.37] & \ttbar charge asymmetry   & \cite{ATLAS:2022waa} \\[3pt]
$C_{Qq}^{(1,1)}$    & [--0.29, 0.29] & $\ttbar Z$ cross-section  & \cite{2312.04450} \\
& [--0.52, 0.28] & \ttbar charge asymmetry   & \cite{ATLAS:2022waa} \\
& [--0.65, 0.67] & \ttbar energy asymmetry   & \cite{ATLAS:2021dqb} \\[3pt]
$C_{Qq}^{(1,8)}$    & ~~~~[--1.2, 0.080] & $\ttbar Z$ cross-section  & \cite{2312.04450} \\
& [--0.25, 0.53] & \ttbar charge asymmetry   & \cite{ATLAS:2022waa} \\
& [--1.72, 2.10] & \ttbar energy asymmetry   & \cite{ATLAS:2021dqb} \\[3pt]
$C_{Qq}^{(3,8)}$    & [--0.68, 0.54] & $\ttbar Z$ cross-section  & \cite{2312.04450} \\
& [--1.23, 0.31] & \ttbar charge asymmetry   & \cite{ATLAS:2022waa} \\[3pt]
$C_{Qu}^{(1)}$      & [--0.47, 0.43] & $\ttbar Z$ cross-section  & \cite{2312.04450} \\
& [--0.31, 0.23] & \ttbar charge asymmetry   & \cite{ATLAS:2022waa} \\[3pt]
$C_{Qu}^{(8)}$      & ~~[--1.6, 0.36]  & $\ttbar Z$ cross-section  & \cite{2312.04450} \\
& [--1.78, 0.27] & \ttbar charge asymmetry   & \cite{ATLAS:2022waa} \\[3pt]
$C_{tq}^{(1)}$      & [--0.42, 0.36] & $\ttbar Z$ cross-section  & \cite{2312.04450} \\
& [--0.20, 0.22] & \ttbar charge asymmetry   & \cite{ATLAS:2022waa} \\
& [--0.69, 0.75] & \ttbar energy asymmetry   & \cite{ATLAS:2021dqb} \\[3pt]
$C_{tq}^{(8)}$      & ~~[--1.5, 0.22]  & $\ttbar Z$ cross-section  & \cite{2312.04450} \\
& [--0.51, 0.58] & \ttbar charge asymmetry   & \cite{ATLAS:2022waa} \\
& [--2.01, 1.43] & \ttbar energy asymmetry   & \cite{ATLAS:2021dqb} \\
& [--0.64, 0.12] & \ttbar cross-section      & \cite{TOPQ-2019-23} \\[3pt]
$C_{td}^{(1)}$      & [--0.56, 0.56] & $\ttbar Z$ cross-section  & \cite{2312.04450} \\
& [--0.60, 0.84] & \ttbar charge asymmetry   & \cite{ATLAS:2022waa} \\[3pt]
$C_{td}^{(8)}$      & [--2.5, 1.2]   & $\ttbar Z$ cross-section  & \cite{2312.04450} \\
& [--1.62, 1.21] & \ttbar charge asymmetry   & \cite{ATLAS:2022waa} \\[3pt]
$C_{tu}^{(1)}$      & [--0.66, 0.64] & $\ttbar Z$ cross-section  & \cite{2312.04450} \\
& [--0.70, 0.31] & \ttbar charge asymmetry   & \cite{ATLAS:2022waa} \\
& [--0.78, 0.81] & \ttbar energy asymmetry   & \cite{ATLAS:2021dqb} \\[3pt]
$C_{tu}^{(8)}$      & ~~[--2.2, 0.38]  & $\ttbar Z$ cross-section  & \cite{2312.04450} \\
& [--0.45, 0.82] & \ttbar charge asymmetry   & \cite{ATLAS:2022waa} \\
& [--1.71, 1.56] & \ttbar energy asymmetry   & \cite{ATLAS:2021dqb} \\[3pt]
$C^1_{QQ}$          & [--3.5, 4.1]   & \tttt cross-section& \cite{ATLAS:2023ajo} \\[3pt]
$C^1_{Qt}$          & [--3.5, 3.0]   & \tttt cross-section& \cite{ATLAS:2023ajo} \\[3pt]
$C^1_{tt}$          & [--1.7, 1.9]   & \tttt cross-section& \cite{ATLAS:2023ajo} \\[3pt]
$C^8_{Qt}$          & [--6.2, 6.9]   & \tttt cross-section& \cite{ATLAS:2023ajo} \\
\bottomrule
\end{tabular}
\label{tab:eft-fourq}
\end{center}
\end{table}

\begin{table}[!tbp]
\caption{Summary of the 95\% CL intervals for top--FCNC Wilson coefficients within SMEFT set by the top-quark \RunTwo measurements and searches. These limits are set for $\Lambda=1$~\TeV and when one coefficient at a time is assumed to be non-zero.}
\begin{center}
\begin{tabular}{cccc}
\toprule
FCNC coefficient & Limits & Input search & Reference \\
\midrule
$|C_{u\phi}^{13,31}|$               & $<0.68$  & $t \rightarrow Hq$       & \cite{ATLAS:2024mih} \\[4pt] %
$|C_{u\phi}^{23,32}|$               & $<0.78$  & $t \rightarrow Hq$       & \cite{ATLAS:2024mih} \\[6pt] %
$|C_{uG}^{13}|$                     & ~~$<0.057$ & $t \rightarrow gq$       & \cite{TOPQ-2018-06} \\[4pt] %
$|C_{uG}^{23}|$                     & $<0.14$  & $t \rightarrow gq$       & \cite{TOPQ-2018-06} \\[6pt] %
$|C_{uW}^{13~*} + C_{uB}^{13~*}|$ & ~~$<0.103$ & $t \rightarrow \gamma q$ & \cite{TOPQ-2019-19} \\[4pt] %
$|C_{uW}^{31} + C_{uB}^{31}|$     & ~~$<0.123$ & $t \rightarrow \gamma q$ & \cite{TOPQ-2019-19} \\[6pt] %
$|C_{uW}^{23~*} + C_{uB}^{23~*}|$ & ~~$<0.227$ & $t \rightarrow \gamma q$ & \cite{TOPQ-2019-19} \\[4pt] %
$|C_{uW}^{32} + C_{uB}^{32}|$     & ~~$<0.235$ & $t \rightarrow \gamma q$ & \cite{TOPQ-2019-19} \\[6pt] %
$|C_{uW}^{13~*} + C_{uB}^{13~*}|$ & $<0.15$  & $t \rightarrow Zq$       & \cite{TOPQ-2019-06} \\[4pt] %
$|C_{uW}^{31} + C_{uB}^{31}|$     & $<0.16$  & $t \rightarrow Zq$       & \cite{TOPQ-2019-06} \\[6pt] %
$|C_{uW}^{23~*} + C_{uB}^{23~*}|$ & $<0.22$  & $t \rightarrow Zq$       & \cite{TOPQ-2019-06} \\[4pt] %
$|C_{uW}^{32} + C_{uB}^{32}|$     & $<0.21$  & $t \rightarrow Zq$       & \cite{TOPQ-2019-06} \\ %
\bottomrule
\end{tabular}
\label{tab:eft-fcnc}
\end{center}
\end{table}


\FloatBarrier
%

%

%
\section{Conclusion}%
\label{sec:conclusion}
By climbing the mountain of $pp$ collision data collected during LHC \RunTwo at a centre-of-mass energy of 13~\TeV, the ATLAS Collaboration has broadened and deepened measurements of the top quark.
Precise measurements now challenge the most accurate theoretical predictions, for example
the inclusive \ttbar cross-section measurement has reached a precision of 1.8\%,
exceeding the precision of the best available predictions.
The abundant statistics allow for differential measurements of both \ttbar and top-quark-associated production processes.
Rarer and rarer production processes, including $tZq$, $tq\gamma$ and \tttt\ processes have been observed and studied for the first time.
Measurements of the \ttZ, \ttW and \tty cross-sections reach a precision of around 10\%, directly probing the top quark's couplings to gauge bosons.
New methods for the precise determination of the top-quark mass have been adopted, alongside those already used in \RunOne.
The unique property of the top quark that it decays before hadronising is probed through the identification and reconstruction of top-quark decay products.
This has resulted in improved measurements of the $W$-boson helicity fractions, the top quark's polarisation and \ttbar spin correlations.
Strong evidence has been seen for the charge asymmetry in \ttbar production and the first observation of quantum entanglement between fundamental quarks has been
achieved using \ttbar events.
Finally, exclusion limits for exotic or anomalous top-quark production and decay mechanisms have been improved significantly, challenging several new physics-model predictions.

In many cases, statistical uncertainties have become subdominant, putting more emphasis on systematic uncertainties.
Systematic uncertainties have been reduced markedly with respect to \RunOne, in particular the precise understanding of detector-related effects has led to a decrease of important systematic uncertainties related to the jet energy scale and the $b$-tagging efficiency.
The dominant uncertainty in many top-quark physics measurements is now related to the modelling of top-quark-pair production in MC simulation.
This motivated several of the measurements discussed in this report that have been or will be used to tune such models, further reducing the impact of systematic uncertainties.
Nevertheless, there are still significant parts of the programme that will benefit from increased statistical power.
This is the case for analyses involving rare processes, such as four- and three-top-quark production.
It occurs also in analyses trying to precisely measure small SM effects, e.g.\ the \ttbar charge asymmetry, and for analyses sensitive to BSM effects, such as predicted in differential cross-section measurements in the tails of kinematic distributions.

While the \RunTwo analyses of top-quark production and properties are not over yet, LHC \RunThr is providing $pp$ collisions at a new record energy of 13.6~\TeV.
This is particularly beneficial for highly-energetic processes, for instance the \tttt\ cross-section is 20\% higher than at 13~\TeV.
There is no doubt that the higher energy will bring further advances in our understanding of the top quark, and in our new physics discovery potential.
%


%
\section*{Acknowledgements}
%

%
%

%
%

We thank CERN for the very successful operation of the LHC and its injectors, as well as the support staff at
CERN and at our institutions worldwide without whom ATLAS could not be operated efficiently.

The crucial computing support from all WLCG partners is acknowledged gratefully, in particular from CERN, the ATLAS Tier-1 facilities at TRIUMF/SFU (Canada), NDGF (Denmark, Norway, Sweden), CC-IN2P3 (France), KIT/GridKA (Germany), INFN-CNAF (Italy), NL-T1 (Netherlands), PIC (Spain), RAL (UK) and BNL (USA), the Tier-2 facilities worldwide and large non-WLCG resource providers. Major contributors of computing resources are listed in Ref.~\cite{ATL-SOFT-PUB-2023-001}.

We gratefully acknowledge the support of ANPCyT, Argentina; YerPhI, Armenia; ARC, Australia; BMWFW and FWF, Austria; ANAS, Azerbaijan; CNPq and FAPESP, Brazil; NSERC, NRC and CFI, Canada; CERN; ANID, Chile; CAS, MOST and NSFC, China; Minciencias, Colombia; MEYS CR, Czech Republic; DNRF and DNSRC, Denmark; IN2P3-CNRS and CEA-DRF/IRFU, France; SRNSFG, Georgia; BMBF, HGF and MPG, Germany; GSRI, Greece; RGC and Hong Kong SAR, China; ISF and Benoziyo Center, Israel; INFN, Italy; MEXT and JSPS, Japan; CNRST, Morocco; NWO, Netherlands; RCN, Norway; MNiSW, Poland; FCT, Portugal; MNE/IFA, Romania; MESTD, Serbia; MSSR, Slovakia; ARIS and MVZI, Slovenia; DSI/NRF, South Africa; MICIU/AEI, Spain; SRC and Wallenberg Foundation, Sweden; SERI, SNSF and Cantons of Bern and Geneva, Switzerland; NSTC, Taipei; TENMAK, T\"urkiye; STFC/UKRI, United Kingdom; DOE and NSF, United States of America.

Individual groups and members have received support from BCKDF, CANARIE, CRC and DRAC, Canada; PRIMUS 21/SCI/017, CERN-CZ and FORTE, Czech Republic; COST, ERC, ERDF, Horizon 2020, ICSC-NextGenerationEU and Marie Sk{\l}odowska-Curie Actions, European Union; Investissements d'Avenir Labex, Investissements d'Avenir Idex and ANR, France; DFG and AvH Foundation, Germany; Herakleitos, Thales and Aristeia programmes co-financed by EU-ESF and the Greek NSRF, Greece; BSF-NSF and MINERVA, Israel; Norwegian Financial Mechanism 2014-2021, Norway; NCN and NAWA, Poland; La Caixa Banking Foundation, CERCA Programme Generalitat de Catalunya and PROMETEO and GenT Programmes Generalitat Valenciana, Spain; G\"{o}ran Gustafssons Stiftelse, Sweden; The Royal Society and Leverhulme Trust, United Kingdom.

In addition, individual members wish to acknowledge support from CERN: European Organization for Nuclear Research (CERN PJAS); Chile: Agencia Nacional de Investigaci\'on y Desarrollo (FONDECYT 1190886, FONDECYT 1210400, FONDECYT 1230812, FONDECYT 1230987); China: Chinese Ministry of Science and Technology (MOST-2023YFA1605700), National Natural Science Foundation of China (NSFC - 12175119, NSFC 12275265, NSFC-12075060); Czech Republic: Czech Science Foundation (GACR - 24-11373S), Ministry of Education Youth and Sports (FORTE CZ.02.01.01/00/22\_008/0004632), PRIMUS Research Programme (PRIMUS/21/SCI/017); EU: H2020 European Research Council (ERC - 101002463); European Union: European Research Council (ERC - 948254, ERC 101089007), Horizon 2020 Framework Programme (MUCCA - CHIST-ERA-19-XAI-00), European Union, Future Artificial Intelligence Research (FAIR-NextGenerationEU PE00000013), Italian Center for High Performance Computing, Big Data and Quantum Computing (ICSC, NextGenerationEU); France: Agence Nationale de la Recherche (ANR-20-CE31-0013, ANR-21-CE31-0013, ANR-21-CE31-0022, ANR-22-EDIR-0002), Investissements d'Avenir Labex (ANR-11-LABX-0012); Germany: Baden-Württemberg Stiftung (BW Stiftung-Postdoc Eliteprogramme), Deutsche Forschungsgemeinschaft (DFG - 469666862, DFG - CR 312/5-2); Italy: Istituto Nazionale di Fisica Nucleare (ICSC, NextGenerationEU), Ministero dell'Università e della Ricerca (PRIN - 20223N7F8K - PNRR M4.C2.1.1); Japan: Japan Society for the Promotion of Science (JSPS KAKENHI JP21H05085, JSPS KAKENHI JP22H01227, JSPS KAKENHI JP22H04944, JSPS KAKENHI JP22KK0227); Netherlands: Netherlands Organisation for Scientific Research (NWO Veni 2020 - VI.Veni.202.179); Norway: Research Council of Norway (RCN-314472); Poland: Polish National Agency for Academic Exchange (PPN/PPO/2020/1/00002/U/00001), Polish National Science Centre (NCN 2021/42/E/ST2/00350, NCN OPUS nr 2022/47/B/ST2/03059, NCN UMO-2019/34/E/ST2/00393, NCN \& H2020 MSCA 945339, UMO-2020/37/B/ST2/01043, UMO-2021/40/C/ST2/00187, UMO-2022/47/O/ST2/00148, UMO-2023/49/B/ST2/04085); Slovenia: Slovenian Research Agency (ARIS grant J1-3010); Spain: Generalitat Valenciana (Artemisa, FEDER, IDIFEDER/2018/048), Ministry of Science and Innovation (MCIN \& NextGenEU PCI2022-135018-2, MICIN \& FEDER PID2021-125273NB, RYC2019-028510-I, RYC2020-030254-I, RYC2021-031273-I, RYC2022-038164-I), PROMETEO and GenT Programmes Generalitat Valenciana (CIDEGENT/2019/023, CIDEGENT/2019/027); Sweden: Swedish Research Council (Swedish Research Council 2023-04654, VR 2018-00482, VR 2022-03845, VR 2022-04683, VR 2023-03403, VR grant 2021-03651), Knut and Alice Wallenberg Foundation (KAW 2018.0157, KAW 2018.0458, KAW 2019.0447, KAW 2022.0358); Switzerland: Swiss National Science Foundation (SNSF - PCEFP2\_194658); United Kingdom: Leverhulme Trust (Leverhulme Trust RPG-2020-004), Royal Society (NIF-R1-231091); United States of America: U.S. Department of Energy (ECA DE-AC02-76SF00515), Neubauer Family Foundation.

%
%


%
%
%
%
%
%

%
%
%
\printbibliography
\clearpage
 
\begin{flushleft}
\hypersetup{urlcolor=black}
{\Large The ATLAS Collaboration}

\bigskip

\AtlasOrcid[0000-0002-6665-4934]{G.~Aad}$^\textrm{\scriptsize 103}$,
\AtlasOrcid[0000-0001-7616-1554]{E.~Aakvaag}$^\textrm{\scriptsize 16}$,
\AtlasOrcid[0000-0002-5888-2734]{B.~Abbott}$^\textrm{\scriptsize 121}$,
\AtlasOrcid[0000-0002-0287-5869]{S.~Abdelhameed}$^\textrm{\scriptsize 117a}$,
\AtlasOrcid[0000-0002-1002-1652]{K.~Abeling}$^\textrm{\scriptsize 55}$,
\AtlasOrcid[0000-0001-5763-2760]{N.J.~Abicht}$^\textrm{\scriptsize 49}$,
\AtlasOrcid[0000-0002-8496-9294]{S.H.~Abidi}$^\textrm{\scriptsize 29}$,
\AtlasOrcid[0009-0003-6578-220X]{M.~Aboelela}$^\textrm{\scriptsize 44}$,
\AtlasOrcid[0000-0002-9987-2292]{A.~Aboulhorma}$^\textrm{\scriptsize 35e}$,
\AtlasOrcid[0000-0001-5329-6640]{H.~Abramowicz}$^\textrm{\scriptsize 153}$
% ...
\bigskip
\\

% ...
$^{16}$Department for Physics and Technology, University of Bergen, Bergen; Norway.\\
% ...
$^{29}$Physics Department, Brookhaven National Laboratory, Upton NY; United States of America.\\
% ...
$^{35}$$^{(a)}$Facult\'e des Sciences Ain Chock, Universit\'e Hassan II de Casablanca;$^{(b)}$Facult\'{e} des Sciences, Universit\'{e} Ibn-Tofail, K\'{e}nitra;$^{(c)}$Facult\'e des Sciences Semlalia, Universit\'e Cadi Ayyad, LPHEA-Marrakech;$^{(d)}$LPMR, Facult\'e des Sciences, Universit\'e Mohamed Premier, Oujda;$^{(e)}$Facult\'e des sciences, Universit\'e Mohammed V, Rabat;$^{(f)}$Institute of Applied Physics, Mohammed VI Polytechnic University, Ben Guerir; Morocco.\\
% ...
$^{44}$Physics Department, Southern Methodist University, Dallas TX; United States of America.\\
% ...
$^{49}$Fakult\"{a}t Physik , Technische Universit{\"a}t Dortmund, Dortmund; Germany.\\
% ...
$^{55}$II. Physikalisches Institut, Georg-August-Universit\"{a}t G\"ottingen, G\"ottingen; Germany.\\
% ...
$^{103}$CPPM, Aix-Marseille Universit\'e, CNRS/IN2P3, Marseille; France.\\
% ...
$^{117}$$^{(a)}$New York University Abu Dhabi, Abu Dhabi;$^{(b)}$United Arab Emirates University, Al Ain; United Arab Emirates.\\
% ...
$^{121}$Homer L. Dodge Department of Physics and Astronomy, University of Oklahoma, Norman OK; United States of America.\\
% ...
$^{153}$Raymond and Beverly Sackler School of Physics and Astronomy, Tel Aviv University, Tel Aviv; Israel.\\
% ...
$^{a}$ Also Affiliated with an institute covered by a cooperation agreement with CERN.\\
$^{b}$ Also at An-Najah National University, Nablus; Palestine.\\
$^{c}$ Also at Borough of Manhattan Community College, City University of New York, New York NY; United States of America.\\
$^{d}$ Also at Center for Interdisciplinary Research and Innovation (CIRI-AUTH), Thessaloniki; Greece.\\
$^{e}$ Also at Centro Studi e Ricerche Enrico Fermi; Italy.\\
$^{f}$ Also at CERN, Geneva; Switzerland.\\
% ...

\end{flushleft}


%

%
%
%
%

%
%
%
%

\end{document}